\documentclass[%
 reprint,
nofootinbib,
nobibnotes,
amsmath,amssymb,twocolumns,
]{revtex4-2}
\usepackage{graphicx}
\usepackage{dcolumn}
\usepackage{bm}
\usepackage[caption=false]{subfig}
\usepackage{comment,xcolor}
\usepackage{paralist, tabularx}
\usepackage{amssymb}
\usepackage{ragged2e}

\newcommand{\avg}[1]{\left\langle #1 \right\rangle}

\usepackage{float}
\usepackage{babel}
\usepackage{hyperref}
\usepackage[noabbrev]{cleveref}

\begin{document}
\crefformat{equation}{(#2#1#3)}
\crefrangeformat{equation}{(#3#1#4) to~(#5#2#6)}
\crefmultiformat{equation}{(#2#1#3)}
{ and~(#2#1#3)}{, (#2#1#3)}{ and~(#2#1#3)}

\title{Coexistence of Competing Microbial Strains under Twofold Environmental Variability and Demographic Fluctuations}

\author{Matthew Asker}\email{mmmwa@leeds.ac.uk}
\author{Llu\'is Hern\'andez-Navarro}
\author{Alastair M. Rucklidge}
\author{Mauro Mobilia}\email{M.Mobilia@leeds.ac.uk}
\affiliation{
 Department of Applied Mathematics, School of Mathematics, University of Leeds, Leeds LS2 9JT, United Kingdom\\
}
\homepage{https://eedfp.com}

\date{\today}

\begin{abstract}
Microbial populations generally evolve in volatile environments, 
under conditions fluctuating between harsh and mild, e.g. as the result of sudden changes in toxin concentration or nutrient abundance. Environmental variability 
thus shapes the long-time population dynamics, notably by  influencing the ability of different strains of microorganisms to coexist.
Inspired by the evolution of antimicrobial resistance, we study  the dynamics 
of a community  consisting of two competing strains subject to twofold environmental variability. The level of toxin varies in time, favouring the growth of one strain 
under low {drug concentration} and the other strain when the toxin level is high. 
We also model time-changing  resource abundance
by a randomly switching carrying capacity that drives the fluctuating size of the community. While one strain dominates in a static environment, we show that species 
coexistence  is possible in the presence of
environmental variability. By  computational and analytical means, we determine the environmental conditions under which long-lived coexistence is possible and when it is almost certain. Notably, we study the circumstances under which environmental and demographic
fluctuations promote, or hinder, the strains coexistence. We also determine how the make-up
of the coexistence phase and the average abundance of each strain
depend on the environmental variability.
\end{abstract}
\maketitle
\section{Introduction}\label{sec:introduction}
Microbial communities evolve in volatile environments that often fluctuate between mild and harsh conditions. For instance,  the  concentration of toxin and the abundance of nutrients
in a community can suddenly and radically change~\cite{vasi_long-term_1994,proft2009microbial,himeoka_dynamics_2020,Tu20}. This results 
in environmental variability (EV) that  shapes the population
evolution~\cite{hooper_effects_2005,fux_survival_2005,brockhurst_cooperation_2007,acar_stochastic_2008,caporaso_moving_2011,Kerr13,Lambert15,rescan_phenotypic_2020,Nguyen21}. In particular, EV greatly influences the ability of species to coexist~\cite{chesson_environmental_1981,Chesson94,Chesson00a,Chesson00b,Barabas18,Abdul21}, which is a characteristic of key importance in biology and ecology, with direct  applications 
in subjects of great societal concern~\cite{eliopoulos_impact_2003,Pennisi09,hubbell_2013,oneill_tackling_2016,dadgostar_antimicrobial_2019,Murugan21} like the maintenance of biodiversity in ecosystems~\cite{kalyuzhny_neutral_2015,Mitri16,Grilli20,meyer_evolutionary_2020,meyer_species_2021,Leibold19,Pinsky19,West22,Gore22} and the evolution of antimicrobial resistance (AMR)~\cite{balaban_bacterial_2004,Yurtsev13,Raymond16,Kussell16,Lopatkin17,coates_antibiotic-induced_2018}.
%

In the absence of detailed
knowledge about the time-variation of external factors, EV is generally
modelled by means of noise terms affecting the species growth and/or death rates~\cite{Ewens,Kimura,Karlin74,Thattai04,Kussel05,Loreau08,He10,Visco10,Assaf13a,Dobra13,
Assaf13b,Chisholm14,ashcroft_fixation_2014,melbinger_emergence_2015,kalyuzhny_neutral_2015,roberts_dynamics_2015,hufton_intrinsic_2016,wienand_evolution_2017,Xue17,Dobra18,marrec_quantifying_2018,wienand_eco-evolutionary_2018,west2020,taitelbaum_population_2020,marrec_resist_2020,shibasaki_exclusion_2021,taitelbaum2023evolutionary}. Demographic noise (DN) is another important source of fluctuations: it can lead to fixation, which is the phenomenon arising when one strain takes over the entire community. The effect of DN is significant in  communities of small size, and becomes  negligible in large populations~\cite{Ewens,Kimura,Nowak,blythe_stochastic_2007,traulsen_stochastic_2008}.
Significantly, the time development  of the size and composition of 
populations are often interdependent~\cite{Roughgarden,chuang_simpsons_2009,chuang_cooperation_2010,melbinger_evolutionary_2010,cremer_growth_2012,sanchez_feedback_2013,Gokhale16,Tilman20,Plotkin23}, with fluctuations of the population size modulating the  strength of DN~\cite{wienand_evolution_2017,wienand_eco-evolutionary_2018,west2020,taitelbaum_population_2020,shibasaki_exclusion_2021,taitelbaum2023evolutionary,LARM23}.
 The interplay between EV and DN is crucial in shaping microbial communities, but the quantitative effects of their coupling
are as yet still mostly unknown. 

{
Environmental and demographic fluctuations play a crucial role in the evolution of AMR~\cite{Kussell16,coates_antibiotic-induced_2018,marrec_quantifying_2018,marrec_resist_2020,Nguyen21}. 
When treatments reduce a microbial community to a
very small size, but fail to eradicate the microorganisms resistant to the drugs, resistant cells
may replicate and restore infection, hence possibly leading to the spread of antibiotic resistance.
Moreover, within a small population, demographic fluctuations may also lead to the extinction of
resistant cells. Understanding how the coexistence of cells resistant and sensitive to
antibiotics is affected by the joint effect of environmental variability and demographic
fluctuations, and how the fraction of resistant cells varies with environmental conditions with the possibility of eradication, are thus central questions in the effort to understand the evolution of AMR~\cite{Yurtsev13,Kussell16,coates_antibiotic-induced_2018,Nguyen21,LARM23}.}


{
It is worth noting that considerable efforts have recently been dedicated to study
the mechanisms underpinning the coexistence of competing species under various scenarios, see e.g.
\cite{Leibold19,Pinsky19,shibasaki_exclusion_2021,Gore22}. 
The influence of
  different kinds of variability (e.g. quenched disorder, heterogeneous rates, ``spillover protection'') on species coexistence in ecosystems~
exhibiting cyclic dominance~\cite{Szolnoki14,Dobra18}
   has notably been investigated  in
\cite{Perc_2007,He10,he2011,Kelsic_2015,Szolnoki_2015,Szolnoki2016}.
} 
Here, inspired by the AMR evolution in a chemostat setup~\cite{shibasaki_exclusion_2021,Abdul21}, 
we study the eco-evolutionary dynamics 
of an idealised microbial community consisting of two competing strains subject to a  time-varying level of toxin, with the growth of one strain favoured 
under low toxin level and a selective advantage to the other strain under high toxin level. We also assume that the resource abundance  varies according to a time-switching 
carrying capacity that drives the fluctuating size of the community. 
In  most of previous works,  EV is either encoded in fluctuating growth rates, with the size or carrying capacity of the population kept constant 
~\cite{Ewens,Kimura,Karlin74,Thattai04,Kussel05,Kussell05b,He10,Visco10,AMR13b,Assaf13a,Dobra13,
Assaf13b,ashcroft_fixation_2014,melbinger_emergence_2015,kalyuzhny_neutral_2015,roberts_dynamics_2015,hufton_intrinsic_2016,Xue17,marrec_quantifying_2018,marrec_resist_2020}, or EV is modelled
by a time-varying carrying capacity that affects the species death rates and drives the population size~\cite{wienand_evolution_2017,wienand_eco-evolutionary_2018,west2020,taitelbaum_population_2020,taitelbaum2023evolutionary} (see also \cite{shibasaki_exclusion_2021}). The distinctive feature of this study is therefore the \emph{twofold environmental variability} 
accounting for fluctuations stemming from the variation of the 
toxin level and the switches of the
carrying capacity resulting in  the coupling of DN and EV; see Fig.~\ref{fig:cartoon}.\\
{As main results,  we obtain} the fixation-coexistence diagrams of the system, and these allow us to determine the environmental conditions under which long-lived coexistence of the strains is possible or certain, and when one strain dominates the other. We also analyse the  make-up
of the population when the strains coexist, and their average abundance. 

The organisation of the paper is as follows: the model is introduced in Sec.~\ref{sec:model}. Sec.~\ref{sec:Moran} is dedicated to the 
study of the case with a constant carrying capacity (subject to a static or varying toxin level) by means of a mean-field analysis and a mapping onto a suitable Moran process. The twofold influence of time-varying fitness and  carrying capacity on the 
coexistence and fixation of the species is analysed in Sec.~\ref{sec:switchingK}.  Sec.~\ref{sec:V} is dedicated to the influence of the EV on the make-up of the coexistence phase and strains abundance. We present our conclusions in Sec.~\ref{sec:VI}. Additional technical details are given in the appendix.

\section{\label{sec:model}Model}
\begin{figure}[t]
    \centering
    \includegraphics[width=0.95\linewidth]{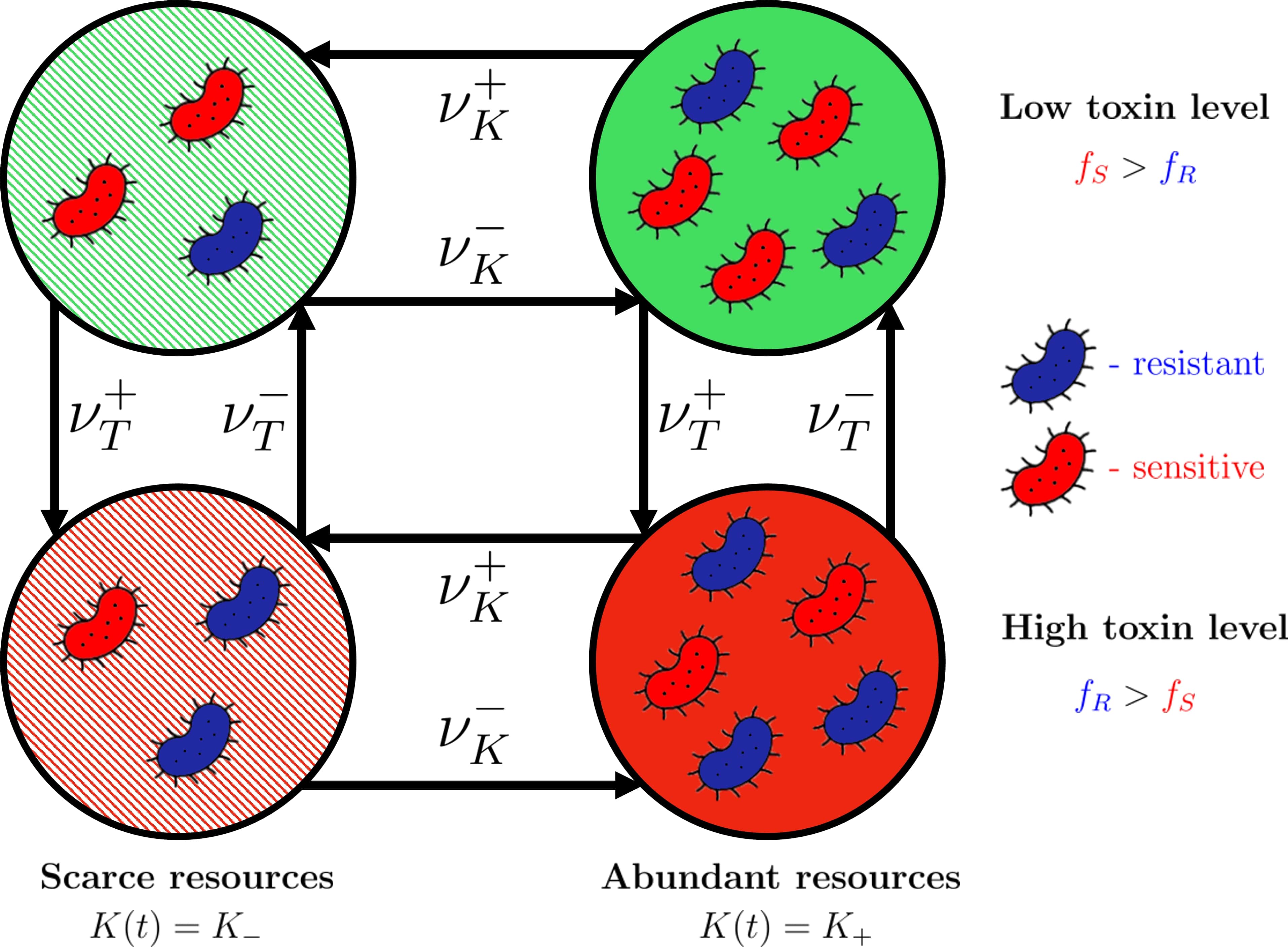}
    \caption{Cartoon of the model characterised by twofold EV. A microbial community 
    consisting of two strains, denoted by $R$ (resistant, blue (charcoal), fitness $f_R$) and $S$ (sensitive, red (grey), fitness $f_S$), 
    evolves in a time-varying environment, as illustrated by the arrows: 
    the level of toxin, $\xi_T$, stochastically switches  with rates $\nu_T^{\pm}$ (vertical arrows) between low ($\xi_T=+1$) and high ($\xi_T=-1$), and the amount of available resources, modelled by the carrying capacity $K(t)$,
    stochastically switches with rates $\nu_K^{\pm}$ (horizontal arrows) between scarce  and abundant. The strain $R$
     fares better than $S$ under high toxin level ($f_R>f_S$), while $S$ grows faster under low   toxin level ($f_R<f_S$). The carrying capacity $K(t)=K_+$ when there are abundant resources, while
     $K(t)=K_-<K_+$ when available nutrients are scarce. The four environmental states characterising the twofold EV 
     are indicated by the coloured background:
     striped / solid green (light grey) refers to low toxin level and scarce / abundant resources, while striped / solid red (grey) indicates high toxin level with scarce / abundant resources. {See the text for the greyscale coding, and the table in Sec.~SM1 of the appendix for detailed notations and definitions.}}
    \label{fig:cartoon}
\end{figure}
We consider a well-mixed population of fluctuating size $N(t)=N_R(t)+N_S(t)$ which, at time~$t$, consists of $N_R$ bacteria of strain $R$ and $N_S$ of type $S$, which compete for the same resources. The former refers to a strain {that pays a metabolic cost to be} resistant to a certain toxin, and the latter to microorganisms sensitive to that toxin. 
 Based on mounting evidence showing  that microbial communities
 generally evolve in volatile environments~\cite{hooper_effects_2005,fux_survival_2005,caporaso_moving_2011,rescan_phenotypic_2020,Murugan21}, we study the eco-evolutionary dynamics of this population under \emph{twofold environmental variability}: external conditions fluctuate between harsh and mild, and affect 
 the level of toxin and  resources that are available in the population; see Fig.~\ref{fig:cartoon}. 
 
 For concreteness, we assume that  
the toxin is biostatic and reduces the growth rate of the sensitive strain, but does not affect the resistant bacteria~\cite{bernatova_following_2013,pankey_clinical_2004,nemeth_bacteriostatic_2015}\footnote{The case where the toxin increases the death rate of the  strain $S$ corresponds to a biocidal toxin, and is not directly considered here. This is not particularly limiting since the same drug can often act as a biostatic or biocidal toxin at low/high concentration~\cite{sanmillan2017fitness}.}.
 In this setting, resistant $R$ bacteria have
 a constant fitness $f_R$, whereas the sensitive $S$ bacteria have an environment-dependent fitness $f_S(\xi_T)$,
 where $\xi_T(t)$ is a time-varying \emph{environmental random variable} encoding the toxin level: $\xi_T>0$ for the low toxin level and $\xi_T < 0$ for the high toxin level. As in previous studies~\cite{danino_fixation_2018,danino_stability_2018,meyer_evolutionary_2020,meyer_species_2021,danino_effect_2016}, we here consider 
 \begin{align}
  \label{eq:fit}
  f_R=1 \qquad\text{and} \qquad  f_S=\exp(s \xi_T),
  \nonumber
 \end{align}
 where $s>0$ denotes the selection bias favouring the strain $S$ when $\xi_T>0$, and strain $R$ when $\xi_T<0$. The parameter $s$ therefore 
 encodes both the selection  and 
 strength of the environmental variability associated with the changes in toxin level ($T$-EV).
  As in many recent theoretical studies \cite{wienand_evolution_2017,hidalgo_species_2017,danino_fixation_2018,
  wienand_eco-evolutionary_2018,taitelbaum_population_2020,west2020,shibasaki_exclusion_2021,hufton_intrinsic_2016}, 
  $T$-EV  is here modelled by coloured
  dichotomous Markov noise (DMN)~\cite{horsthemke_lefever,bena2006,Ridolfi2011}, so that $\xi_T\in\{-1,1\}$; see below. {DMN is an important  example of bounded  noise~\cite{horsthemke_lefever,bena2006,Ridolfi2011,hufton_intrinsic_2016,wienand_eco-evolutionary_2018,taitelbaum_population_2020,west2020,shibasaki_exclusion_2021,hidalgo_species_2017}, with finite correlation time, which 
  allows us to
  efficiently model suddenly changing conditions occurring in bacterial life~\cite{Kussell05b,acar_stochastic_2008,chuang_simpsons_2009,chuang_cooperation_2010,Abdul21}, like the environmental stress resulting from exposure to antibiotics~\cite{sanchez_feedback_2013,Kerr13,Lambert15}. In this context, DMN,  that is easy to simulate accurately  and whose relationships with more complex forms of noise have been studied (see Sec.~SM4 in the appedix and ~\cite{horsthemke_lefever,bena2006,taitelbaum2023evolutionary}),  
  provides us with mathematically-amenable models that can be regarded as the
  theoretical counterparts of commonly-used  laboratory experimental chemostat set-ups~\cite{acar_stochastic_2008,Abdul21,Nguyen21,Lambert15}.}
  
  %
  
 The environmental effect on the level of nutrients ($K$-EV), fluctuating between scarcity and abundance, is  modelled by a binary switching carrying capacity $K(t)\in\{K_-,K_+\}$
 that is driven by the binary random variable (also following a DMN process) $\xi_K(t)\in\{-1,1\}$. The state $\xi_K=-1$ thus corresponds to a harsh state with scare resources, where the carrying capacity is $K_-$, whereas nutrients are abundant in the mild state $\xi_K=+1$ where the carrying capacity is $K_+> K_-\gg 1$. As in \cite{wienand_evolution_2017, wienand_eco-evolutionary_2018, taitelbaum_population_2020,west2020,shibasaki_exclusion_2021,taitelbaum2023evolutionary}, this is encoded in the time-switching carrying capacity
 \begin{equation}
K(t)=\frac{1}{2}\left[K_+ +K_- +\xi_K(t)(K_+ - K_-)\right],
   \label{eq:carryingcapacity}
\end{equation}
which, with $K_0 \equiv \frac{K_+ + K_-}{2}$ and $\gamma \equiv \frac{K_+ - K_-}{2K_0}$,
  can conveniently be  written as \[
 K(t)=K_0[1 +\gamma \xi_K(t)].\] This randomly switching carrying capacity 
 drives the population size $N(t)$ and is hence responsible for its fluctuations, with $K(t)\gg 1$ ensuring that the population dynamics is never solely dominated by demographic fluctuations. {It is worth noting the choice of  $\xi_K$  as a DMN ensures that the carrying  capacity is always bounded and physical, i.e. $K(t)\in \{K_-, K_+\}$ and $K(t)>0$, which is not  the case if EV is modelled by unbounded  (e.g. Gaussian)  noise, see Sec.~SM4 in the appendix.}

 The population thus evolves under twofold EV encoded in the
 environmental states ${\xi_{T}(t),\xi_K(t)}$, see Fig.~\ref{fig:cartoon}, subject to 
time switching according to the reactions  
\begin{equation}
\label{eq:switchingdefinition}
    \xi_{\alpha}=+1 \xrightarrow{\nu_{\alpha}^+} -1, \hspace{0.5cm} \text{and} \hspace{0.5cm} 
    \xi_{\alpha}=-1 \xrightarrow{\nu_{\alpha}^-} +1,
  \nonumber
\end{equation}
where $\nu_{\alpha}^{\pm}$ are the switching rates of the $\alpha$-DMN, with $\alpha\in \{T,K\}$ indicating the relevant environmental noise. 
It is also useful to define the average switching rate 
$\nu_\alpha$ and  switching bias $\delta_\alpha$ for each 
 $\alpha$-DMN as
\[\nu_\alpha\equiv  \frac{\nu_\alpha^- + \nu_\alpha^+}{2}
\hspace{0.5cm} \text{and} \hspace{0.5cm} 
\delta_\alpha \equiv  \frac{\nu_\alpha^- - \nu_\alpha^+}{2\nu_\alpha},
\] such that $\nu_\alpha^\pm\equiv\nu_\alpha(1\mp\delta_\alpha)$.
This means that $\delta_T>0$ corresponds to a bias towards  low toxin level (mild $T$ state, $\xi_T=+1$) favouring the $S$ strain, whereas $\delta_T<0$ indicates a bias towards high toxin level
  (harsh $T$ state, $\xi_T=-1$) where  the growth of $S$ is hampered and the spread of $R$ is favoured, see Fig.~\ref{fig:cartoon}. Similarly, $\delta_K>0$ corresponds to  
bias towards the  environmental state rich in nutrients (where $K=K_+$), while 
$\delta_K<0$ is associated with a bias towards an environment where nutrients are scarce ($K=K_-$). In all cases, we consider  $\alpha$-DMN at stationarity, where 
$\xi_{\alpha}=\pm 1$ with probability $(1\pm \delta_{\alpha})/2$, 
yielding the average $\avg{\xi_{\alpha}}=\delta_{\alpha}$ and  
 autocovariance (autocorrelation up to a constant)  $\avg{\xi_{\alpha}(t)\xi_{\alpha}(t')} - \avg{\xi_\alpha(t)}\avg{\xi_\alpha(t')}  = (1-\delta_{\alpha}^2)\exp(-2\nu_{\alpha}|t-t'|)$, where $\avg{\cdot} $ denotes the $\alpha$-DMN ensemble average. {We notice that 
  the correlation time of the ${\alpha}$-DMN,  $1/(2\nu_{\alpha})$, 
is half of
 the inverse of the average switching rate of $\xi_{\alpha}$}~\cite{horsthemke_lefever,bena2006,Ridolfi2011}. 
From Eq.~\cref{eq:carryingcapacity}, we find that the average carrying capacity is $\avg{K}=K_0(1+\gamma\delta_K)$
and the variance of $K(t)$
is $(K_0\gamma)^2(1-\delta_K^2)$, with the amplitude of $K$-EV thus scaling as $K_0\gamma$, while the variance and amplitude of the $T$-EV increase with $s$; see Sec.~SM2 the appendix. 

 For concreteness, we here assume that 
 $\xi_T$ and $\xi_K$ are totally uncorrelated. In our motivating example, this corresponds to the reasonable assumption that nutrient and antibiotic uptake are independent processes. The case where $\xi_T$ and $\xi_K$ are fully correlated or fully anti-correlated, with $\xi_T=\xi_K=\xi$ or $\xi_T=-\xi_K=\xi$, where $\xi$ is a single DMN process, is briefly discussed in Sec.~SM8 of the appendix.
 
The system considered here best translates to a chemostat setup whereby toxin  and nutrient levels can be maintained at a constant level through time and switched via changing concentrations of medium coming into the system~\cite{shibasaki_exclusion_2021,Abdul21}. The switch $\xi_T\to -\xi_T$ with $\xi_T=-1$ can thus be envisioned as corresponding to switching the concentration of an antibiotic drug from 
above  the  minimum inhibitory concentration (MIC),
where the growth of the sensitive strain is hampered, to a concentration below the MIC where the $S$ strain can spread at the expense of $R$~\cite{Yurtsev13,Kussell16,marrec_resist_2020}.
 
At time $t$ the fraction of $R$-types in the system is $x(t)\equiv \frac{N_R(t)}{N(t)}$ and the average population fitness is 
$\overline{f}(x,\xi_T)\equiv x+(1-x)\exp(s\xi_T)$, which depends on the population composition $x$ and the toxin state $\xi_T$, . We assume that mutation rates between strains are negligible, and seek to characterise the population dynamics by the evolution of its size and composition  according to the multivariate birth-death process~\cite{Gardiner,VanKampen,Pinsky2011-px}
\begin{equation}
\label{eq:populationsizechange}
    N_{R/S} \xrightarrow{T_{R/S}^+} N_{R/S} + 1 \hspace{0.2cm} \text{and} \hspace{0.2cm} N_{R/S} \xrightarrow{T_{R/S}^-} N_{R/S} - 1,
\end{equation}
where
the time-dependent birth and death transition rates are respectively
\begin{equation}
\label{eq:transitionrates}
    T_{R/S}^{+} = \frac{f_{R/S}}{\overline{f}}N_{R/S} \hspace{0.2cm} \text{and} \hspace{0.2cm} T_{R/S}^{-}= \frac{N}{K} N_{R/S}.
\end{equation}
The per-capita birth rates $f_{R/S}/\overline{f}$ (where we normalise with $\overline{f}$ in line with the standard Moran process~\cite{antal_fixation_2006,Nowak,blythe_stochastic_2007,Ewens,traulsen_stochastic_2008}) thus vary with the toxin level 
and population composition, while the logistic-like per-capita death rate 
$N/K$ varies with nutrient level  and population size.
With $\mathbf{N}\equiv(N_R,N_S)$,
the  master equation  
giving the probability $P(\mathbf{N},\xi_T,\xi_K,t)$ for the population to  consist of  $N_R$ and $N_S$
bacteria of type $R$ and $S$, respectively,
in
the environmental state $(\xi_T,\xi_K)$ at time $t$  is
\begin{equation}
\begin{aligned}
\label{eq:ME}
\hspace{-5mm}
&\frac{\partial P(\mathbf{N},\xi_T,\xi_K,t)}{\partial t} =  \left( \mathbb{E}_R^--1\right)\left[T^+_R P(\mathbf{N},\xi_T,\xi_K,t)\right]\\
&+\left( \mathbb{E}_S^--1\right)\left[T^+_S P(\mathbf{N},\xi_T,\xi_K,t)\right]\\
&+
\left( \mathbb{E}_R^+-1\right)\left[T^-_R P(\mathbf{N},\xi_T,\xi_K,t)\right]\\ 
&+\left( \mathbb{E}_S^+-1\right)\left[T^-_S P(\mathbf{N},\xi_T,\xi_K,t)\right]\\
&+ \nu_T^{-\xi_T} P(\mathbf{N},-\xi_T,\xi_K,t)-\nu_T^{\xi_T} P(\mathbf{N},\xi_T,\xi_K,t)\\
&+ \nu_K^{-\xi_K} P(\mathbf{N},\xi_T,-\xi_K,t)-\nu_K^{\xi_K} P(\mathbf{N},\xi_T,\xi_K,t),
\end{aligned}
\end{equation}
where $\mathbb{E}^{\pm}_{R/S}$  are shift operators such that
$\mathbb{E}^{\pm}_R f(N_R,N_S,\xi_T,\xi_K,t) =f(N_R\pm 1,N_S,\xi_T,\xi_K,t)$, and
$\nu_\alpha^{\xi_\alpha}\equiv \nu_\alpha^{\pm}$ when $\xi_\alpha=\pm 1$. We note that $P(\mathbf{N},\xi_T,\xi_K,t)=0$ whenever $N_R<0$ or $N_S<0$, and the last two lines on the right-hand-side of Eq.~\cref{eq:ME}
account for the random environmental switching of toxin ($\xi_T\to -\xi_T$) and carrying capacity ($\xi_K\to -\xi_K$). The dynamics encoded by the multivariate master equation Eq. (\ref{eq:ME}) can be simulated exactly by a stochastic algorithm as explained in \cite{gibson_efficient_2000, anderson_modified_2007}, see Sec. SM4 in the appendix.
Since  $T^{\pm}_{R/S}=0$ whenever $N_{R/S}=0$, there is extinction of $R$ and  fixation
 of $S$ ($N_R=0, N=N_S$), or  fixation of $R$ and extinction of $S$ ($N_S=0, N=N_R$). When one strain fixates and replaces the other, the population composition no longer changes while its size continues to fluctuate\footnote{
 Finally, the population will settle in the absorbing state $N_R=N_S=0$ corresponding to the eventual  extinction  of the 
 entire population. This occurs after a time that grows exponentially with the system size~\cite{Spalding17,wienand_evolution_2017,wienand_eco-evolutionary_2018, taitelbaum_population_2020}. This phenomenon, irrelevant for our purposes (since we always have $K(t)\gg 1$), is not considered here.
 }.
 Fixation of one strain and extinction of the other is expected
 when strains  compete for the same resources (competitive exclusion principle), and always occur in a finite population even when its size fluctuates~\cite{wienand_evolution_2017,wienand_eco-evolutionary_2018, taitelbaum_population_2020}. In stark contrast, here we show that environmental fluctuations can lead to the long-lived coexistence of competing species
 and nontrivially shape the abundance distribution of both strains.

{The notation and definitions of the model parameters and physical quantities of interest are conveniently summarised in the table of Sec.~SM1 in the appendix.}
\section{\label{sec:Moran}
Constant carrying capacity: mean-field analysis and 
Moran process}

Since $\xi_T$ and $\xi_K$ are independent, 
it is useful to first consider the case of a constant
carrying capacity, with environmental variability stemming only from the fluctuations of the toxin level in the birth rates of Eq.~\eqref{eq:transitionrates}.

 In this section, we thus assume that the carrying capacity is constant and large:  $K(t)=K_0\gg 1$. After a short transient the population size
fluctuates about $K_0$, with $N\approx K_0$. When $K_0\gg 1$, we can approximate the population size by $N=K_0$
and make analytical progress by using the well-known results of the Moran process~\cite{antal_fixation_2006,Nowak,blythe_stochastic_2007,Ewens,traulsen_stochastic_2008}.
In this approximation, the population is kept constant, which requires the simultaneous birth and death
of individuals of either species, and the population evolves according to a fitness-dependent Moran process \cite{wienand_evolution_2017,wienand_eco-evolutionary_2018,west2020,taitelbaum_population_2020}, defined in terms of Eqs.~\cref{eq:populationsizechange,eq:transitionrates} by the reactions
\begin{equation}
\begin{aligned}
\label{eq:Moran}
    (N_{R},N_{S})  &\xrightarrow{\widetilde{T}_{R}^+} (N_{R} + 1, N_{S} - 1),\\
    (N_{R},N_{S}) &\xrightarrow{\widetilde{T}_R^-} (N_{R} - 1,N_{S} + 1),
\end{aligned}
\end{equation}
corresponding, respectively, to the simultaneous birth of an $R$ and death of an $S$ with 
rate  $\widetilde{T}_{R}^+$, and death of an $R$ and birth of an $S$ with 
rate  $\widetilde{T}_{R}^-$, where
\begin{equation}
\begin{aligned}
\label{eq:Moranrates}
    \widetilde{T}_{R}^+ &\equiv \frac{T_R^{+} T_S^{-}}{N}=Nx(1-x)\frac{f_R}{\overline{f}(t)},\\
    \widetilde{T}_{R}^-&\equiv\frac{T_R^{-} T_S^{+}}{N}=Nx(1-x)\frac{f_S(t)}{\overline{f}(t)}.
\end{aligned}
\end{equation}
\subsection{Mean-field analysis}
\label{sec:MF}
We now consider the case where $N=K_0\to \infty$, and thus ignore  demographic fluctuations. In this case, the population composition evolves according to the mean-field equation \cite{Gardiner}:

\begin{equation}
 \label{eq:MF}
 \dot{x}= \frac{\widetilde{T}_{R}^+ - \widetilde{T}_{R}^-}{N}
 =x(1-x)\left(\frac{1-e^{s\xi_T}}{x+(1-x)e^{s\xi_T}}\right),
\end{equation}
where the dot denotes the time derivative. It is important to notice that, owing to 
environmental noise
$\xi_T $,
Eq.~\eqref{eq:MF} is a mean-field \emph{stochastic} differential equation that defines a so-called ``piecewise deterministic Markov process'' (PDMP)~\cite{davis_piecewise-deterministic_1984,hufton_intrinsic_2016,wienand_eco-evolutionary_2018,wienand_evolution_2017,taitelbaum_population_2020}.
 According to this PDMP, 
after a switch to an environmental state $\xi_T$,
$x$ evolves deterministically with  
Eq.~\eqref{eq:MF} and a fixed value of  $\xi_T$,
until a switch $\xi_T\to-\xi_T$ occurs, see Sec.~\ref{sec:Va}. 

We consider Eq.~\eqref{eq:MF} in the  regimes of (i) low, (ii) high, and (iii) intermediate  switching rate $\nu_T$:\\
(i) Under low switching rate, $\nu_T\to 0$, the population settles in its final state without experiencing any $T$-switches. In this regime, 
the population reaches its final state in its initial toxin level $\xi_T(0)$, i.e.  $\xi_T(0)=\xi_T(\infty)=\pm 1$ with probability $(1\pm \delta_T)/2$. In this regime,
Eq.~\eqref{eq:MF} thus boils down to
\begin{equation}
 \dot{x}
 = \begin{cases}
    -\frac{x(1-x) (e^{s}-1)}{x+(1-x)e^{s}}& \text{ with probability } \frac{1+\delta_T}{2}\\
    \frac{x(1-x)(1-e^{-s})}{x+(1-x)e^{-s}}& \text{ with probability } \frac{1-\delta_T}{2}.
   \end{cases}
   \nonumber
\end{equation}
Since $s>0$, 
with a probability $(1+ \delta_T)/2$ we have $\xi_T(0)=\xi_T(\infty)=+1$ and $x\to 0$
 ($R$ vanishes), while with a probability $(1- \delta_T)/2$ we have $\xi_T(0)=\xi_T(\infty)=-1$ and  $x\to 1$  ($S$ vanishes). In either case, the mean-field dynamics are characterised by the dominance of one of the strains. Therefore, in the absence of demographic fluctuations, there is \emph{never long-lived coexistence of 
 the strains $R$ and $S$  under low switching rate of the toxin level.}
\\
(ii) Under high switching rate, $\nu_T\gg 1$, the population experiences a large number of $T$-switches before relaxing into its final state; see below. In this case 
the $T$-DN \emph{self-averages}~\cite{wienand_evolution_2017,wienand_eco-evolutionary_2018,taitelbaum_population_2020,taitelbaum2023evolutionary,ashcroft_fixation_2014,meyer_evolutionary_2020} and we are left with a Moran process defined by the effective rates
$\widetilde{T}_{R}^{\pm} \to \avg{\widetilde{T}_{R}^{\pm}}$ obtained by averaging $\xi_T$ over its stationary distribution, yielding
\begin{equation}
\hspace{-3mm}
\begin{aligned}
\label{eq:fastswitchrates}
\avg{\widetilde{T}_{R}^+}  &=\frac{Nx(1-x)}{2}\left(\frac{1+\delta_T}{x+(1-x)e^s} +\frac{1-\delta_T}{x+(1-x)e^{-s}}\right),\\
\avg{\widetilde{T}_{R}^-}  &=\frac{Nx(1-x)}{2}\left(\frac{(1+\delta_T)e^s}{x+(1-x)e^s}+\frac{(1-\delta_T)e^{-s}}{x+(1-x)e^{-s}}\right).
\end{aligned}
\end{equation}
When $N\to \infty$, the mean-field (MF) rate equation associated with this effective Moran process 
thus reads~\cite{blythe_stochastic_2007,Gardiner,VanKampen}:
\begin{align}
\label{eq:fastmeanfieldODE}
    \dot{x}&=\frac{\avg{\widetilde{T}_{R}^+} -\avg{ \widetilde{T}_{R}^-} }{N}\\
    &=\frac{x(1-x)}{2}\left[\frac{(1+\delta_T) (1-e^{s})}{x+(1-x)e^{s}} + \frac{(1-\delta_T)(1-e^{-s})}{x+(1-x)e^{-s}}\right]\nonumber,
\end{align}
where the right-hand-side (RHS) can be interpreted as the RHS of Eq.~\eqref{eq:MF} averaged over $\xi_T$. In addition to the trivial fixed points $x=0, 1$, Eq.~\eqref{eq:fastmeanfieldODE} admits a \emph{coexistence equilibrium}
\begin{equation}
\label{eq:mean_field_coex}
    x^*=\frac{1}{2}-\frac{\delta_T}{2}\coth{\frac{s}{2}},
\end{equation}
when $-\tanh{\frac{s}{2}}<\delta_T<\tanh{\frac{s}{2}}$.
This equilibrium stems from the $T$-DMN and thus is a \emph{fluctuation-induced coexistence point}. In the case of large $s$ we have that $\coth{\frac{s}{2}}\rightarrow1$, and $x^*$ exists ($0<x^*<1$) for all values of $\delta_T$. 
Since $\left.\frac{\text{d}\dot{x}}{\text{d}x}\right|_{x^*}=-\frac{4}{1-\delta_T^2}\tanh^2\left(\frac{s}{2}\right)\left(1-x^*\right)<0$

is negative, linear stability analysis 
reveals that $x^*$ is the sole asymptotically stable equilibrium of Eq.~\eqref{eq:fastmeanfieldODE} when it exists ($x=0,1$ are thus unstable).
When $s\ll 1$, $\coth{\frac{s}{2}}\rightarrow \frac{2}{s}$, 
and $x^*$ exists only for $-\frac{s}{2}<\delta_T<\frac{s}{2}$.
This means that for $s\ll 1$, coexistence is essentially possible only under  symmetric switching ($\delta_T=0$), see Sec.~SM2 the appendix. In what follows, we  focus 
on the less restrictive case $s={\cal O}(1)$, for which coexistence is possible for a broad range of parameters $(\nu_T,\delta_T)$.

(iii) In the regime of intermediate switching rate, where $\nu_T\sim 1$, the population experiences
a finite number of $T$-switches prior to settling in its final state. Depending 
on this number, as well as the selection strength $s$ and the population size, 
the dynamics may be closer to either the low or high $\nu_T$ regime, with dominance or coexistence possible but{, in general,} not certain; see Fig.~\ref{fig:moranheatmap} below.

\subsection{\label{sec:finite}Finite populations - fixation and long-lived coexistence}
\begin{figure}
\includegraphics[width=0.67\linewidth]{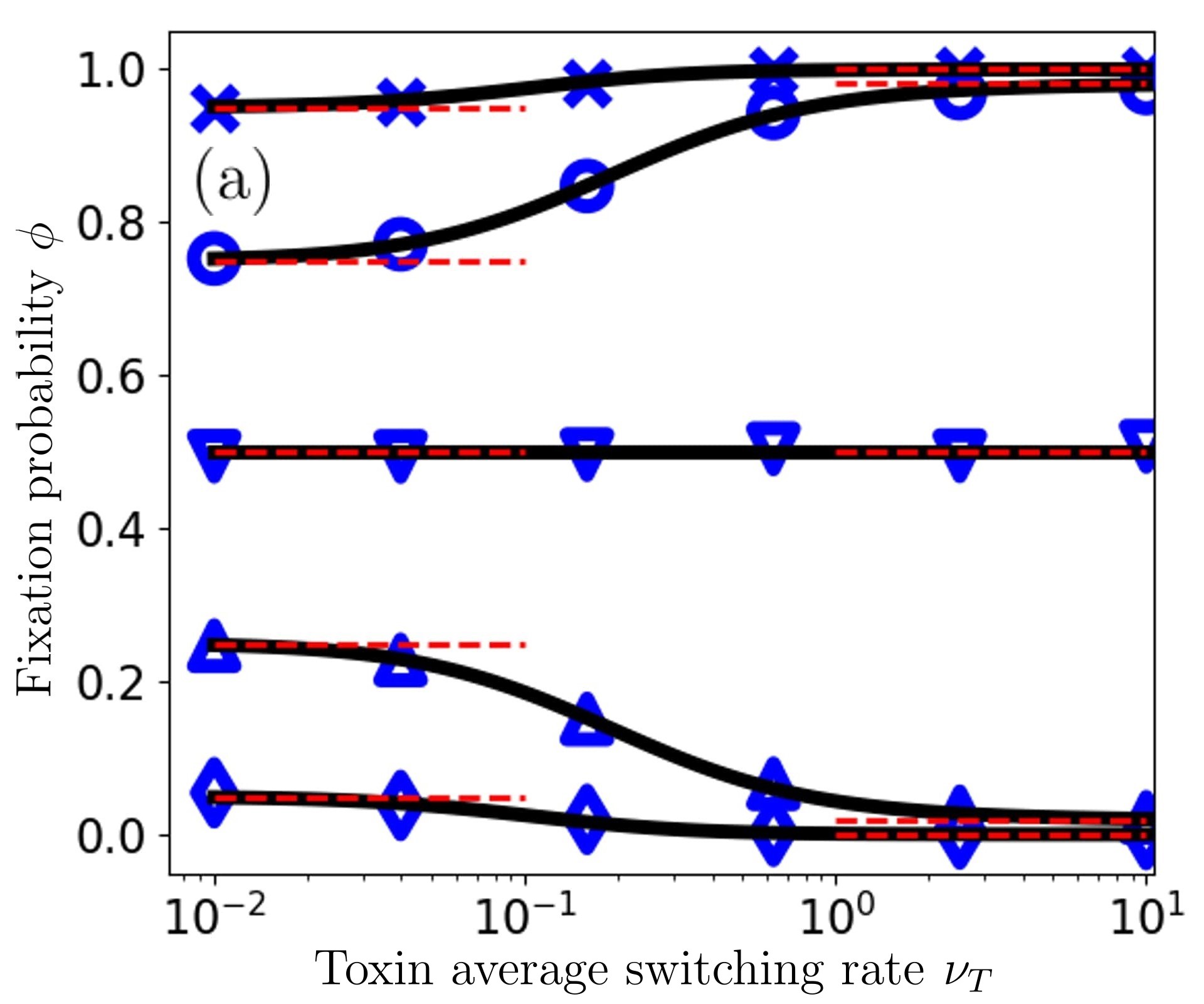}\\
\includegraphics[width=0.67\linewidth]{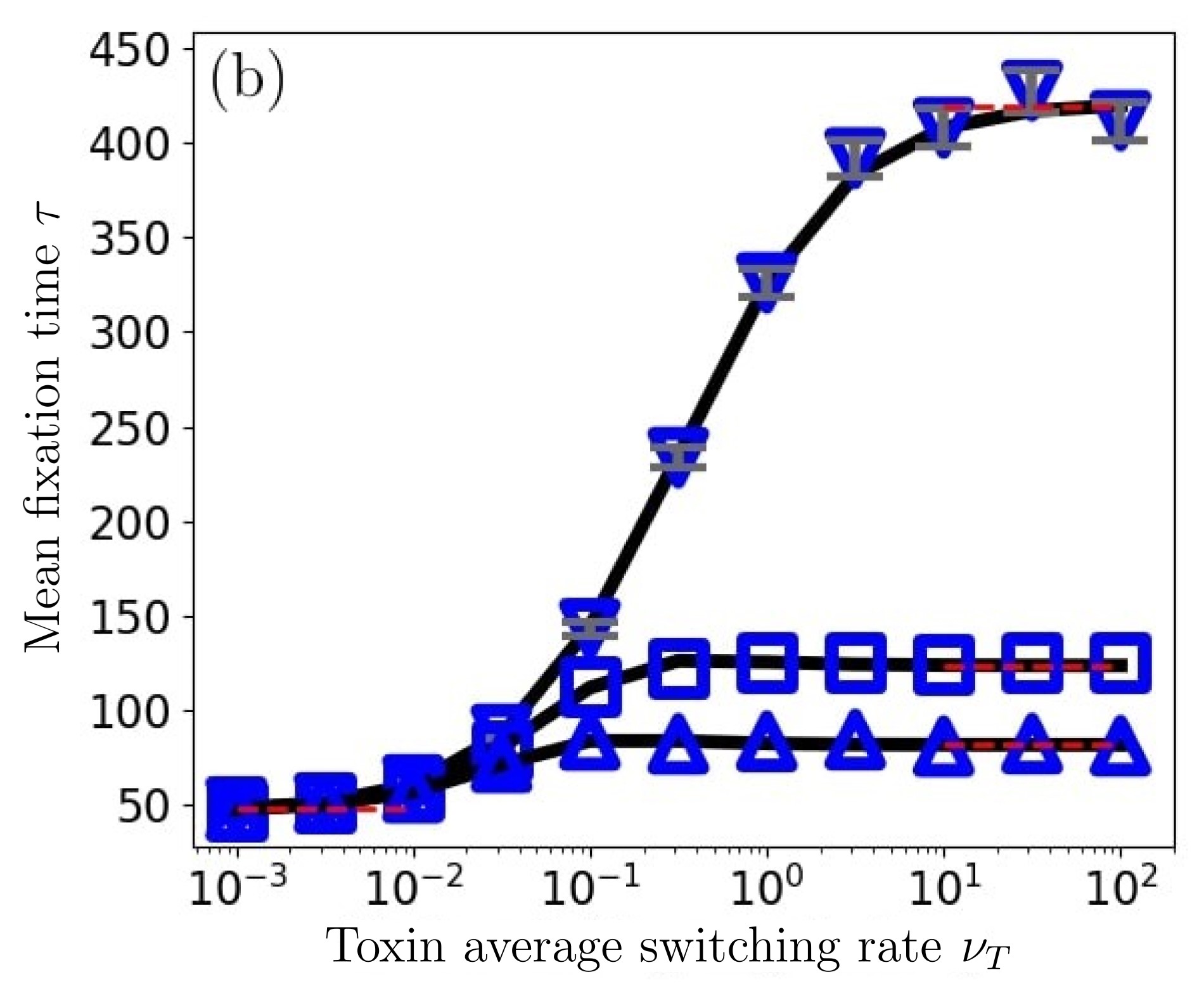}\\
\includegraphics[width=0.67\linewidth]{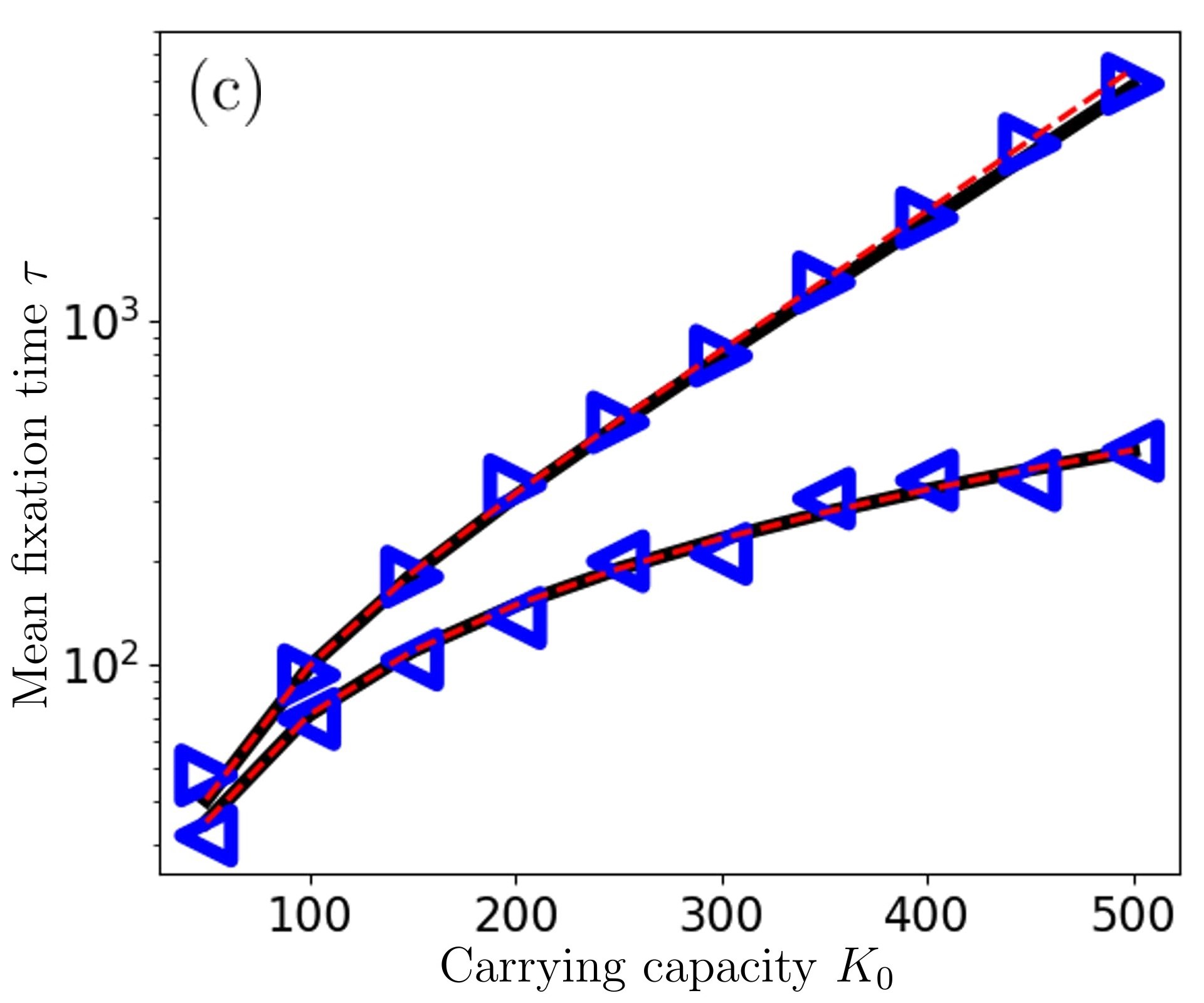}
\caption{
$R$ fixation probability $\phi$ and MFT  $\tau$ under constant carrying capacity  and $T$-EV.
Symbols are from stochastic simulations ($10^3$ realisations) with 
carrying capacity $K_0$ (and non-constant $N$), and full black lines are 
for the Moran approximation with $N=K_0$, based on Eq.~\eqref{eq:firststepphi} (a) 
and 
Eq.~\eqref{eq:firststeptau} (b,c). 
(a) $\phi$
 against  $\nu_T$, for $\delta_T=-0.9$ ($\times$), $-0.5$ ($\bigcirc$), $0$ ($\bigtriangledown$), $0.5$ ($\bigtriangleup$), $0.9$ ($\lozenge$), from top to bottom, 
 with 
$s=0.3$ and $K_0=50$. Red dashed lines are  predictions of
Eq.~\eqref{eq:lowswitchphi} and Eq.~\eqref{eq:fastswitchphi}.
(b)  $\tau$ against $\nu_T$, for
$\delta_T=0$ ($\bigtriangledown$), $0.3$ ($\square$), $0.5$ ($\bigtriangleup$)
from top to bottom, with 
 $s=0.1$ and $K_0=500$. Any bias $\delta_T\neq0$, reduces $\tau$. {Error bars (dark grey) are overlaid for the case $\delta_T=0$ and almost indistinguishable from the symbols; see text and SM4 the appendix.}
 Red dashed lines are analytical predictions in the limiting regimes $\nu_T\to 0,\infty$ (Eq.~(S4) the appendix with same time-averaging process of rates as in Sec. \ref{sec:MF}).
 (c)~
$\tau$ against  $K_0$ under  fast $T$-EV in the coexistence regime for $\delta_T=0$, $s=0.1$ (\rotatebox[origin=c]{90}{$\bigtriangleup$}) and $s=0.3$ (\rotatebox[origin=c]{270}{$\bigtriangleup$}). Here, there is long-lived coexistence of the strains
at the (meta-)stable equilibrium $x^*=x_0=0.5$ prior to fixation.
The MFT grows exponentially when 
 $K_0 s^2\gg 1$, see Sec.~SM2 the appendix. Red dashed lines show the analytical predictions
 for the MFT when $\nu_T\to\infty$ (Eq.~(S4) the appendix with Eq.~\eqref{eq:fastswitchrates}),  compared with the predictions of Eq.~\eqref{eq:firststeptau} (black lines) and simulation results (markers)
 for $\nu_T=100$.}
\label{fig:limitmatching}
\end{figure}
From the MF analysis, we have found that when $N\to \infty$ species coexistence is feasible under fast $T$-EV switching, whereas only {$R$ or $S$} dominance occurs under slow switching. {We now} study how these results  nontrivially morph when the population is fixed and finite.

Since the model is defined as a finite Markov chain with absorbing boundaries, see Eqs.~\eqref{eq:Moran} and \eqref{eq:Moranrates}, its final state  unavoidably 
corresponds 
to the fixation of one strain and the extinction of the other, i.e. the population eventually ends up in either the state $(N_R,N_S)=(N,0)$ or $(N_R,N_S)=(0,N)$
~\cite{Pinsky2011-px,antal_fixation_2006,blythe_stochastic_2007,Ewens}. This means that, strictly, the finite population does not admit stable coexistence: when it exists, $x^*$ is   metastable~\cite{assaf2017,MA10,AM10}. In fact, 
while it  is guaranteed that eventually only one of the strains will finally survive, fixation can occur after a very long time and can follow a long-term coexistence of the strains, as suggested by the MF analysis of the regime with $\nu_T\gg 1$. It is thus relevant to study under which circumstances there is \emph{long-lived coexistence} of the strains. 

The evolutionary dynamics is characterised by the  fixation probability of the strain $R$, here denoted by~$\phi$. This is the probability that a population,  consisting initially of a fraction $x_0$ of $R$ bacteria, is eventually taken over by the strain $R$.\\
A related quantity is the unconditional mean fixation time (MFT), here denoted by $\tau$, which is the average time for the fixation of either species to occur. 

In what follows, we study how the $R$ fixation probability $\phi(\nu_T)$ and the MFT $\tau(\nu_T)$ vary with the average switching rate of the $T$-EV for different values of $K_0$, $\delta_T$, and $s$ (treated as parameters), and determine when there is long-lived coexistence of the strains.

In the limits $\nu_T\hspace{-0.2cm}\to\hspace{-0.2cm}0, \infty$, 
we can use the well-known properties of the Moran model~\cite{antal_fixation_2006,Ewens,traulsen_stochastic_2008} to 
obtain  $\phi(\nu_T)$ and $\tau(\nu_T)$ from  their Moran approximation (MA) counterparts{, denoted by}  
$\phi_{\rm MA}(N)$ and $\tau_{\rm MA}(N)$, which are respectively the $R$ fixation probability and mean fixation time in the associated Moran process for a population of constant size $N$; see Sec.~SM3 the appendix.

For a given initial resistant fraction\footnote{In all our examples we set $x_0=0.5$.}, the fixation probability in the slow-switching regime, $\phi(\nu_T\to0)$
is obtained by averaging $\phi_{{\rm MA}}(N){\big|}_{\xi_T}$,
 denoting
the $R$ fixation probability in the realm of the MA in static environment $\xi_T$, over the stationary distribution of $\xi_T$~\cite{wienand_evolution_2017, wienand_eco-evolutionary_2018,taitelbaum_population_2020,taitelbaum2023evolutionary}:
\begin{align}
\label{eq:lowswitchphi}
\phi(\nu_T\to0)&=\left(\frac{1+\delta_T}{2}\right)\phi_{{\rm MA}}(N){\big|}_{\xi_T=+1} \nonumber\\
&+\left(\frac{1-\delta_T}{2}\right)\phi_{{\rm MA}}(N){\big|}_{\xi_T=-1}.
\end{align}
When $N\gg1$ and $\xi_T=+1$ the strain $S$ is always favoured and $\phi_{{\rm MA}}(N){\big|}_{\xi_T=+1}\approx 0$, whereas $R$ is favoured when 
$\xi_T=-1$ and in this case $\phi_{{\rm MA}}(N){\big|}_{\xi_T=-1}\approx 1$. 
Since $\xi_T(0) = - 1$
with probability $(1- \delta_T)/2$, this coincides with the 
$R$ fixation probability: 
$\phi(\nu_T\to0)\approx (1- \delta_T)/2$. The probability that $S$ fixates
when $\nu_T\to0$ is thus $1-\phi(\nu_T\to0)\approx (1+ \delta_T)/2$ 

In the fast-switching regime the fixation probability is that of the Moran process defined by the effective rates in
Eq.~\eqref{eq:fastswitchrates}. Using Eq.~\eqref{eq:fastswitchrates} with $x=n/N$, we thus find~\cite{Ewens,antal_fixation_2006}: 
\begin{equation}
\label{eq:fastswitchphi}
    \phi(\nu_T\to\infty)=\frac{1+\sum_{k=1}^{Nx_0-1}\prod_{n=1}^{k}\frac{\avg{ \widetilde{T}_{R}^-\left(n/N\right)} }{\avg{ \widetilde{T}_{R}^+\left(n/N\right)} }}{1+\sum_{k=1}^{N-1}\prod_{n=1}^{k}\frac{\avg{ \widetilde{T}_{R}^-\left(n/N\right)} }{\avg{\widetilde{T}_{R}^+\left(n/N\right)}}};
\end{equation}
see Sec.~SM3.A the appendix. A similar analysis can also be carried out for $\tau$, see Sec.~SM3.B the appendix.
Results reported in Fig.~\ref{fig:limitmatching} show that Eqs.~\eqref{eq:lowswitchphi}  and \eqref{eq:fastswitchphi} accurately capture the behaviour of  $\phi$ in the limiting regimes $\nu_T\to0,\infty$, see Fig.~\ref{fig:limitmatching}(a).  Fig.~\ref{fig:limitmatching}(b) shows that the predictions for $\tau$ when $\nu_T\to0,\infty$, are also in good agreement with simulation results, with a much larger MFT under high $\nu_T$ than under low switching rate (at fixed $\delta_T$). 
In Fig.~\ref{fig:limitmatching}(b), the MFT when $\nu_T\gg 1$ for  $\delta_T=0$  is significantly larger than under $\delta_T\neq 0$.  This stems from 
$x^*=x_0=1/2$ being the 
attractor of Eq.~\eqref{eq:fastmeanfieldODE} when $\delta_T=0$, but not being an equilibrium when $\delta_T=0.3$ or $\delta_T=0.5$. Fig.~\ref{fig:limitmatching}(a,b) also illustrate the excellent agreement between the predictions of the MA with $N=K_0$ and those obtained from stochastic simulations with $K=K_0$ constant. {Note that typical error bars are shown for $\delta_T=0$ in Fig.~\ref{fig:limitmatching}(b). These are found to be small and almost coincide with the markers. For the sake of readability, we have thus omitted similar error bars from the other panels and figures.}

The MF analysis and results of Fig.~\ref{fig:limitmatching} suggest that under sufficiently high switching rate $\nu_T$
there is long-lived coexistence of the strains. We can rationalise this picture by noting that in the regime of dominance of one strain the MFT scales sublinearly with the population size $N$, while the MFT grows superlinearly (exponentially when $N=K_0\gg1$, see Fig.~\ref{fig:limitmatching}(c)) in the regime 
of long-lived coexistence~\cite{antal_fixation_2006,cremer_edge_2009,reich2007,he2011}. The dominance and long-lived coexistence scenarios are separated by a 
regime where the MFT scales with the population size, i.e. $\tau\sim N$, where the dynamics is governed by random fluctuations. This leads us to
consider that  long-lived coexistence of the $R$ and $S$ strains arises whenever
the MFT exceeds $2\langle N \rangle$, i.e. when $\tau>2\langle N \rangle$, 
where $\langle N \rangle$ is the mean population size at (quasi-)stationarity; see below\footnote{The factor 2 has been  chosen arbitrarily to   prevent $\tau\sim \avg{N}$ from appearing as coexistence. Other choices are of course possible, and would have only modest effects on the crossover regime between the phases of dominance and coexistence.}. This is illustrated
in the provided videos of \cite{videos} commented in Sec.~SM7 of the appendix. When, as in this section,  
$N=K_0$  or $N$ fluctuates about the constant carrying capacity $K_0$ ($N\approx K_0$), we simply have $\langle N \rangle=K_0$. The criterion $\tau>2\langle N \rangle=2K_0$ thus  prescribes that  long-lived coexistence occurs when the MFT scales superlinearly with   $K_0$ and hence exceeds 
the double of the average population size,
$2K_0\gg 1$. 

\begin{figure*}
    \centering
    \includegraphics[width=\textwidth]{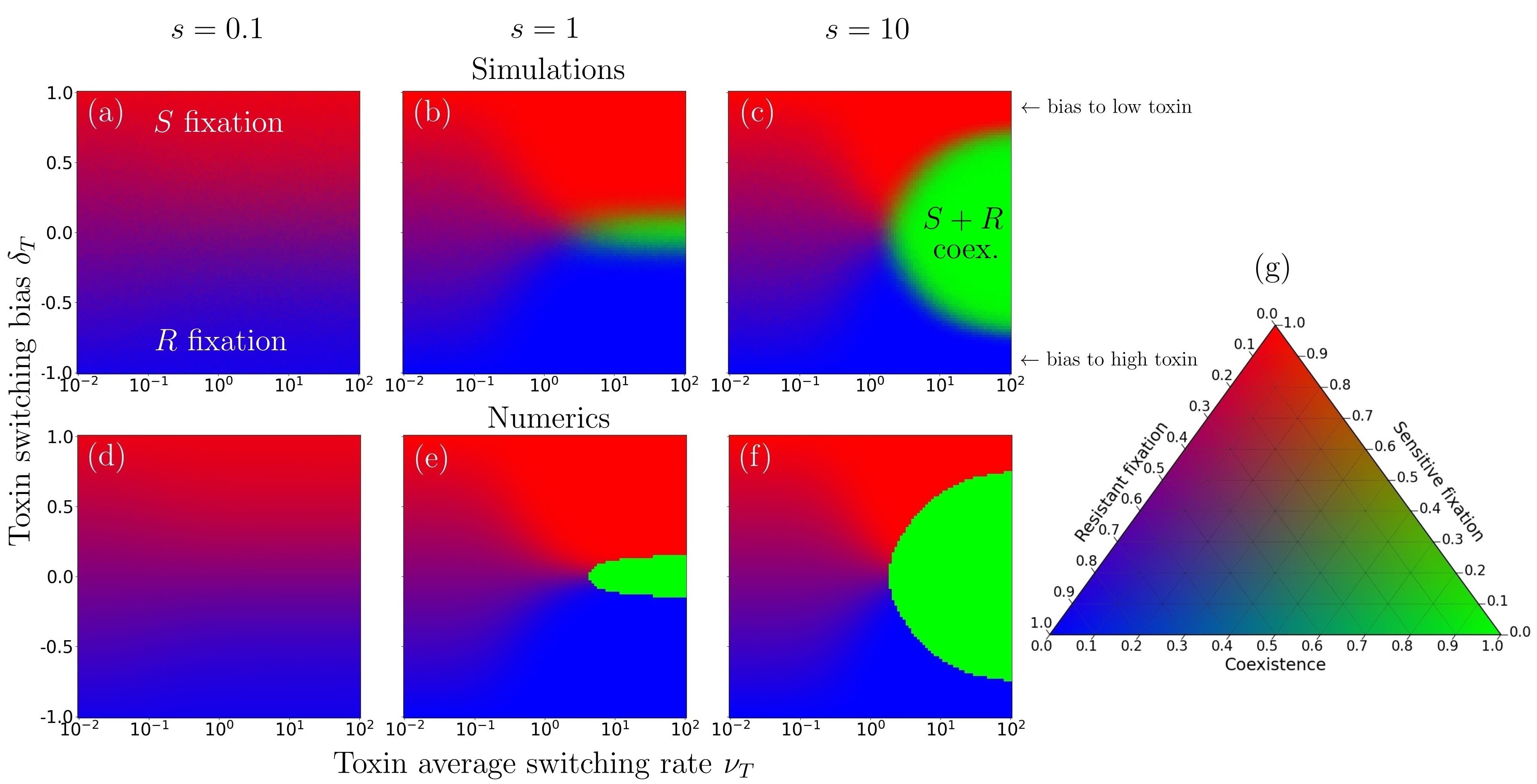}
    \caption{Fixation/coexistence  diagrams 
    under $T$-EV
    in the $(\nu_T,\delta_T)$ parameter space for a small system
    of constant carrying capacity $K_0=50$, with selection bias  $s=0.1$ (a,d), $s=1$ (b,e), and $s=10$ (c,f), after a time $t=2K_0$.
    (a)-(c): phase diagrams obtained from stochastic simulations of  the model with a constant  $K_0$ ($N$ fluctuates about $K_0$) over $10^3$ realisations
    and     coded
    according to the RGB colourmap of panel (g): red / blue {(grey / charcoal)} corresponds to the likely fixation of $S/R$ (red {(grey)}: $S$ dominance; blue {(charcoal)}: $R$ dominance), {regions indicated as ``$S/R$ fixation'' in panel (a}); {magenta (dark grey) indicates where fixation of $R$ or $S$ is likely, area between ``$S/R$ fixation'' regions in panel (a)}; green {(light grey)} indicates where long-lived coexistence is most likely, {area highlighted as ``$S+R$ coex.''} in panel (c).
    (d)-(f): same as in (a)-(c) but from numerically exact solutions of Eq.~\eqref{eq:marginalisedsolution} for the corresponding Moran process with a constant population size $N=K_0$, see text.}
    \label{fig:moranheatmap}
\end{figure*}

The conditions under which the long-lived coexistence criterion, $\tau>2\langle N \rangle$, is satisfied can be estimated by noting that, from the MF analysis, we expect coexistence to occur when $\xi_T$ self-averages  
under sufficiently high switching rate $\nu_T$. Since the average number of $T$-switches by 
$t=2\langle N \rangle$ scales as $\nu_T \langle N \rangle$, self-averaging  occurs when $\nu_T \langle N \rangle\gg 1$. We thus consider that there is fast $T$-EV switching when  $\nu_T \gg 1/\langle N \rangle$,
while $\nu_T \ll 1/\langle N \rangle$ corresponds to the slow $T$-EV regime.
To ensure long-lived coexistence, the necessary condition $\nu_T \gg 1/\langle N \rangle$ is supplemented by the requirement that $s\sim 1$. This ensures enough environmental variability 
and a regime of coexistence  where 
the MFT  is generally $\tau \sim e^{c\langle N \rangle}$ (where $c$ is some positive constant) when $s={\cal O}(1)$~\cite{antal_fixation_2006,cremer_edge_2009,Ewens,Kimura,MA10,AM10,AM11} guaranteeing  
$\tau>2\langle N \rangle$.

Hence, the expected conditions for long-lived coexistence 
are $\nu_T \gg 1/\langle N \rangle$ (fast $T$-switching) and $s={\cal O}(1)$
(enough EV), 
which are satisfied in the examples considered here
when $\nu_T \sim 1, s\sim 1$ or greater, and $\avg{N}\gg 1$. 

We have studied the  influence of $T$-EV on the fixation and coexistence 
properties of the model with constant carrying capacity $K=K_0$ and selection strength $s$ by 
running a large number of computer simulations up to a time $t=2K_0$ across the $\nu_T-\delta_T$ parameter space.
When just after $t=2K_0$ both species are present, the run for $(\nu_T,\delta_T,s)$ is characterised by long-lived coexistence which is RGB coded $(0, 1, 0)$.
 There is no long-lived  coexistence for the run $(\nu_T,\delta_T)$
if one of the species fixates by $t\leq 2K_0$: either the strain $R$, which is RGB coded $(1,0,0)$, or the strain $S$, which is coded by $(0,0,1)$. This procedure is repeated for $10^3$ realisations for different $(\nu_T,\delta_T)$ and, after sample averaging, yields the RGB-diagram of Fig.~\ref{fig:moranheatmap}(a-c); see Sec.~SM4 the appendix. {In greyscale, the RGB coding translates into red $\rightarrow$ grey, blue $\rightarrow$ charcoal, and green $\rightarrow$ light grey. In what follows, the  
crossover regimes are coded in magenta and faint green with dark grey and faint light grey as their respective greyscale counterparts, see below.}


It is also useful to study the effect of the  $T$-EV 
in the realm of the MA by means of numerically exact results. For this,
with the transition rates Eq.~\eqref{eq:Moranrates}, 
we notice that when $N=K_0$ 
is constant and there are initially $n$ cells of type $R$,
the $R$-strain fixation probability,  $\phi_n^{\xi_T}$,
 in the environmental state $\xi_T$ satisfies
the first-step analysis equation~\cite{Gardiner,Pinsky2011-px,ashcroft_fixation_2014}
\begin{align}
\label{eq:firststepphi}
(\widetilde{T}_{R}^+(n) + \widetilde{T}_{R}^-(n)+\nu_T^{\xi_T})\phi_n^{\xi_T}=~&\widetilde{T}_{R}^+(n)\phi_{n+1}^{\xi_T}
+ \widetilde{T}_{R}^-\phi_{n-1}^{\xi_T} \nonumber\\&+ \nu_T^{\xi_T}\phi_{n}^{-{\xi_T}},
\end{align}
subject to the boundary conditions $\phi_0^{\xi_T} = 0$ and $\phi_N^{\xi_T} = 1$. The mean fixation time in the environmental state $\xi_T$, $\tau_n^{\xi_T}$,  
satisfies a similar equation:
\begin{align}
\label{eq:firststeptau}
(\widetilde{T}_{R}^+(n) + \widetilde{T}_{R}^-(n)+\nu_T^{\xi_T})\tau_n^{\xi_T}=~&1+\widetilde{T}_{R}^+(n)\tau_{n+1}^{\xi_T}
+ \widetilde{T}_{R}^-(n)\tau_{n-1}^{\xi_T}\nonumber\\ &+ \nu_T^{\xi_T} \tau_{n}^{-{\xi_T}},
\end{align}
with boundary conditions $\tau_0^{\xi_T}=\tau_N^{\xi_T}=0$. Eqs.~\eqref{eq:firststepphi} and \eqref{eq:firststeptau} 
are thus solved numerically, and the fixation probability and MFT are obtained 
after averaging over the stationary distribution of $\xi_T$, yielding
\begin{equation}
\begin{aligned}
\phi_n &= \left(\frac{1+\delta_T}{2}\right) \phi_n^{+} +   \left(\frac{1-\delta_T}{2}\right) \phi_n^{-},\quad \text{and} \\ \tau_n &= \left(\frac{1+\delta_T}{2}\right) \tau_n^{+} +  \left(\frac{1-\delta_T}{2}\right) \tau_n^{-}.
\end{aligned}
\label{eq:marginalisedsolution}
\end{equation}
In our examples, we always consider $x_0=1/2$, and henceforth  write  $\phi_{{\rm MA}}(N)\equiv\phi_{N/2}$  for the $R$-fixation probability and 
$\tau_{{\rm MA}}(N)\equiv\tau_{N/2}$
for the MFT in the realm of the MA.
For each triple $(\nu_T,\delta_T,s)$, we numerically 
solved Eq.~\eqref{eq:firststeptau} and, in the region of the 
the parameter space where $\tau_{{\rm MA}}(N)> 2K_0$,
there is long-lived coexistence, which is coded by $(0,1,0)$  in the RGB-diagram of Fig.~\ref{fig:moranheatmap}(d-f). When $\tau_{{\rm MA}}(N)\leq 2K_0$,  there is dominance of one of the species, characterised by the fixation probabilities $\phi_{{\rm MA}}(N)$ and $1-\phi_{{\rm MA}}(N)$
of  $R$ and $S$, respectively, obtained from Eq.~\eqref{eq:firststepphi} and
coded by $(\phi_{{\rm MA}}(N),0,1-\phi_{{\rm MA}}(N))$  in Fig.~\ref{fig:moranheatmap}(d-f). 

Exact numerical results for the MA with $N=K_0$
in Fig.~\ref{fig:moranheatmap}(d-f) are in excellent agreement with those 
of simulations obtained  
for $K=K_0$  in Fig.~\ref{fig:moranheatmap}(a-c).
In line with the MF analysis, we find that long-lived coexistence, occurs  for $T$-EV of sufficiently large magnitude, i.e. $s\sim 1$ or higher,
and under high enough switching rate, i.e. $\nu_T\sim 1$ or higher, shown as green {(light grey)} areas in Fig.~\ref{fig:moranheatmap}. The region of coexistence separates regimes dominated by either species, especially at high $\nu_T$ when $\phi\approx 0$ when $\delta_T\>0$ while $\phi\approx 1$ 
when $\delta_T<0$. {In Fig.~\ref{fig:moranheatmap},} the boundaries between the regimes of  
 $R/S$ dominance, coded in blue {(charcoal)} / red {(grey)},  and coexistence, areas in green {(light grey)}), 
 are interspersed by crossover regimes where both species are likely to fixate (magenta {(dark grey)} in  Fig.~\ref{fig:moranheatmap}), or 
 coexist with probability between 0 and 1 (faint green {(faint light grey)} in Fig.~\ref{fig:moranheatmap}), as coded in Fig.~\ref{fig:moranheatmap}(g).

\section{\label{sec:switchingK}Twofold Environmental Variability: Coexistence  and fixation under time-varying fitness and switching carrying capacity}

We have seen that under a constant carrying capacity, 
long-lived coexistence of the strains is feasible when 
$s$ and $\nu_T$ are of order $1$ or higher (enough $T$-EV and fast $T$-switching). We now study how this picture morphs when, in addition to the time-variation of $f_S$ and $\overline{f}$, the  carrying capacity $K(t)$ switches according to Eq.~\eqref{eq:carryingcapacity} and drives the fluctuating population size $N$. EV is thus twofold, and the population evolves under the joint effect of $T$-EV and $K$-EV. 

We  consider  $K(t)\in\{K_-,K_+\}$ with
$1\ll K_-<K_+$, and in the first instance assume that $N$ is always sufficiently large to
allow us to neglect  the DN, yielding  
\cite{davis_piecewise-deterministic_1984,wienand_evolution_2017,wienand_eco-evolutionary_2018,west2020,taitelbaum_population_2020,taitelbaum2023evolutionary,Gardiner}

\begin{subequations}
\begin{align}
\label{eq:meanfieldtotala}
  \dot{N}=T_R^+-T_R^-+T_S^+-T_S^-=N\left(1-\frac{N}{K}\right),\\
  \label{eq:meanfieldtotalb}
    \dot{x}=\frac{T_R^+-T_R^-}{N}-x\frac{\dot{N}}{N}=x\left(\frac{f_R}{\overline{f}}-1\right),
\end{align}
\end{subequations}
where $N$ is independent of $s$ and affected only by $K$-EV via ${\xi_K}$ in Eq.~\eqref{eq:carryingcapacity}, while 
the evolution of $x$ in Eq.~\eqref{eq:meanfieldtotalb}
is impacted by $\xi_K$, $\xi_T$, and $s$ via $x=N_R/N$
and $\overline{f}(t)=x+(1-x)\exp(s\xi_T$). The population composition  is hence coupled 
to the evolution of the population size (eco-evolutionary dynamics), while the  statistics of $N$, like its average, denoted by $\langle N\rangle$,
are obtained by ensemble averaging over $\xi_K$.
The stochastic logistic differential equation Eq.~\eqref{eq:meanfieldtotala} defines an
$N$-PDMP
whose properties allow us to characterise the distribution of $N$~\cite{davis_piecewise-deterministic_1984,
hufton_intrinsic_2016,wienand_evolution_2017,wienand_eco-evolutionary_2018}.

\begin{figure}
    \centering
    \includegraphics[width=1.0\linewidth]{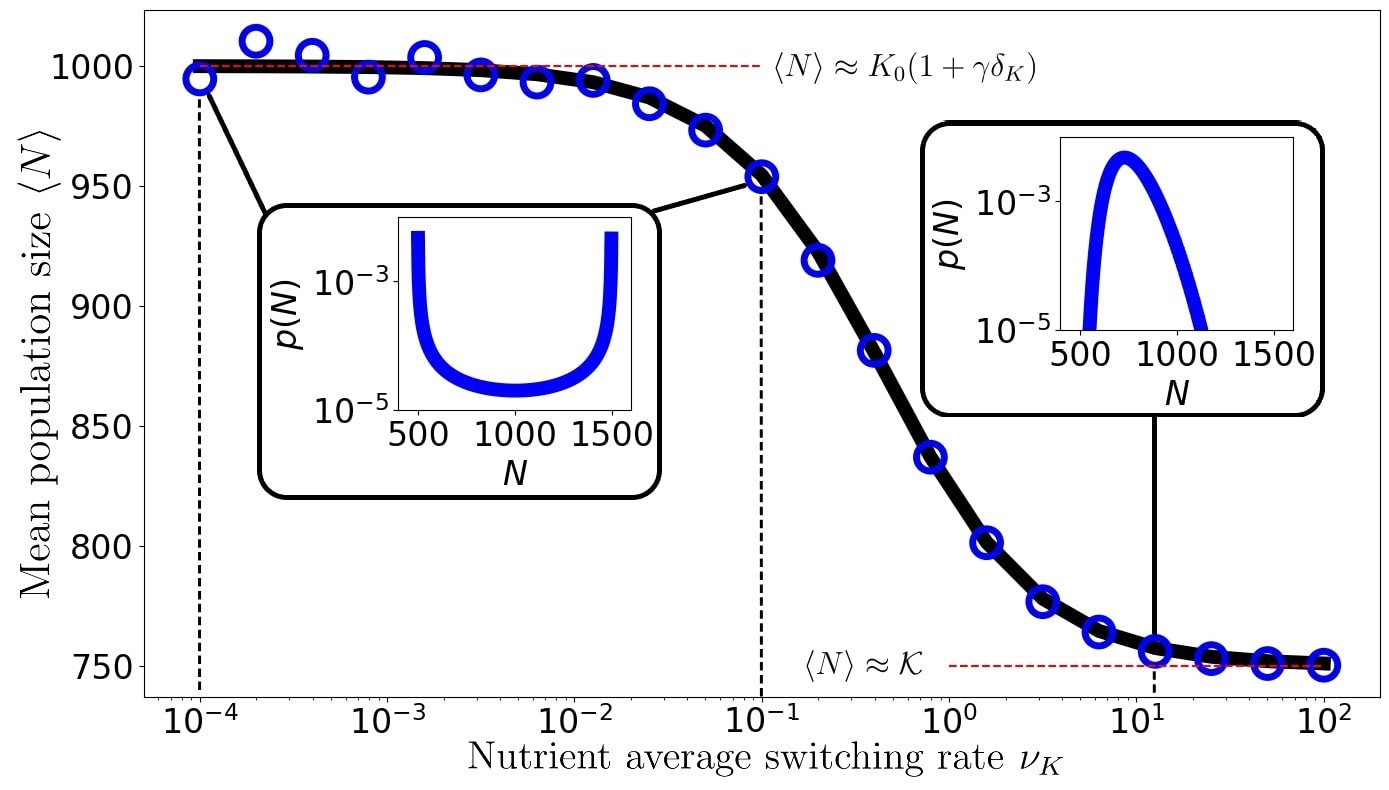}
    \caption{Long-time average population size $\langle N\rangle$ versus the $K$-EV average switching rate $\nu_K$.
    Symbols are from simulations (averaged over $10^3$ realisations at time $t=2\langle K\rangle$), the solid black line shows the PDMP prediction Eq.~\eqref{eq:mean_N}. 
For fixed $\delta_K$, $\langle N\rangle$ decreases with $\nu_K$, and 
$\langle N\rangle\approx \langle K\rangle=K_0(1+\gamma\delta_K)$ when $\nu_K\ll 1$
while $\langle N\rangle\approx \mathcal{K}$ when  $\nu_K\gg 1$ (red horizontal dashed lines). 
Insets show the PDMP PDF Eq.~\eqref{eq:N_PDMP} under low  switching rate $\nu_K$ (left) and for $\nu_K\gg 1$ (right); dashed vertical lines are eye-guides indicating the corresponding values of $\nu_K$, where for low $\nu_K$ the PDMP PDF is indistinguishable in either case.
    Parameters are:  $K_0=1000$, $\gamma=0.5$, and $\delta_K=0.0$.
     See text for details.}
    \label{fig:N_vs_nu_K}
\end{figure}

As discussed in \cite{wienand_evolution_2017,wienand_eco-evolutionary_2018,west2020,taitelbaum_population_2020},
the (marginal) stationary probability density function (PDF) $p(N)$ of the $N$-PDMP  Eq.~\eqref{eq:meanfieldtotala}, while ignoring DN, provides a useful approximation of the actual quasi-stationary population size distribution (QPSD). Here, the stationary PDF is~\cite{horsthemke_lefever,bena2006,Ridolfi2011,wienand_evolution_2017,wienand_eco-evolutionary_2018,west2020,taitelbaum_population_2020}
\begin{equation}
\label{eq:N_PDMP}
p(N) = \frac{\mathcal{Z}}{N^2} \left(\frac{K_+ - N}{N}\right)^{\nu_K(1-\delta_K)-1} \left(\frac{N-K_-}{N}\right)^{\nu_K(1+\delta_K)-1},
\end{equation}
of support $[K_-,K_+]$, with the normalisation
 constant $\mathcal{Z}$ ensuring that $\int_{K_-}^{K_+} p(N)~dN=1$.

\subsection{$N$-PDMP approximation}
 
\label{sec:IVA}
 
The PDF $p(N)$ captures well the main effects of the $K$-EV on the QPSD,
which is bimodal under low $\nu_K$ and becomes unimodal when $\nu_K\gg 1$, see Fig.~\ref{fig:N_vs_nu_K}. In the realm of the $N$-PDMP approximation, 
$p(N)$ aptly reproduces the location of the QPSD peaks and the
transition from a bimodal to unimodal distribution as $\nu_K$ increases: 
The distribution is sharply peaked around 
$N\approx K_{\pm}$
 when $\nu_K \to 0$ (with probability $(1\mp \delta_K)/2$),
flattens when $\nu_K\sim 1$, and then sharpens 
about 
$N\approx \mathcal{K}\equiv K_0(1-\gamma^2)/(1-\gamma\delta_K)$
when $\nu_K \to \infty$. Since $p(N)$ ignores DN, it cannot 
capture the width of the QPSD about the peaks, but
it provides an accurate description of the mean population size, see Fig.~\ref{fig:N_vs_nu_K},
that is well approximated by 
\begin{equation}
\label{eq:mean_N}
    \avg{N}=\int_{K_-}^{K_+} Np(N)~\text{d}N,
\end{equation}
with $\langle N\rangle\approx K_0(1+\gamma\delta_K)$ when $\nu_K\ll 1$
and $\langle N\rangle\approx \mathcal{K}$ when $\nu_K\gg 1$~\cite{wienand_evolution_2017,wienand_eco-evolutionary_2018,west2020,taitelbaum_population_2020}, as shown in Fig.~\ref{fig:N_vs_nu_K}.

 The  PDF $p(N)$ is particularly useful to obtain the fixation/coexistence diagrams under
the effects of both  $T$-EV and $K$-EV. Theoretical/numerical
predictions of the fixation and coexistence probabilities can indeed be derived in the  vein of \cite{wienand_evolution_2017,wienand_eco-evolutionary_2018,west2020,taitelbaum_population_2020,taitelbaum2023evolutionary} by focusing on situations where coexistence occurs when $x$ and $N$ relax on similar timescales. Long-lived coexistence of the strains thus arises when
$s\sim 1$, with fixation typically occurring after $N$
has settled in the QPSD, see Sec.~SM7 the appendix and videos in \cite{videos}. Hence, for the analytial description of the fixation/coexistence diagrams the $R$ fixation probability (with $x_0=1/2$) 
can be suitably approximated by averaging $\phi_{{\rm MA}}(N)$ over $p(N)$ as follows:
\begin{equation}
\label{eq:overallfixationprobability}
    \phi \simeq \int_{K_-}^{K_+} \phi_{{\rm MA}}(N) p(N) ~\text{d}N,
\end{equation}
where  $\phi_{{\rm MA}}(N)$ is  obtained from solving the corresponding Eq.~\eqref{eq:firststepphi} for the
$R$ fixation probability of the associated Moran process, as seen in Sec.~\ref{sec:finite}.

We can  use the PDF $p(N)$ and the results for the mean fixation time
$\tau_{{\rm MA}}(N)$, obtained from solving  Eq.~\eqref{eq:firststeptau}, to determine the probability of coexistence in the realm of the 
$N$-PDMP approximation.
For this, we first solve 
Eq.\eqref{eq:firststeptau} for
$\tau_{{\rm MA}}(N^*)=2\langle N\rangle$, where $\langle N\rangle$ is given by Eq.~\eqref{eq:mean_N}. Since $\tau_{{\rm MA}}$ is an increasing function of $N$,
see Fig.~\ref{fig:limitmatching}(c), we have 
$\tau_{{\rm MA}}(N)>2\langle N\rangle$ for all $N>N^*$, which is the long-lived coexistence condition. Within the 
$N$-PDMP approximation, the lowest possible value of $N^*$ is $K_-$ (since $N\in [K_-,K_+]$). 
We then determine the probability 
$\eta$ that this condition is satisfied 
by integrating $p(N)$ over $[{\rm max}(N^*,K_-),K_+]$:
\begin{equation}
\label{eq:coexistenceprobability}
    \eta \equiv \text{Prob.}\left\{\tau_{{\rm MA}}(N)>2\avg{N}\right\}=\int_{{\rm max}(N^*,K_-)}^{K_+} p(N)~\text{d}N,
\end{equation}
where $N^*$ depends on both $T$-EV and $K$-EV (via  $\avg{N}$), while
the integrand depends only on  $K$-EV. Clearly, $\eta\to 1$ when $N^*\to K_-$. Hence, long-lived coexistence is almost certain when $N^*\approx K_-$, i.e. whenever the mean-fixation time of the population of fixed size $N=K_-$ exceeds $2\avg{N}$. Based on the results of Sec.~\ref{sec:finite},
$\eta$ increases with $\nu_T$ and $s$, and thus for sufficiently large $\nu_T$ and $s$ (but not too large $\delta_T$), we expect $N^*\to K_-$ and $\eta\to 1$.

\subsection{Fixation/coexistence diagrams under   $T$-EV and $K$-EV}

\label{sec:IVB}

\begin{figure*}
    \centering
    \includegraphics[width=\linewidth]{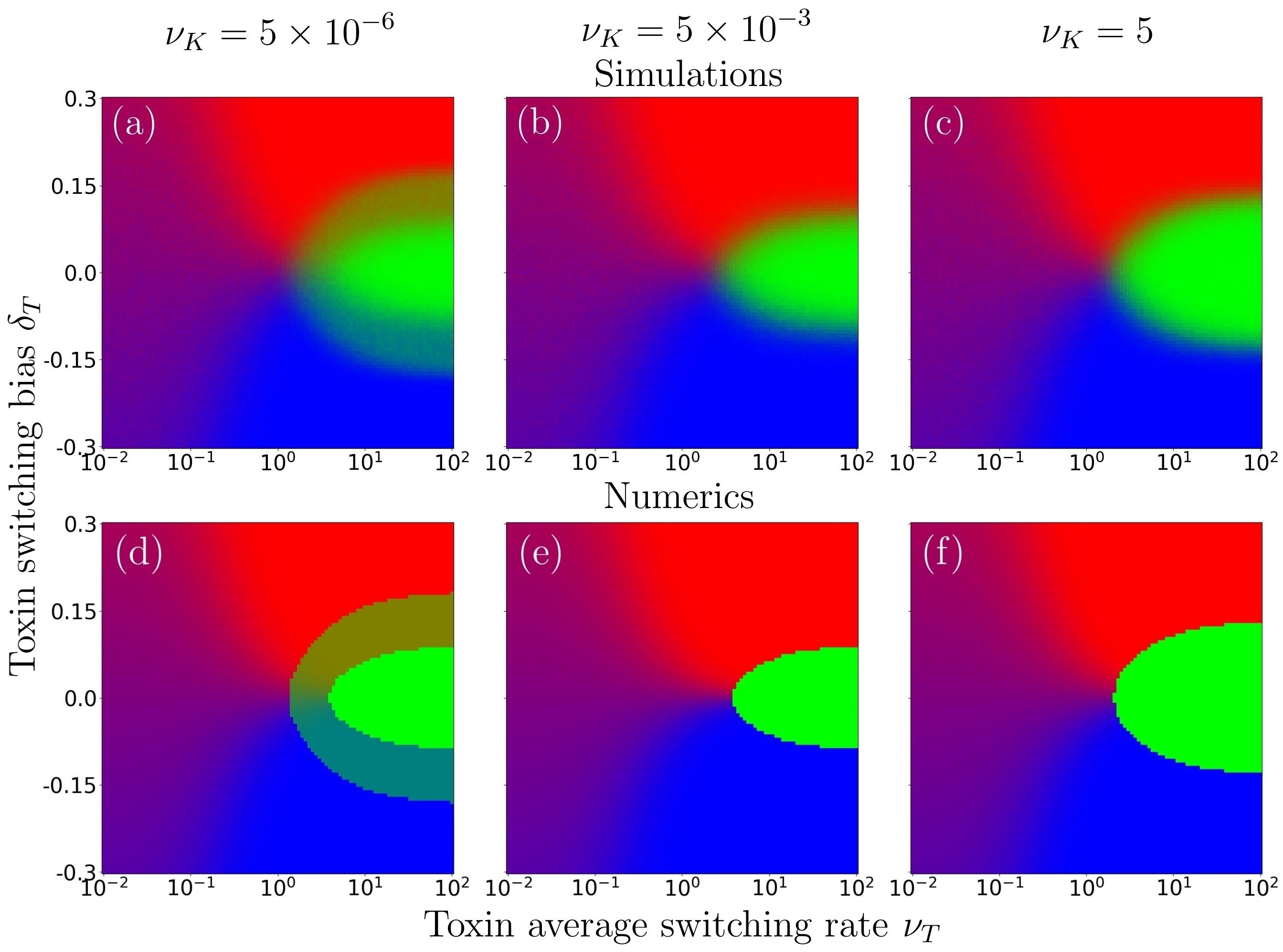}
    \caption{
    Fixation/coexistence  diagrams 
    under $T$-EV and $K$-EV
    in the $(\nu_T,\delta_T)$ parameter space
     showing the influence of K-switching rate $\nu_K$ on the fixation and coexistence probabilities, 
    for $\nu_K=5\times 10^{-6}$ (a,d), $\nu_K=5\times 10^{-3}$ (b,e), and $\nu_K=5$ (c,f). Other parameters are 
    $K_0=1200$,  $s=0.5$, $\gamma=2/3$, $\delta_K=0$. 
    (a)-(c): phase diagrams obtained from stochastic simulations
    data collected at $t=2\langle N\rangle$ over $10^3$ realisations.
    (d)-(f): same as in (a)-(c) but from the theoretical predictions 
    based on Eqs.~\eqref{eq:overallfixationprobability} and \eqref{eq:coexistenceprobability}, see text for details. All diagrams are colour-coded {(greyscale-coded)} as in Fig.~\ref{fig:moranheatmap}.}
    \label{fig:rgb_nu_K}
\end{figure*}

The fixation/coexistence diagrams under joint effect of  $T$-EV and $K$-EV
are obtained as in Sec.~\ref{sec:finite}, with the difference that
long-lived coexistence arises when $t>2\avg{N}$, a condition that depends on $(\nu_K,\delta_K)$, see Fig.~\ref{fig:limitmatching}(c).
In our simulations, we considered different values of $\nu_K$ (letting $\delta_K=0$ for simplicity), and ran  simulations until $t=2\avg{N}$. Each run in which both species still coexist just after 
$t=2\avg{N}$ are RGB {(greyscale)} coded $(0,1,0)$, whereas those in which $R$  or $S$  fixates
by $t\leq 2\avg{N}$ 
are respectively RGB {(greyscale)} coded $(1,0,0)$ or $(0,0,1)$. The RGB {(greyscale)} fixation/coexistence diagrams of Fig.~\ref{fig:rgb_nu_K}(a-c) are obtained after sample-averaging the outcome of this procedure,
repeated $10^3$ times for each pair $(\nu_T,\delta_T)$ and different values of $\nu_K$. 

Theoretical RGB {(greyscale)}  diagrams  are obtained from the $N$-PDMP based approximation built on Eqs.~\eqref{eq:overallfixationprobability}
and \eqref{eq:coexistenceprobability}: for a given $\nu_K$,
we  allocate the RGB  {(greyscale)} value $((1-\eta)(1-\phi), \eta, (1-\eta)\phi)$ obtained for each pair $(\nu_T,\delta_T)$ of the diagram, see Fig.~\ref{fig:rgb_nu_K}(d-f). This triple corresponds to the probability of having, by $t=2\avg{N}$, either no long-lived coexistence (with probability $1-\eta$) and 
fixation of $R$ or $S$ (with respective probabilities $\phi$ and $1-\phi$), or long-lived coexistence (with probability $\eta$). 
In practice, Eqs.~\eqref{eq:overallfixationprobability}
and \eqref{eq:coexistenceprobability} have been used for relatively small systems whereas an equivalent, but more efficient, method was used for large systems, see Sec.~SM4 the appendix. 

The comparison of the top and bottom rows of  Fig.~\ref{fig:rgb_nu_K} shows that the theoretical RGB {(greyscale)} diagrams quantitatively reproduce the features of those obtained from simulations. In general, we find that coexistence regions 
are brighter in  diagrams obtained from $N$-PDMP based approximation than in those stemming from simulations. This difference stems from 
former ignoring demographic fluctuations which slightly broaden  the crossover (magenta {(dark grey)} and faint green {(faint light grey)}) regimes in the latter.

The regions of  Fig.~\ref{fig:rgb_nu_K} where $|\delta_T|\to 1$
are characterised by dominance of one of the strains, and essentially reduces
to the model studied in \cite{wienand_evolution_2017,wienand_eco-evolutionary_2018,taitelbaum_population_2020}, and we can therefore focus 
on characterising the coexistence phase.

When $K_0$ is large, under sufficient environmental variability ($s=0.5$, $\gamma=2/3$ in Fig.~\ref{fig:rgb_nu_K}), the joint effect of $T$-EV and $K$-EV
on the phase of long-lived coexistence in the RGB {(greyscale)} diagrams of Fig.~\ref{fig:rgb_nu_K}
can be summarised as follows: (i) when $\nu_K\to 0$, a (bright green) region where $\eta\approx 1$ and coexistence is almost certain is surrounded by a faint green {(faint light grey)}``outer shell'' where coexistence is possible but not certain ($0<\eta<1$), see  Fig.~\ref{fig:rgb_nu_K}(a,d); (ii) at low, but non-vanishingly small, values of $\nu_K$,
the outer-shell where $0<\eta<1$ fades, and there is essentially only a bright green {(bright light grey)}
region of coexistence where $\eta\approx 1$, see  Fig.~\ref{fig:rgb_nu_K}(b,e); (iii) when $\nu_K\gg 1$, the coexistence region corresponds essentially to $\eta\approx 1$ (bright green {/ bright light grey}) and is broader than under low $\nu_K$, Fig.~\ref{fig:rgb_nu_K}(c,f). 
In all scenarios (i)-(iii), 
$\eta$ increases with $\nu_T\gtrsim 1$ (for not too large $\delta_T$) and hence all the green {(bright and faint light grey)} coexistence phases in  Fig.~\ref{fig:rgb_nu_K}) become brighter as $\nu_T$ is raised  and $\eta \to 1$. 

These different scenarios can be explained by the dependence of the QPSD on $\nu_K$, well captured by the PDF Eq.~\eqref{eq:N_PDMP}. In regime (i) where $\nu_K \ll 1/K_0$, the QPSD and $p(N)$ are bimodal,
 $N\approx K_{\pm}$ with probability $(1\pm \delta_K)/2$, 
and  any $K$-switches by $t=2\avg N \sim K_0$ are unlikely, yielding the faint green {(faint light grey)} outer shell of 
Fig.~\ref{fig:rgb_nu_K}(a,d) corresponding to long-lived coexistence arising only when $N\approx K_+$,  with a probability 
$\eta\approx (1+\delta_K)/2$. In regime (ii), where $1/K_0\ll \nu_K \ll 1$, the QPSD and $p(N)$ are still bimodal but some 
$K$-switches occur by $t\sim K_0$, resulting in long-lived coexistence 
arising almost  only when $\nu_T$ is high enough  
to ensure  $\eta\approx 1$
when $N\approx K_-$. In regime (iii), where $\nu_K\gg 1$
the QPSD and $p(N)$ are unimodal with average $\avg N\approx {\cal K}\geq K_-$, which results in a long-lived coexistence 
region where $\eta\approx 1$ that is broader than in  (i) and (ii), Fig.~\ref{fig:rgb_nu_K}(c,f).
The size of the coexistence region in regime (iii) actually depends nontrivially on $\nu_K$, as revealed by the modal value
of the PDF  Eq.~\eqref{eq:N_PDMP} when $\nu_K(1-|\delta_K|)>1$, which reads
\begin{equation}
\begin{aligned}
\label{eq:N_mode}
\hat{N}=~&\frac{K_0}{2}\left[1+\nu_K(1-\gamma\delta_K)\right]\\
&-\frac{K_0}{2}\sqrt{(1+\nu_K(1-\gamma\delta_K))^2-4\nu_K(1-\gamma^2)},
\end{aligned}
\end{equation}
with $\lim_{\nu_K\to \infty} \hat{N}=\avg{N}=\mathcal{K}$.
We notice that $\hat{N}$ is an increasing function of $\nu_K$ when $\gamma>\delta_K$, and it decreases if $\gamma<\delta_K$ (remaining constant when $\gamma=\delta_K$). As a consequence,
the long-lived coexistence region under high $K$ switching rate grows with $\nu_K$ when $\gamma>\delta_K$, as in Fig.~\ref{fig:rgb_nu_K}(c,f), and, when $\gamma<\delta_K$, shrinks as $\nu_K$ is increased, see Sec.~SM6 Fig.~(S3) the appendix.

\subsection{Influence of the $K$-EV amplitude on coexistence}
 
\label{sec:IVC}

\begin{figure*}[t!]
    \centering
    \includegraphics[width=\linewidth]{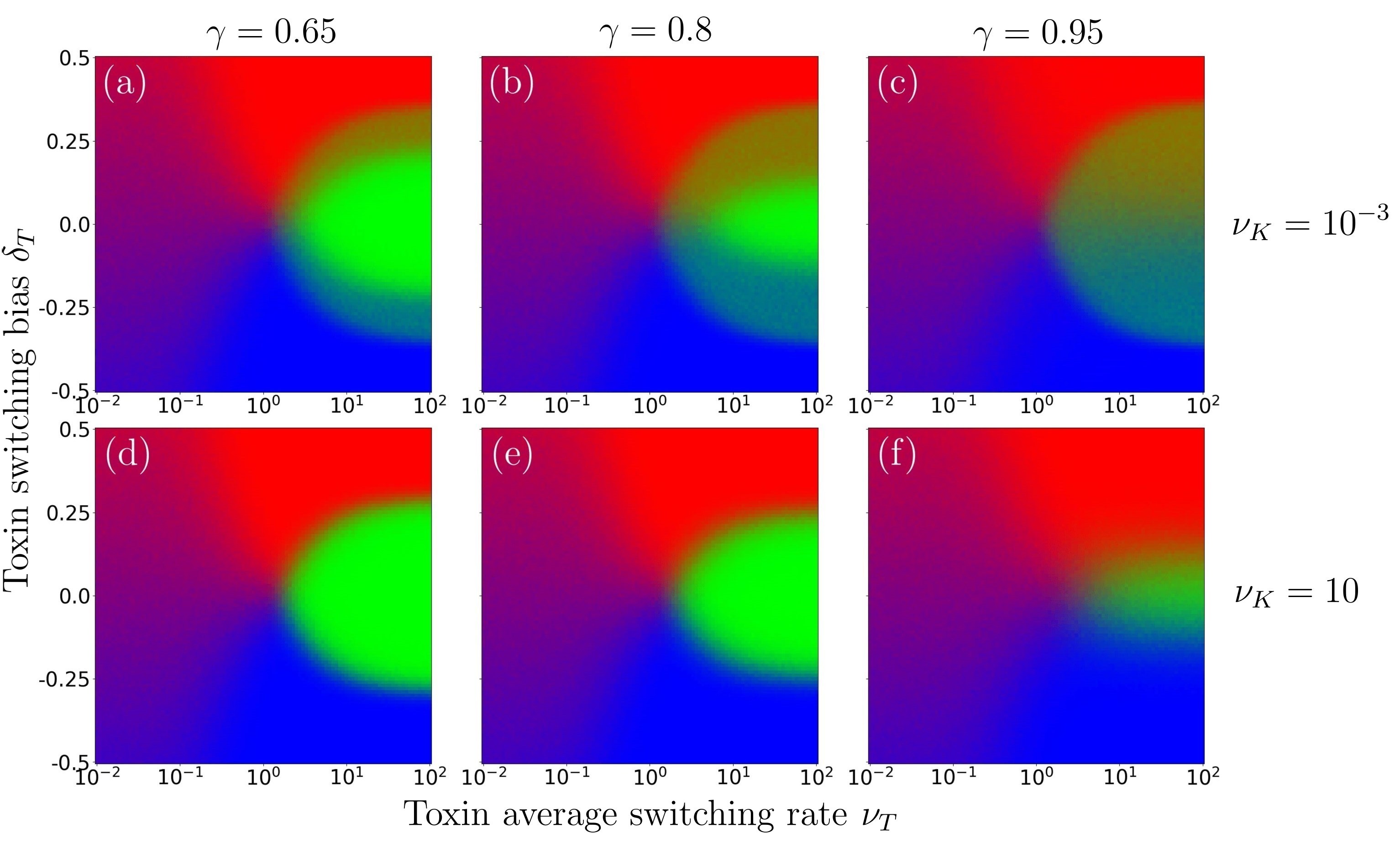}
    \caption{
    Fixation/coexistence  diagrams 
    under $T$-EV and $K$-EV
    in the $(\nu_T,\delta_T)$ parameter space  showing the effect of the amplitude of $K$-EV $\gamma$ on the fixation and coexistence probabilities {when $K_0=500$ is kept fixed},   
    for   $\gamma=0.65$ $\left(K_-=175,\right.$ $\left.K_+=825\right)$ (a,d), $\gamma=0.8$ $\left(K_-=100,\right.$ $\left.K_+=900\right)$ (b,e), and $\gamma=0.95$ $\left(K_-=25,\right.$ $\left.K_+=975\right)$ (c,f). Other parameters are 
     $s=1.0$, $\delta_K=0$, $\nu_K=10^{-3}$ in (a)-(c) and $\nu_K=10$ in (d)-(f).
    Phase diagrams obtained from stochastic simulations
    data collected at $t=2\langle N\rangle$ over $10^3$ realisations.
    All diagrams are colour-coded {(greyscale-coded)}  as in Fig.~\ref{fig:moranheatmap}.
    }
    \label{fig:gamma_change}
\end{figure*}

We have seen that increasing the selection bias $s$, raises the amplitude of the $T$-EV and facilitates the emergence of long-lived coexistence. Here, 
{by keeping $K_0$ constant},
we investigate the influence of the parameter $\gamma$, which controls the amplitude of $K$-EV, on the fixation/coexistence diagrams. When $\gamma \to 1$ and $K_0\gg 1$, there is  $K$-EV of large amplitude, with the population subject to a harsh population bottleneck ($K_-\to 0$) accompanied by strong demographic fluctuations. 
The latter facilitate fixation of either strain and counter the 
effect of $T$-EV that drives the community to 
coexistence. Results of Fig.~\ref{fig:gamma_change} illustrate the influence of $\gamma$ under low and high $K$-switching rate ($\delta_K=0$):
\begin{enumerate}
 \item[-] Under low $\nu_K$, the 
 probability of long-lived coexistence $\eta$ 
 decreases together with the value of $K_-=K_0(1-\gamma)$
 when $\gamma$ is increased (all other parameters being kept fixed). As a consequence, 
 the  bright green {(bright light grey)}  region 
 in Fig.~\ref{fig:gamma_change}(a) where 
  long-lived coexistence is almost certain ($\eta\approx 1$) shrinks with $\gamma$ and is gradually replaced by a faint green {(faint light grey)} area where coexistence occurs with a lower probability ($\eta=(1+\delta_K)/2< 1$), see  Fig.~\ref{fig:gamma_change}(b,c).
   \item[-] Under high $\nu_K$, we have  $N\approx \mathcal{K}$ and the effect of $\gamma$ is encoded in the expression of $\mathcal{K}= K_0 (1-\gamma^2)/(1-\gamma\delta_K)$. When $\delta_K\leq 0$, 
 $\mathcal{K}$ and $\eta$ decrease with $\gamma$, and as  a result the 
  bright green {(bright light grey)}  region 
 in Fig.~\ref{fig:gamma_change}(d) shrinks and is eventually replaced by a smaller faint green {(faint light grey)}  region where coexistence is possible but not certain ($0<\eta< 1$), see  Fig.~\ref{fig:gamma_change}(e,f).
 When $\delta_K> 0$,
 there is a bias towards $K=K_+$ and 
 $\mathcal{K}$ increases with $\gamma$ until 
 $\gamma=\bar{\gamma}\equiv(1-\sqrt{1-\delta_K^2})/\delta_K$ and then decreases, with $\mathcal{K}<K_0$, when $\gamma>\delta_K$. This results in a non-monotonic dependence of the coexistence region where $\eta\approx 1$: under  $\nu_K\gg 1$ and $\delta_K>0$, the long-lived (bright-green) coexistence region grows with $\gamma$ up to $\bar{\gamma}$ and shrinks when $\gamma> \bar{\gamma}$. 
\end{enumerate}
 
We have thus found that the 
 environmental fluctuations have 
 opposite effects on the species coexistence:  increasing the amplitude of $T$-EV (by raising $s$) prolongs the coexistence of the strains and expands the coexistence region, but raising the amplitude of  $K$-EV (by raising $\gamma$)
 can  significantly reduce the probability of long-lived coexistence for all values of $\nu_K$.

\section{Make-up of the coexistence phase and strains average  abundance}
\label{sec:V}
Having characterised in detail the conditions under which long-lived coexistence and fixation occur, we now study the make-up of the coexistence phase and then use this result to determine the stationary average abundance of each strain. 

\subsection{Coexistence phase make-up}
\label{sec:Va}
\begin{figure}[t!]
    \centering
    \includegraphics[width=1.0\linewidth]{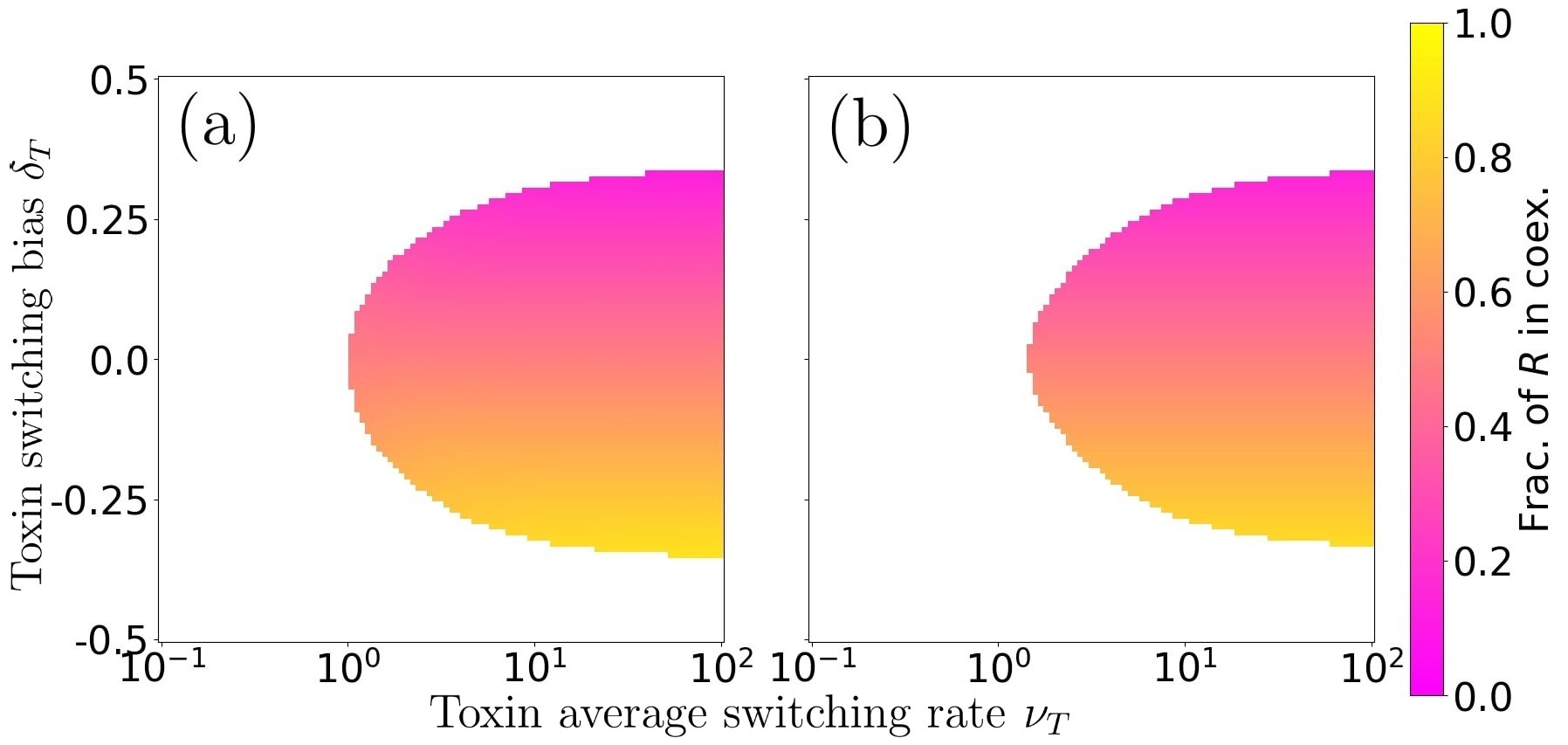}
    \caption{Make-up of  the coexistence state: fraction of the resistant strain $R$ in the coexistence phase as function of $\nu_T$ and $\delta_T$;     from simulation results (a) and  predictions of
     Eqs.~\eqref{eq:mean_field_coex} and \eqref{eq:coexistenceprobability} in (b),   {respectively for $x^*$ and $\eta$}. 
     The colourbar {(brightness bar)}  gives the characteristic fraction of the $R$ strain in the region of the $\nu_T$-$\delta_T$ parameter space where $\eta>0.01$.  Simulation results have been obtained just after $t=2\langle N\rangle$ and averaged over $10^4$ realisations. 
     Parameters  are $K_0=500$, $\gamma=0.5$, $\nu_K=100$, and $s=1$.}
    \label{fig:xcomposition}
\end{figure}

We are interested in the characteristic fraction of the resistant strain  $R$ in the coexistence phase, here defined as $x^*$. This is the fraction of $R$ expected, given that we have coexistence at $t=2\avg{N}$. According to the mean-field theory, the fraction of the  strain $R$ in the coexistence phase is given by the expression Eq.~\eqref{eq:mean_field_coex}  of $x^*$. It turns out that 
deep into the coexistence region whereby $\eta \approx 1$ and $\nu_T$ is sufficiently high, there is good agreement between theory and simulations, see Fig.~\ref{fig:xcomposition}(a,b). In addition, even when $\eta<1$, the {mean-field prediction $x^*$} remains remarkably  {close to the value of fraction of $R$ measured in the coexistence state obtained in simulations, with small deviations arising} as $\eta$ approaches 0. We notice that the characteristic fraction of $R$, for given $\delta_T$, is almost independent of $\nu_T$. 

We can also predict the fraction of $R$ regardless of coexistence or fixation, here denoted by $\avg{x}$. 
The quantity $\avg{x}$ thus characterises the fraction of $R$ in the coexistence, fixation, and crossover regime  where both coexistence and fixation are possible, with respective probabilities $\eta$ and $\phi$, but neither is certain.
Making use of Eqs.~\eqref{eq:mean_field_coex}, \eqref{eq:overallfixationprobability}, and \eqref{eq:coexistenceprobability} we thus define $\avg{x}$ as
\begin{equation}
\label{eq:avgx}
    \avg{x}=\eta x^* + (1-\eta)\phi.
\end{equation}
This captures well the dependence of $\avg{x}$ on $\nu_T$ and reduces to the fraction of $R$ in  the coexistence phase, $\avg{x}=x^*$, when $\eta\approx 1$ and long-lived coexistence is almost certain (see SM5, Fig.~S2). As shown in Sec.~SM5 of the appendix, a closed-form alternative to $\avg{x}$ is provided by the modal value of the stationary PDF of the PDMP defined by Eq.~\eqref{eq:MF}, which, while less accurate than $\avg{x}$, matches qualitatively well to simulations.
\subsection{Strain average abundance}
\label{sec:Vb}
\begin{figure}
    \centering
    \includegraphics[width=0.8\linewidth]{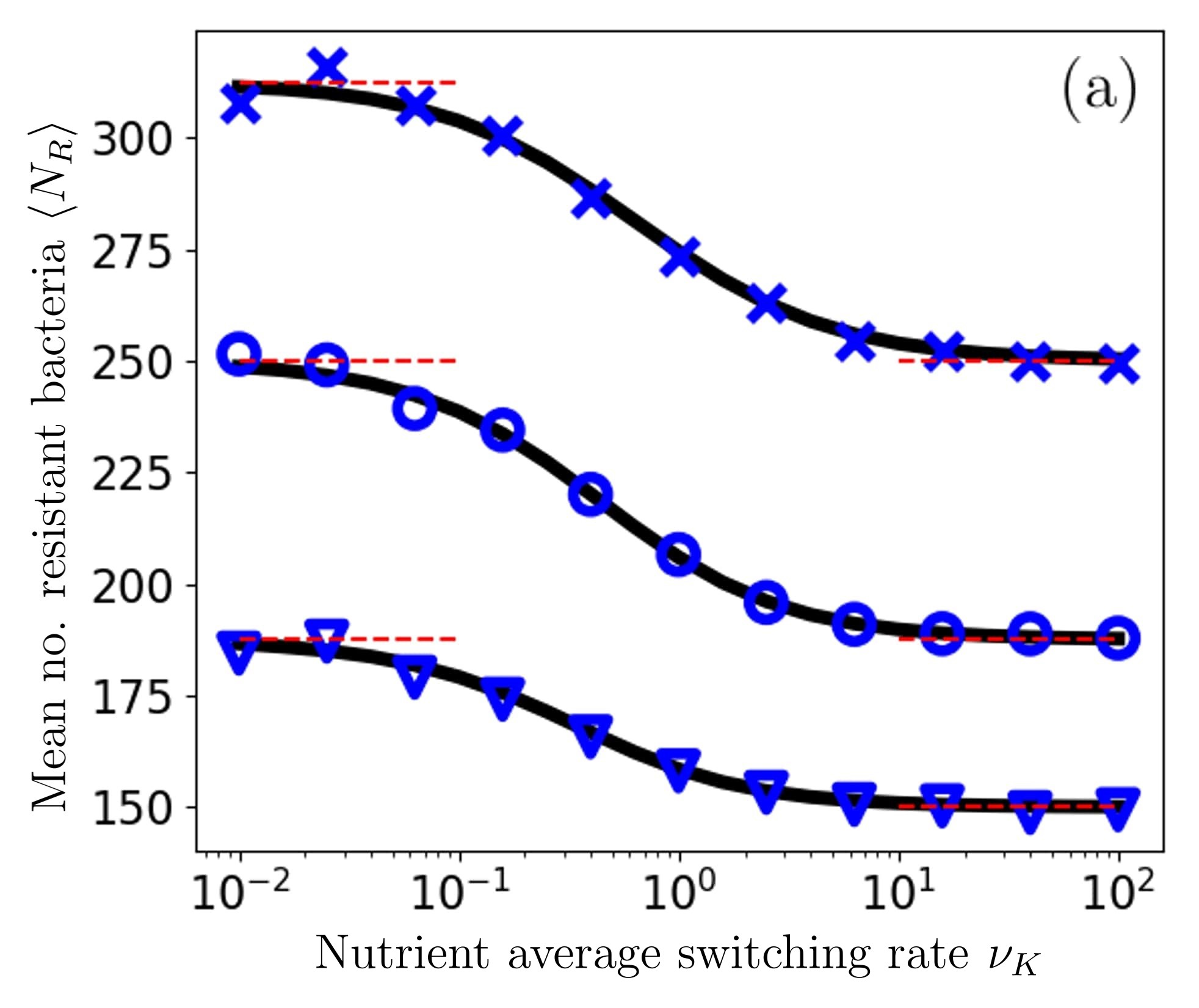}\\
    \includegraphics[width=0.8\linewidth]{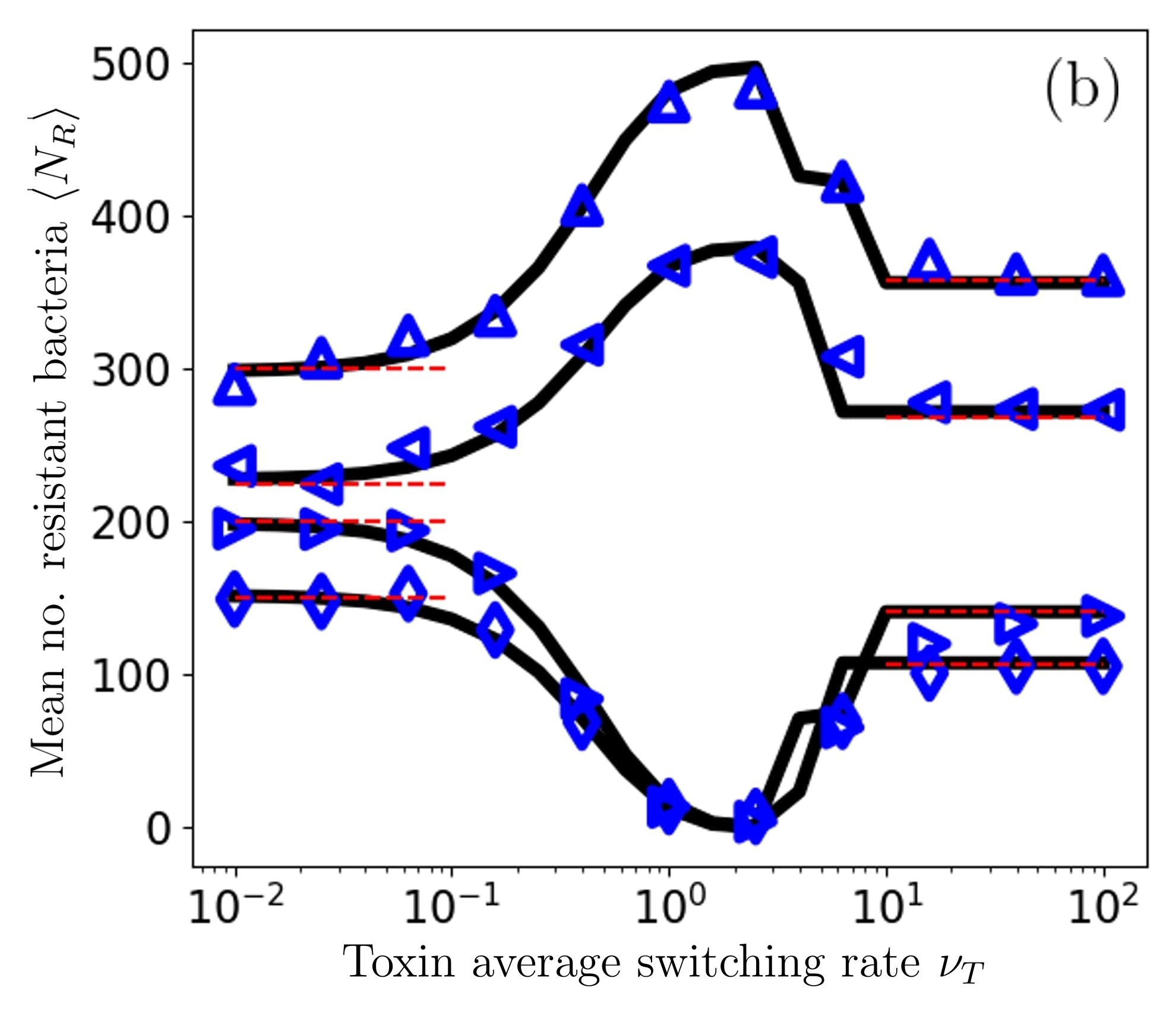}
    \caption{Long-time average $R$ abundance $\langle N_R \rangle$
    as function of the {average} switching rate of the $T/K$-EV. Solid lines are theoretical predictions of Eq.~\eqref{eq:N_R_average} with $x^*$, $\langle N\rangle$, $\phi$ and $\eta$
    given by Eqs.~\eqref{eq:mean_field_coex},  \eqref{eq:mean_N}-\eqref{eq:coexistenceprobability};
     symbols are from simulation data.
    (a)  $\langle N_R \rangle$ versus $\nu_K$ for $\delta_K=0.5$ ($\mathbf{\times}$), $\delta_K=0$ ($\mathbf{\bigcirc}$) and $\delta_K=-0.5$ ($\mathbf{\bigtriangledown}$). Limiting value are plotted for $\nu_K\rightarrow0,\infty$ as dotted lines. Other parameters are:  $K_0=500$, $\gamma=0.5$, $s=1.0$, $\nu_T=10$, and $\delta_T=0.0$.
    (b)  $\langle N_R \rangle$ versus $\nu_T$ for $(\delta_T,\nu_K)=(-0.2,0.01)$ ($\bigtriangleup$), $(\delta_T,\nu_K)=(-0.2,10)$ (\rotatebox[origin=c]{90}{$\bigtriangleup$}),$(\delta_T,\nu_K)=(0.2,0.01)$ (\rotatebox[origin=c]{270}{$\bigtriangleup$}), $(\delta_T,\nu_K)=(0.2,10)$ ($\lozenge$). Limiting value are plotted for $\nu_T\rightarrow0,\infty$ as dotted lines.
    Other parameters are:
    $K_0=500$, $\gamma=0.5$, $s=1.0$, and $\delta_K=0.0$.
    Simulations data have been collected and sample-averaged after $t=2\langle K\rangle$ over $10^2$ realisations. See text for details.}
    \label{fig:N_R varying}
\end{figure}
In this section, we study the (quasi-)stationary average abundance of the strains $R$ and $S$, respectively denoted by $\avg{N_R}$ and  $\avg{N_S}$.
Since $\avg{N_S}=\avg{N}-\avg{N_R}$, and $\avg{N}$ is well described by Eq.~\eqref{eq:mean_N}, see Fig.~\ref{fig:N_vs_nu_K}, we only need to focus on studying $\avg{N_R}$. 

In fact, while the evolution of $N$
is governed by $K$-EV and is well-captured by the stochastic logistic equation Eq.~\eqref{eq:meanfieldtotala} and the corresponding $N$-PDMP, the dynamics of the abundance of the $R$ strain depends on both $T$-EV  and $K$-EV. In the  mean-field limit, where demographic fluctuations are neglected, we indeed have \cite{Gardiner}
\begin{align}
\nonumber
    \dot{N}_R=T_R^+-T_R^-&=\left(\frac{1}{\overline{f}(t)}-\frac{N}{K(t)}\right)N_R\\ &=\left(\frac{1}{x+(1-x)e^{s\xi_T}}-\frac{N}{K_0(1+\gamma\xi_K)}\right)N_R, \nonumber
\end{align}
which is a stochastic differential equation depending on both 
 $\xi_K$ and $\xi_T$, and coupled to
the $N$- and $x$-PDMPs defined respectively by Eqs.~\eqref{eq:meanfieldtotala} and \eqref{eq:meanfieldtotalb}. In the dominance regimes, $\avg{N_R} \approx 0$ ($S$ dominance) or $\avg{N_R} \approx\avg{N}$ ($R$ dominance), 
which can be obtained from Eq.~\eqref{eq:mean_N}. However, finding  $\avg{N_R}$ in the coexistence phase is a nontrivial task. Progress can be made 
by noticing that, $\xi_K$ and $\xi_T$ being independent, we can write 
\begin{equation}
\label{eq:N_R_average}
    \avg{N_R} \approx \avg{N}\avg{x}\equiv\avg{N}(\eta x^* + (1-\eta)\phi),
\end{equation}
where $\avg{N}\eta x^*$ is the contribution to $\avg{N_R}$ when there is coexistence (with probability $\eta$), and $\avg{N}(1-\eta)\phi$ is the contribution arising when there is fixation of the strain $R$ (with probability $(1-\eta)\phi$).
In our theoretical analysis, $x^*, \avg{N}, \phi$ and $\eta$
are obtained from Eqs.~\eqref{eq:mean_field_coex} and \eqref{eq:mean_N}-\eqref{eq:coexistenceprobability}. Eq.~\eqref{eq:N_R_average}
thus captures the behaviour of $\avg{N_R}$ in each regime: the dominance regime where $\eta\approx 0$ and we have $\avg{N_R}\approx\avg{N}\phi$,
deep in the coexistence phase where we have  $\eta\approx 1$ and 
$\avg{N_R}\approx \avg{N} x^*$, and where $0<\eta< 1$ and coexistence is possible but not certain where we have  
$\avg{N_R} \approx \avg{N}\avg{x}$.

In Fig.~\ref{fig:N_R varying}, we find that the theoretical predictions based on Eq.~\eqref{eq:N_R_average} agree well with simulation results over a broad range of  $\nu_K$ and $\nu_T$, and for  different values of $\delta_K$ and $\delta_T$. The dependence of 
$\avg{N_R}$ on $\nu_K$ reflects that of $\avg{N}$ shown  in Fig.~\ref{fig:N_vs_nu_K}: 
$\avg{N_R}$ 
decreases with $\nu_K$ at fixed $\delta_K$, see Fig.~\ref{fig:N_R varying}(a), and we have $\avg{N}\approx {\cal K}$ when 
$\nu_K\to \infty$ yielding $\avg{N_R}\approx {\cal K}x^*$ 
deep in the coexistence phase where $\nu_T\gg 1$, 
and similarly $\avg{N}\approx K_0 (1+\gamma\delta_K)$ when $\nu_K\to 0$ yields  $\avg{N_R}\approx K_0 (1+\gamma\delta_K) x^*$. Not shown in Fig.~8(a) is the case of $\nu_T\ll1$, whereby we have only dominance such that $\avg{N_R}\approx\avg{N}(1-\delta_T)/2$.
Fig.~\ref{fig:N_R varying}(b) shows that the dependence of 
$\avg{N_R}$ on $\nu_T$ can be non-monotonic and exhibit an extreme dip ($\delta_T>0$) or peak ($\delta_T<0$) at intermediate $T$-switching rate, $\nu_T\sim 1$. This behaviour can be understood by referring to the diagrams of Fig.~\ref{fig:gamma_change}:
 as $\nu_T$ is raised from $\nu_T=0$ with $\delta_T<0$ kept fixed,  the $R$ fixation 
probability  first slowly increases across the slightly $R$-dominant phase  where coexistence is unlikely ($\eta\approx0$) and $\avg{N_R}\approx \avg{N}\phi$. 
When $\nu_T$ is increased further and  $R$ is the strongly dominant species (blue {(charcoal)} phases in Fig.~\ref{fig:gamma_change}), with $\phi\approx 1$ and $\avg{N_R}\approx \avg{N}$ is maximal;
coexistence then becomes first possible ($0<\eta<1$, faint green {(faint light gray)} in Fig.~\ref{fig:gamma_change}) and then almost certain ($\eta\approx 1$, bright green {(bright light gray)} in Fig.~\ref{fig:gamma_change}) when $\nu_T$ is increased further, which results in a reduction of the $R$ abundance to $\avg{N_R}\approx \avg{N}x^*<\avg{N}$. A similar reasoning holds for the $S$ strain when $\delta_T>0$ and results in a maximal value $\avg{N_S}\approx \avg{N}$
and therefore a dip of the $R$ abundance, with a minimal value $\avg{N_R}\approx 0$, 
 when $\nu_T\sim 1$.

The results of this section hence show that the twofold EV 
has nontrivial effects on the  make-up of the coexistence phase, and on the average number of cells of each strain, as shown by Fig.~\ref{fig:xcomposition}
and the nonmonotonic dependence of $\avg{N_R}$ on $\nu_T$ in Fig.~\ref{fig:N_R varying}.
 
\section{Conclusion}
\label{sec:VI}
Microorganisms live in environments that unavoidably fluctuate between mild and harsh conditions. Environmental variability can cause endless changes in the  concentration of toxins and amount of available nutrients, and thus shapes the eco-evolutionary properties of microbial communities including the ability of species to coexist.
Understanding under which  circumstances various microbial species can coexist, and how their coexistence and abundance vary with environmental factors, is crucial to shed further light on the mechanisms promoting  biodiversity in ecosystems and  to elucidate the evolution of antimicrobial resistance (AMR).

Motivated by these considerations, and inspired by the antimicrobial resistance evolution in a chemostat setup, 
we have studied the eco-evolutionary dynamics 
of an idealised microbial community of fluctuating size consisting  of two strains competing for the same resources under \emph{twofold environmental variability} ($T$-EV and $K$-EV): the level of toxin and the abundance of nutrients in the community both vary in time.
One of the strains is resistant while the other is sensitive to the drug present in the community, and both  compete for the same resources.

Environmental variability is thus assumed to affect the strains growth and death rates, and is modelled by means of binary randomly time-switching fitness ($T$-EV) and carrying capacity ($K$-EV). 
Under harsh conditions, the level of toxin is high and resources are scarce, while environmental conditions are mild when the level of toxin is low and resources are abundant.
In this setting, the strain resistant to the drug has a selective advantage under high toxin-level, whereas it is outgrown by the  sensitive strain when the level of toxin is low. Moreover, the time-switching carrying capacity  drives the fluctuating size of the microbial community, which in turn modulates the amplitude of the demographic fluctuations, resulting in their coupling with the variation of the available resources.
 When the environment is static, there is no lasting coexistence since one species dominates and rapidly fixates the entire population.\\
 Here, we have shown that this picture changes radically in fluctuating environments: we have indeed found that long-lived species coexistence is possible in the presence of environmental fluctuations. Using  stochastic simulations  and  the properties of suitable piecewise-deterministic 
 and Moran processes, we have  computationally and analytically
 obtained the  fixation-coexistence phase diagrams of the system. These have allowed us to 
 identify the detailed environmental conditions under which species coexist almost certainly for extended period of times, and the phases
 where one species dominates, as well as the crossover regimes where   both coexistence and fixation are possible but not guaranteed. We have found that long-lived coexistence requires sufficient variation of the toxin level, while resource variability  can oppose coexistence when strong $K$-EV leads to population bottlenecks responsible for large demographic fluctuations facilitating fixation. More generally, our analysis has allowed us to assess the influence of the population size distribution, whose shape changes greatly with the rate of $K$-EV,   on the fixation-coexistence phase diagram.
 We have also determined 
 how the  make-up of the coexistence phase and average abundance of each strain depend on the rates of environmental change.

 Environmental variability generally comes about in many forms in a variety of settings throughout biology and ecology, and the conundrum of coexistence within a system is impacted by it, alongside demographic fluctuations. This leads to complex   eco-evolutionary dynamics. 
 In particular, how microbial communities evolve {subject} to environmental variability is vital when considering the issue of AMR, so that the effectiveness of treatments can be maximised, while minimising the harmful effects. 
 In considering  twofold environmental variations, we have shown that these can have qualitative effects on the population evolution as they can either promote or jeopardise lasting species coexistence. 

In summary, our analysis allows us to understand under which circumstances environmental variability,  together with demographic fluctuations, favours or hinders the long-lived coexistence of competing species, and how it affects the  fraction and abundance of each strain in the community.
{This work hence contributes to further elucidate  the role of  fluctuations on the maintenance of biodiversity in complex ecosystems.}

In particular, our findings  demonstrate the influence of environmental fluctuations on  biodiversity in microbial communities, and may thus have potential impacts on
 numerous  applications. For instance, the model studied here 
 is well suited to describe the {\it in vitro}
 evolution of antimicrobial resistance 
  in a chemostat setup where the level of antibiotics would fluctuate below and above the minimum inhibitory concentration. In this context, the model is able to predict, under a broad range of external constraints, the best conditions to avoid the fixation of the strain resistant to the drug and when both strains coexist.  A more realistic model of AMR evolution would take into account that the drug resistance is often mediated by a form of public goods~\cite{Yurtsev13,LARM23}, {and that there may exist more than two competing species and various toxins. The eco-evolutionary dynamics  of communities consisting of multiple  species, resources and  toxins can generally not be simply inferred from those of  two-species eco-systems, even though in some cases simple models can  be  
illuminating~\cite{shibasaki_exclusion_2021}.}
  {Another potential application of 
  the model considered here, with varying drug levels, concerns 
  the so-called adaptive therapy used in cancer treatment to 
   prevent or delay the cancer from becoming completely drug resistant \cite{adaptive_therapy}.}
  \\

\section*{Data accessibility}
{The data that support the findings of this study are openly available, see \cite{data}.}\\

\section*{Author Contributions}
\noindent
{\bf  Matthew Asker:} Conceptualisation (supporting), Methodology, Formal Analysis (lead), Software, Writing - Original Draft, Writing - Review \& Editing, Visualisation, Investigation, Validation.\\ {\bf Llu\'is Hern\'andez-Navarro:} Formal Analysis (supporting),  Writing - Review \& Editing (supporting), Supervision (supporting).\\  {\bf Alastair M. Rucklidge:} Writing - Review \& Editing (supporting), Supervision (supporting),  Funding acquisition (supporting).\\ {\bf Mauro Mobilia:} Conceptualisation (lead), Methodology (lead), Formal Analysis (supporting), Writing - Original Draft, Writing - Review \& Editing,  Visualisation, Supervision (lead), Project administration, Funding acquisition (lead).\\

\begin{acknowledgments}
We would like to thank K. Distefano, J. Jiménez, S. Muñoz-Montero, M. Pleimling, M. Swailem, and U.~C. T\"auber for  helpful discussions. L.H.N., A.M.R and M.M. gratefully acknowledge funding from the U.K.
Engineering and Physical Sciences Research Council (EPSRC)
under the grant No. EP/V014439/1
for the project `DMS-EPSRC Eco-Evolutionary Dynamics of Fluctuating Populations’. The support of a Ph.D. scholarship to M.A.
by the EPSRC grant No. EP/T517860/1  is also thankfully acknowledged.  This work was undertaken on ARC4,
part of the High Performance Computing facilities at the
University of Leeds, UK.
\end{acknowledgments}
\bibliography{bibliography.bib}

\begin{thebibliography}{114}%
\makeatletter
\providecommand \@ifxundefined [1]{%
 \@ifx{#1\undefined}
}%
\providecommand \@ifnum [1]{%
 \ifnum #1\expandafter \@firstoftwo
 \else \expandafter \@secondoftwo
 \fi
}%
\providecommand \@ifx [1]{%
 \ifx #1\expandafter \@firstoftwo
 \else \expandafter \@secondoftwo
 \fi
}%
\providecommand \natexlab [1]{#1}%
\providecommand \enquote  [1]{``#1''}%
\providecommand \bibnamefont  [1]{#1}%
\providecommand \bibfnamefont [1]{#1}%
\providecommand \citenamefont [1]{#1}%
\providecommand \href@noop [0]{\@secondoftwo}%
\providecommand \href [0]{\begingroup \@sanitize@url \@href}%
\providecommand \@href[1]{\@@startlink{#1}\@@href}%
\providecommand \@@href[1]{\endgroup#1\@@endlink}%
\providecommand \@sanitize@url [0]{\catcode `\\12\catcode `\$12\catcode
  `\&12\catcode `\#12\catcode `\^12\catcode `\_12\catcode `\%12\relax}%
\providecommand \@@startlink[1]{}%
\providecommand \@@endlink[0]{}%
\providecommand \url  [0]{\begingroup\@sanitize@url \@url }%
\providecommand \@url [1]{\endgroup\@href {#1}{\urlprefix }}%
\providecommand \urlprefix  [0]{URL }%
\providecommand \Eprint [0]{\href }%
\providecommand \doibase [0]{https://doi.org/}%
\providecommand \selectlanguage [0]{\@gobble}%
\providecommand \bibinfo  [0]{\@secondoftwo}%
\providecommand \bibfield  [0]{\@secondoftwo}%
\providecommand \translation [1]{[#1]}%
\providecommand \BibitemOpen [0]{}%
\providecommand \bibitemStop [0]{}%
\providecommand \bibitemNoStop [0]{.\EOS\space}%
\providecommand \EOS [0]{\spacefactor3000\relax}%
\providecommand \BibitemShut  [1]{\csname bibitem#1\endcsname}%
\let\auto@bib@innerbib\@empty
\bibitem [{\citenamefont {Vasi}\ \emph {et~al.}(1994)\citenamefont {Vasi},
  \citenamefont {Travisano},\ and\ \citenamefont
  {Lenski}}]{vasi_long-term_1994}%
  \BibitemOpen
  \bibfield  {author} {\bibinfo {author} {\bibfnamefont {F.}~\bibnamefont
  {Vasi}}, \bibinfo {author} {\bibfnamefont {M.}~\bibnamefont {Travisano}},\
  and\ \bibinfo {author} {\bibfnamefont {R.~E.}\ \bibnamefont {Lenski}},\
  }\href {https://doi.org/10.1086/285685} {\bibfield  {journal} {\bibinfo
  {journal} {Am. Nat.}\ }\textbf {\bibinfo {volume} {144}},\ \bibinfo {pages}
  {432} (\bibinfo {year} {1994})}\BibitemShut {NoStop}%
\bibitem [{\citenamefont {Proft}(2009)}]{proft2009microbial}%
  \BibitemOpen
  \bibfield  {author} {\bibinfo {author} {\bibfnamefont {T.}~\bibnamefont
  {Proft}},\ }\href@noop {} {\emph {\bibinfo {title} {Microbial Toxins: Current
  Research and Future Trends}}}\ (\bibinfo  {publisher} {Caister Academic
  Press},\ \bibinfo {year} {2009})\BibitemShut {NoStop}%
\bibitem [{\citenamefont {Himeoka}\ and\ \citenamefont
  {Mitarai}(2020)}]{himeoka_dynamics_2020}%
  \BibitemOpen
  \bibfield  {author} {\bibinfo {author} {\bibfnamefont {Y.}~\bibnamefont
  {Himeoka}}\ and\ \bibinfo {author} {\bibfnamefont {N.}~\bibnamefont
  {Mitarai}},\ }\href {https://doi.org/10.1103/PhysRevResearch.2.013372}
  {\bibfield  {journal} {\bibinfo  {journal} {Phys. Rev. Res.}\ }\textbf
  {\bibinfo {volume} {2}},\ \bibinfo {pages} {013372} (\bibinfo {year}
  {2020})}\BibitemShut {NoStop}%
\bibitem [{\citenamefont {Tu}\ \emph {et~al.}(2020)\citenamefont {Tu},
  \citenamefont {Chi}, \citenamefont {Bodnar}, \citenamefont {Zhang},
  \citenamefont {Gao}, \citenamefont {Bian}, \citenamefont {Stewart},
  \citenamefont {Fry},\ and\ \citenamefont {Lu}}]{Tu20}%
  \BibitemOpen
  \bibfield  {author} {\bibinfo {author} {\bibfnamefont {P.}~\bibnamefont
  {Tu}}, \bibinfo {author} {\bibfnamefont {L.}~\bibnamefont {Chi}}, \bibinfo
  {author} {\bibfnamefont {W.}~\bibnamefont {Bodnar}}, \bibinfo {author}
  {\bibfnamefont {Z.}~\bibnamefont {Zhang}}, \bibinfo {author} {\bibfnamefont
  {B.}~\bibnamefont {Gao}}, \bibinfo {author} {\bibfnamefont {X.}~\bibnamefont
  {Bian}}, \bibinfo {author} {\bibfnamefont {J.}~\bibnamefont {Stewart}},
  \bibinfo {author} {\bibfnamefont {R.}~\bibnamefont {Fry}},\ and\ \bibinfo
  {author} {\bibfnamefont {K.}~\bibnamefont {Lu}},\ }\href
  {https://doi.org/10.3390/toxics8010019} {\bibfield  {journal} {\bibinfo
  {journal} {Toxics}\ }\textbf {\bibinfo {volume} {8}},\ \bibinfo {pages} {19}
  (\bibinfo {year} {2020})}\BibitemShut {NoStop}%
\bibitem [{\citenamefont {Hooper}\ \emph {et~al.}(2005)\citenamefont {Hooper},
  \citenamefont {Chapin~III}, \citenamefont {Ewel}, \citenamefont {Hector},
  \citenamefont {Inchausti}, \citenamefont {Lavorel}, \citenamefont {Lawton},
  \citenamefont {Lodge}, \citenamefont {Loreau}, \citenamefont {Naeem},
  \citenamefont {Schmid}, \citenamefont {Setälä}, \citenamefont {Symstad},
  \citenamefont {Vandermeer},\ and\ \citenamefont
  {Wardle}}]{hooper_effects_2005}%
  \BibitemOpen
  \bibfield  {author} {\bibinfo {author} {\bibfnamefont {D.~U.}\ \bibnamefont
  {Hooper}}, \bibinfo {author} {\bibfnamefont {F.~S.}\ \bibnamefont
  {Chapin~III}}, \bibinfo {author} {\bibfnamefont {J.~J.}\ \bibnamefont
  {Ewel}}, \bibinfo {author} {\bibfnamefont {A.}~\bibnamefont {Hector}},
  \bibinfo {author} {\bibfnamefont {P.}~\bibnamefont {Inchausti}}, \bibinfo
  {author} {\bibfnamefont {S.}~\bibnamefont {Lavorel}}, \bibinfo {author}
  {\bibfnamefont {J.~H.}\ \bibnamefont {Lawton}}, \bibinfo {author}
  {\bibfnamefont {D.~M.}\ \bibnamefont {Lodge}}, \bibinfo {author}
  {\bibfnamefont {M.}~\bibnamefont {Loreau}}, \bibinfo {author} {\bibfnamefont
  {S.}~\bibnamefont {Naeem}}, \bibinfo {author} {\bibfnamefont
  {B.}~\bibnamefont {Schmid}}, \bibinfo {author} {\bibfnamefont
  {H.}~\bibnamefont {Setälä}}, \bibinfo {author} {\bibfnamefont {A.~J.}\
  \bibnamefont {Symstad}}, \bibinfo {author} {\bibfnamefont {J.}~\bibnamefont
  {Vandermeer}},\ and\ \bibinfo {author} {\bibfnamefont {D.~A.}\ \bibnamefont
  {Wardle}},\ }\href {https://doi.org/10.1890/04-0922} {\bibfield  {journal}
  {\bibinfo  {journal} {Ecol. Monogr.}\ }\textbf {\bibinfo {volume} {75}},\
  \bibinfo {pages} {3} (\bibinfo {year} {2005})}\BibitemShut {NoStop}%
\bibitem [{\citenamefont {Fux}\ \emph {et~al.}(2005)\citenamefont {Fux},
  \citenamefont {Costerton}, \citenamefont {Stewart},\ and\ \citenamefont
  {Stoodley}}]{fux_survival_2005}%
  \BibitemOpen
  \bibfield  {author} {\bibinfo {author} {\bibfnamefont {C.~A.}\ \bibnamefont
  {Fux}}, \bibinfo {author} {\bibfnamefont {J.~W.}\ \bibnamefont {Costerton}},
  \bibinfo {author} {\bibfnamefont {P.~S.}\ \bibnamefont {Stewart}},\ and\
  \bibinfo {author} {\bibfnamefont {P.}~\bibnamefont {Stoodley}},\ }\href
  {https://doi.org/10.1016/j.tim.2004.11.010} {\bibfield  {journal} {\bibinfo
  {journal} {Trends Microbiol.}\ }\textbf {\bibinfo {volume} {13}},\ \bibinfo
  {pages} {34} (\bibinfo {year} {2005})}\BibitemShut {NoStop}%
\bibitem [{\citenamefont {Brockhurst}\ \emph {et~al.}(2007)\citenamefont
  {Brockhurst}, \citenamefont {Buckling},\ and\ \citenamefont
  {Gardner}}]{brockhurst_cooperation_2007}%
  \BibitemOpen
  \bibfield  {author} {\bibinfo {author} {\bibfnamefont {M.~A.}\ \bibnamefont
  {Brockhurst}}, \bibinfo {author} {\bibfnamefont {A.}~\bibnamefont
  {Buckling}},\ and\ \bibinfo {author} {\bibfnamefont {A.}~\bibnamefont
  {Gardner}},\ }\href {https://doi.org/10.1016/j.cub.2007.02.057} {\bibfield
  {journal} {\bibinfo  {journal} {Curr. Biol.}\ }\textbf {\bibinfo {volume}
  {17}},\ \bibinfo {pages} {761} (\bibinfo {year} {2007})}\BibitemShut
  {NoStop}%
\bibitem [{\citenamefont {Acar}\ \emph {et~al.}(2008)\citenamefont {Acar},
  \citenamefont {Mettetal},\ and\ \citenamefont {van
  Oudenaarden}}]{acar_stochastic_2008}%
  \BibitemOpen
  \bibfield  {author} {\bibinfo {author} {\bibfnamefont {M.}~\bibnamefont
  {Acar}}, \bibinfo {author} {\bibfnamefont {J.~T.}\ \bibnamefont {Mettetal}},\
  and\ \bibinfo {author} {\bibfnamefont {A.}~\bibnamefont {van Oudenaarden}},\
  }\href {https://doi.org/10.1038/ng.110} {\bibfield  {journal} {\bibinfo
  {journal} {Nat. Genet.}\ }\textbf {\bibinfo {volume} {40}},\ \bibinfo {pages}
  {471} (\bibinfo {year} {2008})}\BibitemShut {NoStop}%
\bibitem [{\citenamefont {Caporaso}\ \emph {et~al.}(2011)\citenamefont
  {Caporaso}, \citenamefont {Lauber}, \citenamefont {Costello}, \citenamefont
  {Berg-Lyons}, \citenamefont {Gonzalez}, \citenamefont {Stombaugh},
  \citenamefont {Knights}, \citenamefont {Gajer}, \citenamefont {Ravel},
  \citenamefont {Fierer}, \citenamefont {Gordon},\ and\ \citenamefont
  {Knight}}]{caporaso_moving_2011}%
  \BibitemOpen
  \bibfield  {author} {\bibinfo {author} {\bibfnamefont {J.~G.}\ \bibnamefont
  {Caporaso}}, \bibinfo {author} {\bibfnamefont {C.~L.}\ \bibnamefont
  {Lauber}}, \bibinfo {author} {\bibfnamefont {E.~K.}\ \bibnamefont
  {Costello}}, \bibinfo {author} {\bibfnamefont {D.}~\bibnamefont
  {Berg-Lyons}}, \bibinfo {author} {\bibfnamefont {A.}~\bibnamefont
  {Gonzalez}}, \bibinfo {author} {\bibfnamefont {J.}~\bibnamefont {Stombaugh}},
  \bibinfo {author} {\bibfnamefont {D.}~\bibnamefont {Knights}}, \bibinfo
  {author} {\bibfnamefont {P.}~\bibnamefont {Gajer}}, \bibinfo {author}
  {\bibfnamefont {J.}~\bibnamefont {Ravel}}, \bibinfo {author} {\bibfnamefont
  {N.}~\bibnamefont {Fierer}}, \bibinfo {author} {\bibfnamefont {J.~I.}\
  \bibnamefont {Gordon}},\ and\ \bibinfo {author} {\bibfnamefont
  {R.}~\bibnamefont {Knight}},\ }\href
  {https://doi.org/10.1186/gb-2011-12-5-r50} {\bibfield  {journal} {\bibinfo
  {journal} {Genome Biol.}\ }\textbf {\bibinfo {volume} {12}},\ \bibinfo
  {pages} {R50} (\bibinfo {year} {2011})}\BibitemShut {NoStop}%
\bibitem [{\citenamefont {Lindsey}\ \emph {et~al.}(2013)\citenamefont
  {Lindsey}, \citenamefont {Gallie}, \citenamefont {Taylor},\ and\
  \citenamefont {Kerr}}]{Kerr13}%
  \BibitemOpen
  \bibfield  {author} {\bibinfo {author} {\bibfnamefont {H.~A.}\ \bibnamefont
  {Lindsey}}, \bibinfo {author} {\bibfnamefont {J.}~\bibnamefont {Gallie}},
  \bibinfo {author} {\bibfnamefont {S.}~\bibnamefont {Taylor}},\ and\ \bibinfo
  {author} {\bibfnamefont {B.}~\bibnamefont {Kerr}},\ }\href
  {https://doi.org/10.1038/nature11879} {\bibfield  {journal} {\bibinfo
  {journal} {Nature}\ }\textbf {\bibinfo {volume} {494}},\ \bibinfo {pages}
  {463} (\bibinfo {year} {2013})}\BibitemShut {NoStop}%
\bibitem [{\citenamefont {Lambert}\ and\ \citenamefont
  {Kussell}(2015)}]{Lambert15}%
  \BibitemOpen
  \bibfield  {author} {\bibinfo {author} {\bibfnamefont {G.}~\bibnamefont
  {Lambert}}\ and\ \bibinfo {author} {\bibfnamefont {E.}~\bibnamefont
  {Kussell}},\ }\href {https://doi.org/10.1103/PhysRevX.5.011016} {\bibfield
  {journal} {\bibinfo  {journal} {Phys. Rev. X}\ }\textbf {\bibinfo {volume}
  {5}},\ \bibinfo {pages} {011016} (\bibinfo {year} {2015})}\BibitemShut
  {NoStop}%
\bibitem [{\citenamefont {Rescan}\ \emph {et~al.}(2020)\citenamefont {Rescan},
  \citenamefont {Grulois}, \citenamefont {Ortega-Aboud},\ and\ \citenamefont
  {Chevin}}]{rescan_phenotypic_2020}%
  \BibitemOpen
  \bibfield  {author} {\bibinfo {author} {\bibfnamefont {M.}~\bibnamefont
  {Rescan}}, \bibinfo {author} {\bibfnamefont {D.}~\bibnamefont {Grulois}},
  \bibinfo {author} {\bibfnamefont {E.}~\bibnamefont {Ortega-Aboud}},\ and\
  \bibinfo {author} {\bibfnamefont {L.-M.}\ \bibnamefont {Chevin}},\ }\href
  {https://doi.org/10.1038/s41559-019-1089-6} {\bibfield  {journal} {\bibinfo
  {journal} {Nat. Ecol. Evol.}\ }\textbf {\bibinfo {volume} {4}},\ \bibinfo
  {pages} {193} (\bibinfo {year} {2020})}\BibitemShut {NoStop}%
\bibitem [{\citenamefont {Nguyen}\ \emph {et~al.}(2021)\citenamefont {Nguyen},
  \citenamefont {Lara-Gutiérrez},\ and\ \citenamefont {Stocker}}]{Nguyen21}%
  \BibitemOpen
  \bibfield  {author} {\bibinfo {author} {\bibfnamefont {T.}~\bibnamefont
  {Nguyen}}, \bibinfo {author} {\bibfnamefont {J.}~\bibnamefont
  {Lara-Gutiérrez}},\ and\ \bibinfo {author} {\bibfnamefont {R.}~\bibnamefont
  {Stocker}},\ }\href {https://doi.org/10.1093/femsre/fuaa068} {\bibfield
  {journal} {\bibinfo  {journal} {FEMS Microbiol. Rev.}\ }\textbf {\bibinfo
  {volume} {45}},\ \bibinfo {pages} {fuaa068} (\bibinfo {year}
  {2021})}\BibitemShut {NoStop}%
\bibitem [{\citenamefont {Chesson}\ and\ \citenamefont
  {Warner}(1981)}]{chesson_environmental_1981}%
  \BibitemOpen
  \bibfield  {author} {\bibinfo {author} {\bibfnamefont {P.}~\bibnamefont
  {Chesson}}\ and\ \bibinfo {author} {\bibfnamefont {R.~R.}\ \bibnamefont
  {Warner}},\ }\href {https://doi.org/10.1086/283778} {\bibfield  {journal}
  {\bibinfo  {journal} {Am. Nat.}\ }\textbf {\bibinfo {volume} {117}},\
  \bibinfo {pages} {923} (\bibinfo {year} {1981})}\BibitemShut {NoStop}%
\bibitem [{\citenamefont {Chesson}(1994)}]{Chesson94}%
  \BibitemOpen
  \bibfield  {author} {\bibinfo {author} {\bibfnamefont {P.}~\bibnamefont
  {Chesson}},\ }\href {https://doi.org/10.1006/tpbi.1994.1013} {\bibfield
  {journal} {\bibinfo  {journal} {Theor. Popul. Biol.}\ }\textbf {\bibinfo
  {volume} {45}},\ \bibinfo {pages} {227} (\bibinfo {year} {1994})}\BibitemShut
  {NoStop}%
\bibitem [{\citenamefont {Chesson}(2000{\natexlab{a}})}]{Chesson00a}%
  \BibitemOpen
  \bibfield  {author} {\bibinfo {author} {\bibfnamefont {P.}~\bibnamefont
  {Chesson}},\ }\href {https://doi.org/10.1146/annurev.ecolsys.31.1.343}
  {\bibfield  {journal} {\bibinfo  {journal} {Annu. Rev. Ecol. Syst}\ }\textbf
  {\bibinfo {volume} {31}},\ \bibinfo {pages} {343} (\bibinfo {year}
  {2000}{\natexlab{a}})}\BibitemShut {NoStop}%
\bibitem [{\citenamefont {Chesson}(2000{\natexlab{b}})}]{Chesson00b}%
  \BibitemOpen
  \bibfield  {author} {\bibinfo {author} {\bibfnamefont {P.}~\bibnamefont
  {Chesson}},\ }\href {https://doi.org/10.1006/tpbi.2000.1486} {\bibfield
  {journal} {\bibinfo  {journal} {Theor. Popul. Biol.}\ }\textbf {\bibinfo
  {volume} {58}},\ \bibinfo {pages} {211} (\bibinfo {year}
  {2000}{\natexlab{b}})}\BibitemShut {NoStop}%
\bibitem [{\citenamefont {Barab\'as}\ \emph {et~al.}(2018)\citenamefont
  {Barab\'as}, \citenamefont {D'Andrea},\ and\ \citenamefont
  {Stump}}]{Barabas18}%
  \BibitemOpen
  \bibfield  {author} {\bibinfo {author} {\bibfnamefont {G.}~\bibnamefont
  {Barab\'as}}, \bibinfo {author} {\bibfnamefont {R.}~\bibnamefont
  {D'Andrea}},\ and\ \bibinfo {author} {\bibfnamefont {S.}~\bibnamefont
  {Stump}},\ }\href {https://doi.org/10.1002/ecm.1302} {\bibfield  {journal}
  {\bibinfo  {journal} {Ecol. Monogr.}\ }\textbf {\bibinfo {volume} {88}},\
  \bibinfo {pages} {1} (\bibinfo {year} {2018})}\BibitemShut {NoStop}%
\bibitem [{\citenamefont {Abdul-Rahman}\ \emph {et~al.}(2021)\citenamefont
  {Abdul-Rahman}, \citenamefont {Tranchina},\ and\ \citenamefont
  {Gresham}}]{Abdul21}%
  \BibitemOpen
  \bibfield  {author} {\bibinfo {author} {\bibfnamefont {F.}~\bibnamefont
  {Abdul-Rahman}}, \bibinfo {author} {\bibfnamefont {D.}~\bibnamefont
  {Tranchina}},\ and\ \bibinfo {author} {\bibfnamefont {D.}~\bibnamefont
  {Gresham}},\ }\href {https://doi.org/10.1093/molbev/msab173} {\bibfield
  {journal} {\bibinfo  {journal} {Mol. Biol. Evol.}\ }\textbf {\bibinfo
  {volume} {38}},\ \bibinfo {pages} {msab173} (\bibinfo {year}
  {2021})}\BibitemShut {NoStop}%
\bibitem [{\citenamefont {Eliopoulos}\ \emph {et~al.}(2003)\citenamefont
  {Eliopoulos}, \citenamefont {Cosgrove},\ and\ \citenamefont
  {Carmeli}}]{eliopoulos_impact_2003}%
  \BibitemOpen
  \bibfield  {author} {\bibinfo {author} {\bibfnamefont {G.~M.}\ \bibnamefont
  {Eliopoulos}}, \bibinfo {author} {\bibfnamefont {S.~E.}\ \bibnamefont
  {Cosgrove}},\ and\ \bibinfo {author} {\bibfnamefont {Y.}~\bibnamefont
  {Carmeli}},\ }\href {https://doi.org/10.1086/375081} {\bibfield  {journal}
  {\bibinfo  {journal} {Clin. Infect. Dis.}\ }\textbf {\bibinfo {volume}
  {36}},\ \bibinfo {pages} {1433} (\bibinfo {year} {2003})}\BibitemShut
  {NoStop}%
\bibitem [{\citenamefont {Pennisi}(2009)}]{Pennisi09}%
  \BibitemOpen
  \bibfield  {author} {\bibinfo {author} {\bibfnamefont {E.}~\bibnamefont
  {Pennisi}},\ }\href {https://doi.org/10.1126/science.309.5731.90} {\bibfield
  {journal} {\bibinfo  {journal} {Science}\ }\textbf {\bibinfo {volume}
  {309}},\ \bibinfo {pages} {90} (\bibinfo {year} {2009})}\BibitemShut
  {NoStop}%
\bibitem [{\citenamefont {Hubbell}(2013)}]{hubbell_2013}%
  \BibitemOpen
  \bibfield  {author} {\bibinfo {author} {\bibfnamefont {S.~P.}\ \bibnamefont
  {Hubbell}},\ }\href@noop {} {\emph {\bibinfo {title} {The Unified Neutral
  Theory of biodiversity and Biogeography}}}\ (\bibinfo  {publisher} {Princeton
  University Press},\ \bibinfo {year} {2013})\BibitemShut {NoStop}%
\bibitem [{\citenamefont {O'Neill}(2016)}]{oneill_tackling_2016}%
  \BibitemOpen
  \bibfield  {author} {\bibinfo {author} {\bibfnamefont {J.}~\bibnamefont
  {O'Neill}},\ }\href {https://apo.org.au/node/63983} {{\selectlanguage
  {English}\emph {\bibinfo {title} {Tackling drug-resistant infections
  globally: final report and recommendations}}}},\ \bibinfo {type} {Report}\
  (\bibinfo  {institution} {Government of the United Kingdom},\ \bibinfo {year}
  {2016})\BibitemShut {NoStop}%
\bibitem [{\citenamefont {Dadgostar}(2019)}]{dadgostar_antimicrobial_2019}%
  \BibitemOpen
  \bibfield  {author} {\bibinfo {author} {\bibfnamefont {P.}~\bibnamefont
  {Dadgostar}},\ }\href {https://doi.org/10.2147/IDR.S234610} {\bibfield
  {journal} {\bibinfo  {journal} {Infect. Drug Resist.}\ }\textbf {\bibinfo
  {volume} {12}},\ \bibinfo {pages} {3903} (\bibinfo {year}
  {2019})}\BibitemShut {NoStop}%
\bibitem [{\citenamefont {Murugan}\ \emph {et~al.}(2021)\citenamefont
  {Murugan}, \citenamefont {Husain},\ and\ \citenamefont {Rust~{\it et
  al.}}}]{Murugan21}%
  \BibitemOpen
  \bibfield  {author} {\bibinfo {author} {\bibfnamefont {A.}~\bibnamefont
  {Murugan}}, \bibinfo {author} {\bibfnamefont {K.}~\bibnamefont {Husain}},\
  and\ \bibinfo {author} {\bibfnamefont {M.~J.}\ \bibnamefont {Rust~{\it et
  al.}}},\ }\href {https://doi.org/10.1088/1478-3975/abde8d} {\bibfield
  {journal} {\bibinfo  {journal} {Phys. Biol.}\ }\textbf {\bibinfo {volume}
  {18}},\ \bibinfo {pages} {041502} (\bibinfo {year} {2021})}\BibitemShut
  {NoStop}%
\bibitem [{\citenamefont {Kalyuzhny}\ \emph {et~al.}(2015)\citenamefont
  {Kalyuzhny}, \citenamefont {Kadmon},\ and\ \citenamefont
  {Shnerb}}]{kalyuzhny_neutral_2015}%
  \BibitemOpen
  \bibfield  {author} {\bibinfo {author} {\bibfnamefont {M.}~\bibnamefont
  {Kalyuzhny}}, \bibinfo {author} {\bibfnamefont {R.}~\bibnamefont {Kadmon}},\
  and\ \bibinfo {author} {\bibfnamefont {N.~M.}\ \bibnamefont {Shnerb}},\
  }\href {https://doi.org/10.1111/ele.12439} {\bibfield  {journal} {\bibinfo
  {journal} {Ecol. Lett.}\ }\textbf {\bibinfo {volume} {18}},\ \bibinfo {pages}
  {572} (\bibinfo {year} {2015})}\BibitemShut {NoStop}%
\bibitem [{\citenamefont {Ghoul}\ and\ \citenamefont {Mitri}(2016)}]{Mitri16}%
  \BibitemOpen
  \bibfield  {author} {\bibinfo {author} {\bibfnamefont {M.}~\bibnamefont
  {Ghoul}}\ and\ \bibinfo {author} {\bibfnamefont {S.}~\bibnamefont {Mitri}},\
  }\href {https://doi.org/10.1016/j.tim.2016.06.011} {\bibfield  {journal}
  {\bibinfo  {journal} {Trends Microbiol.}\ }\textbf {\bibinfo {volume} {24}},\
  \bibinfo {pages} {833} (\bibinfo {year} {2016})}\BibitemShut {NoStop}%
\bibitem [{\citenamefont {Grilli}(2020)}]{Grilli20}%
  \BibitemOpen
  \bibfield  {author} {\bibinfo {author} {\bibfnamefont {J.}~\bibnamefont
  {Grilli}},\ }\href {https://doi.org/10.1038/s41467-020-18529-y} {\bibfield
  {journal} {\bibinfo  {journal} {Nat. Commun.}\ }\textbf {\bibinfo {volume}
  {11}},\ \bibinfo {pages} {4743} (\bibinfo {year} {2020})}\BibitemShut
  {NoStop}%
\bibitem [{\citenamefont {Meyer}\ and\ \citenamefont
  {Shnerb}(2020)}]{meyer_evolutionary_2020}%
  \BibitemOpen
  \bibfield  {author} {\bibinfo {author} {\bibfnamefont {I.}~\bibnamefont
  {Meyer}}\ and\ \bibinfo {author} {\bibfnamefont {N.~M.}\ \bibnamefont
  {Shnerb}},\ }\href {https://doi.org/10.1103/PhysRevResearch.2.023308}
  {\bibfield  {journal} {\bibinfo  {journal} {Phys. Rev. Res.}\ }\textbf
  {\bibinfo {volume} {2}},\ \bibinfo {pages} {023308} (\bibinfo {year}
  {2020})}\BibitemShut {NoStop}%
\bibitem [{\citenamefont {Meyer}\ \emph {et~al.}(2021)\citenamefont {Meyer},
  \citenamefont {Steinmetz},\ and\ \citenamefont
  {Shnerb}}]{meyer_species_2021}%
  \BibitemOpen
  \bibfield  {author} {\bibinfo {author} {\bibfnamefont {I.}~\bibnamefont
  {Meyer}}, \bibinfo {author} {\bibfnamefont {B.}~\bibnamefont {Steinmetz}},\
  and\ \bibinfo {author} {\bibfnamefont {N.~M.}\ \bibnamefont {Shnerb}},\
  }\bibfield  {journal} {\bibinfo  {journal} {bioRxiv}\ }\href
  {https://doi.org/10.1101/2021.04.20.440706} {10.1101/2021.04.20.440706}
  (\bibinfo {year} {2021})\BibitemShut {NoStop}%
\bibitem [{\citenamefont {Leibold}\ \emph {et~al.}(2019)\citenamefont
  {Leibold}, \citenamefont {Urban}, \citenamefont {De~Meester}, \citenamefont
  {Klausmeier},\ and\ \citenamefont {Vanoverbeke}}]{Leibold19}%
  \BibitemOpen
  \bibfield  {author} {\bibinfo {author} {\bibfnamefont {M.}~\bibnamefont
  {Leibold}}, \bibinfo {author} {\bibfnamefont {M.}~\bibnamefont {Urban}},
  \bibinfo {author} {\bibfnamefont {L.}~\bibnamefont {De~Meester}}, \bibinfo
  {author} {\bibfnamefont {C.}~\bibnamefont {Klausmeier}},\ and\ \bibinfo
  {author} {\bibfnamefont {J.}~\bibnamefont {Vanoverbeke}},\ }\href
  {https://doi.org/10.1073/pnas.1808615116} {\bibfield  {journal} {\bibinfo
  {journal} {Proc. Natl Acad. Sci. USA}\ }\textbf {\bibinfo {volume} {116}},\
  \bibinfo {pages} {2612} (\bibinfo {year} {2019})}\BibitemShut {NoStop}%
\bibitem [{\citenamefont {Pinsky}(2019)}]{Pinsky19}%
  \BibitemOpen
  \bibfield  {author} {\bibinfo {author} {\bibfnamefont {M.}~\bibnamefont
  {Pinsky}},\ }\href {https://doi.org/10.1073/pnas.18220911} {\bibfield
  {journal} {\bibinfo  {journal} {Proc. Natl Acad. Sci. USA}\ }\textbf
  {\bibinfo {volume} {116}},\ \bibinfo {pages} {2407} (\bibinfo {year}
  {2019})}\BibitemShut {NoStop}%
\bibitem [{\citenamefont {West}\ and\ \citenamefont {Shnerb}(2022)}]{West22}%
  \BibitemOpen
  \bibfield  {author} {\bibinfo {author} {\bibfnamefont {R.}~\bibnamefont
  {West}}\ and\ \bibinfo {author} {\bibfnamefont {N.~M.}\ \bibnamefont
  {Shnerb}},\ }\href {https://doi.org/10.1086/720665} {\bibfield  {journal}
  {\bibinfo  {journal} {Am. Nat.}\ }\textbf {\bibinfo {volume} {88}},\ \bibinfo
  {pages} {E160} (\bibinfo {year} {2022})}\BibitemShut {NoStop}%
\bibitem [{\citenamefont {Hu}\ \emph {et~al.}(2022)\citenamefont {Hu},
  \citenamefont {Amor}, \citenamefont {Barbier}, \citenamefont {Bunin},\ and\
  \citenamefont {Gore}}]{Gore22}%
  \BibitemOpen
  \bibfield  {author} {\bibinfo {author} {\bibfnamefont {J.}~\bibnamefont
  {Hu}}, \bibinfo {author} {\bibfnamefont {D.~R.}\ \bibnamefont {Amor}},
  \bibinfo {author} {\bibfnamefont {M.}~\bibnamefont {Barbier}}, \bibinfo
  {author} {\bibfnamefont {G.}~\bibnamefont {Bunin}},\ and\ \bibinfo {author}
  {\bibfnamefont {J.}~\bibnamefont {Gore}},\ }\href
  {https://www.science.org/doi/10.1126/science.abm7841} {\bibfield  {journal}
  {\bibinfo  {journal} {Science}\ }\textbf {\bibinfo {volume} {378}},\ \bibinfo
  {pages} {85} (\bibinfo {year} {2022})}\BibitemShut {NoStop}%
\bibitem [{\citenamefont {Balaban}\ \emph {et~al.}(2004)\citenamefont
  {Balaban}, \citenamefont {Merrin}, \citenamefont {Chait}, \citenamefont
  {Kowalik},\ and\ \citenamefont {Leibler}}]{balaban_bacterial_2004}%
  \BibitemOpen
  \bibfield  {author} {\bibinfo {author} {\bibfnamefont {N.~Q.}\ \bibnamefont
  {Balaban}}, \bibinfo {author} {\bibfnamefont {J.}~\bibnamefont {Merrin}},
  \bibinfo {author} {\bibfnamefont {R.}~\bibnamefont {Chait}}, \bibinfo
  {author} {\bibfnamefont {L.}~\bibnamefont {Kowalik}},\ and\ \bibinfo {author}
  {\bibfnamefont {S.}~\bibnamefont {Leibler}},\ }\href
  {https://doi.org/10.1126/science.1099390} {\bibfield  {journal} {\bibinfo
  {journal} {Science}\ }\textbf {\bibinfo {volume} {305}},\ \bibinfo {pages}
  {1622} (\bibinfo {year} {2004})}\BibitemShut {NoStop}%
\bibitem [{\citenamefont {Yurtsev}\ \emph {et~al.}(2013)\citenamefont
  {Yurtsev}, \citenamefont {Chao}, \citenamefont {Datta}, \citenamefont
  {Artemova},\ and\ \citenamefont {Gore}}]{Yurtsev13}%
  \BibitemOpen
  \bibfield  {author} {\bibinfo {author} {\bibfnamefont {E.~A.}\ \bibnamefont
  {Yurtsev}}, \bibinfo {author} {\bibfnamefont {H.~X.}\ \bibnamefont {Chao}},
  \bibinfo {author} {\bibfnamefont {M.~S.}\ \bibnamefont {Datta}}, \bibinfo
  {author} {\bibfnamefont {T.}~\bibnamefont {Artemova}},\ and\ \bibinfo
  {author} {\bibfnamefont {J.}~\bibnamefont {Gore}},\ }\href
  {https://doi.org/10.1038/msb.2013.39} {\bibfield  {journal} {\bibinfo
  {journal} {Mol. Syst. Biol.}\ }\textbf {\bibinfo {volume} {9}},\ \bibinfo
  {pages} {683} (\bibinfo {year} {2013})}\BibitemShut {NoStop}%
\bibitem [{\citenamefont {Raymond~{\it et al.}}(2016)}]{Raymond16}%
  \BibitemOpen
  \bibfield  {author} {\bibinfo {author} {\bibfnamefont {F.}~\bibnamefont
  {Raymond~{\it et al.}}},\ }\href {https://doi.org/10.1038/ismej.2015.148}
  {\bibfield  {journal} {\bibinfo  {journal} {ISME Journal}\ }\textbf {\bibinfo
  {volume} {10}},\ \bibinfo {pages} {707} (\bibinfo {year} {2016})}\BibitemShut
  {NoStop}%
\bibitem [{\citenamefont {Lin}\ and\ \citenamefont
  {Kussell}(2016)}]{Kussell16}%
  \BibitemOpen
  \bibfield  {author} {\bibinfo {author} {\bibfnamefont {W.-H.}\ \bibnamefont
  {Lin}}\ and\ \bibinfo {author} {\bibfnamefont {E.}~\bibnamefont {Kussell}},\
  }\href {https://doi.org/10.1016/j.cub.2016.04.015} {\bibfield  {journal}
  {\bibinfo  {journal} {Curr. Biol.}\ }\textbf {\bibinfo {volume} {26}},\
  \bibinfo {pages} {1486} (\bibinfo {year} {2016})}\BibitemShut {NoStop}%
\bibitem [{\citenamefont {Lopatkin}\ \emph {et~al.}(2017)\citenamefont
  {Lopatkin}, \citenamefont {Meredith}, \citenamefont {Srimani}, \citenamefont
  {Pfeiffer}, \citenamefont {Durrett},\ and\ \citenamefont {You}}]{Lopatkin17}%
  \BibitemOpen
  \bibfield  {author} {\bibinfo {author} {\bibfnamefont {A.}~\bibnamefont
  {Lopatkin}}, \bibinfo {author} {\bibfnamefont {H.}~\bibnamefont {Meredith}},
  \bibinfo {author} {\bibfnamefont {J.}~\bibnamefont {Srimani}}, \bibinfo
  {author} {\bibfnamefont {C.}~\bibnamefont {Pfeiffer}}, \bibinfo {author}
  {\bibfnamefont {R.}~\bibnamefont {Durrett}},\ and\ \bibinfo {author}
  {\bibfnamefont {L.}~\bibnamefont {You}},\ }\href
  {https://doi.org/10.1038/s41467-017-01532-1} {\bibfield  {journal} {\bibinfo
  {journal} {Nat. Commun.}\ }\textbf {\bibinfo {volume} {8}},\ \bibinfo {pages}
  {1689} (\bibinfo {year} {2017})}\BibitemShut {NoStop}%
\bibitem [{\citenamefont {Coates}\ \emph {et~al.}(2018)\citenamefont {Coates},
  \citenamefont {Park}, \citenamefont {Le}, \citenamefont {Şimşek},
  \citenamefont {Chaudhry},\ and\ \citenamefont
  {Kim}}]{coates_antibiotic-induced_2018}%
  \BibitemOpen
  \bibfield  {author} {\bibinfo {author} {\bibfnamefont {J.}~\bibnamefont
  {Coates}}, \bibinfo {author} {\bibfnamefont {B.~R.}\ \bibnamefont {Park}},
  \bibinfo {author} {\bibfnamefont {D.}~\bibnamefont {Le}}, \bibinfo {author}
  {\bibfnamefont {E.}~\bibnamefont {Şimşek}}, \bibinfo {author}
  {\bibfnamefont {W.}~\bibnamefont {Chaudhry}},\ and\ \bibinfo {author}
  {\bibfnamefont {M.}~\bibnamefont {Kim}},\ }\href
  {https://doi.org/10.7554/eLife.32976} {\bibfield  {journal} {\bibinfo
  {journal} {eLife}\ }\textbf {\bibinfo {volume} {7}},\ \bibinfo {pages}
  {e32976} (\bibinfo {year} {2018})}\BibitemShut {NoStop}%
\bibitem [{\citenamefont {Ewens}(2004)}]{Ewens}%
  \BibitemOpen
  \bibfield  {author} {\bibinfo {author} {\bibfnamefont {W.~J.}\ \bibnamefont
  {Ewens}},\ }\href@noop {} {\emph {\bibinfo {title} {Mathematical Population
  Genetics}}}\ (\bibinfo  {publisher} {Springer, New York, USA},\ \bibinfo
  {year} {2004})\BibitemShut {NoStop}%
\bibitem [{\citenamefont {Crow}\ and\ \citenamefont {Kimura}(2009)}]{Kimura}%
  \BibitemOpen
  \bibfield  {author} {\bibinfo {author} {\bibfnamefont {J.~F.~F.}\
  \bibnamefont {Crow}}\ and\ \bibinfo {author} {\bibfnamefont {M.}~\bibnamefont
  {Kimura}},\ }\href@noop {} {\emph {\bibinfo {title} {An Introduction to
  Population Genetics Theory}}}\ (\bibinfo  {publisher} {Blackburn Press,
  Caldwell, NJ, USA.},\ \bibinfo {year} {2009})\BibitemShut {NoStop}%
\bibitem [{\citenamefont {Karlin}\ and\ \citenamefont
  {Levikson}(1974)}]{Karlin74}%
  \BibitemOpen
  \bibfield  {author} {\bibinfo {author} {\bibfnamefont {S.}~\bibnamefont
  {Karlin}}\ and\ \bibinfo {author} {\bibfnamefont {B.}~\bibnamefont
  {Levikson}},\ }\href {https://doi.org/10.1016/0040-5809(74)90017-3}
  {\bibfield  {journal} {\bibinfo  {journal} {Theor. Popul. Biol.}\ }\textbf
  {\bibinfo {volume} {6}},\ \bibinfo {pages} {383} (\bibinfo {year}
  {1974})}\BibitemShut {NoStop}%
\bibitem [{\citenamefont {Thattai}\ and\ \citenamefont
  {Van~Oudenaarden}(2004)}]{Thattai04}%
  \BibitemOpen
  \bibfield  {author} {\bibinfo {author} {\bibfnamefont {M.}~\bibnamefont
  {Thattai}}\ and\ \bibinfo {author} {\bibfnamefont {A.}~\bibnamefont
  {Van~Oudenaarden}},\ }\href {https://doi.org/10.1534/genetics.167.1.523}
  {\bibfield  {journal} {\bibinfo  {journal} {Genetics}\ }\textbf {\bibinfo
  {volume} {167}},\ \bibinfo {pages} {523} (\bibinfo {year}
  {2004})}\BibitemShut {NoStop}%
\bibitem [{\citenamefont {Kussell}\ and\ \citenamefont
  {Leibler}(2005)}]{Kussel05}%
  \BibitemOpen
  \bibfield  {author} {\bibinfo {author} {\bibfnamefont {E.}~\bibnamefont
  {Kussell}}\ and\ \bibinfo {author} {\bibfnamefont {S.}~\bibnamefont
  {Leibler}},\ }\href {https://doi.org/10.1126/science.1114383} {\bibfield
  {journal} {\bibinfo  {journal} {Science}\ }\textbf {\bibinfo {volume}
  {309}},\ \bibinfo {pages} {2075} (\bibinfo {year} {2005})}\BibitemShut
  {NoStop}%
\bibitem [{\citenamefont {Loreau}\ and\ \citenamefont
  {de~Mazancourt}(2008)}]{Loreau08}%
  \BibitemOpen
  \bibfield  {author} {\bibinfo {author} {\bibfnamefont {M.}~\bibnamefont
  {Loreau}}\ and\ \bibinfo {author} {\bibfnamefont {C.}~\bibnamefont
  {de~Mazancourt}},\ }\href {https://doi.org/10.1086/589746} {\bibfield
  {journal} {\bibinfo  {journal} {Am.~Nat.}\ }\textbf {\bibinfo {volume}
  {172}},\ \bibinfo {pages} {E48} (\bibinfo {year} {2008})}\BibitemShut
  {NoStop}%
\bibitem [{\citenamefont {He}\ \emph {et~al.}(2010)\citenamefont {He},
  \citenamefont {Mobilia},\ and\ \citenamefont {T\"auber}}]{He10}%
  \BibitemOpen
  \bibfield  {author} {\bibinfo {author} {\bibfnamefont {Q.}~\bibnamefont
  {He}}, \bibinfo {author} {\bibfnamefont {M.}~\bibnamefont {Mobilia}},\ and\
  \bibinfo {author} {\bibfnamefont {U.~C.}\ \bibnamefont {T\"auber}},\ }\href
  {https://doi.org/10.1103/PhysRevE.82.051909} {\bibfield  {journal} {\bibinfo
  {journal} {Phys. Rev. E}\ }\textbf {\bibinfo {volume} {82}},\ \bibinfo
  {pages} {051909} (\bibinfo {year} {2010})}\BibitemShut {NoStop}%
\bibitem [{\citenamefont {Visco}\ \emph {et~al.}(2010)\citenamefont {Visco},
  \citenamefont {Allen}, \citenamefont {Majumdar},\ and\ \citenamefont
  {Evans}}]{Visco10}%
  \BibitemOpen
  \bibfield  {author} {\bibinfo {author} {\bibfnamefont {P.}~\bibnamefont
  {Visco}}, \bibinfo {author} {\bibfnamefont {J.}~\bibnamefont {Allen}},
  \bibinfo {author} {\bibfnamefont {S.~N.}\ \bibnamefont {Majumdar}},\ and\
  \bibinfo {author} {\bibfnamefont {M.~R.}\ \bibnamefont {Evans}},\ }\href
  {https://doi.org/10.1016/j.bpj.2009.11.049} {\bibfield  {journal} {\bibinfo
  {journal} {Biophys. J.}\ }\textbf {\bibinfo {volume} {98}},\ \bibinfo {pages}
  {1099} (\bibinfo {year} {2010})}\BibitemShut {NoStop}%
\bibitem [{\citenamefont {Assaf}\ \emph
  {et~al.}(2013{\natexlab{a}})\citenamefont {Assaf}, \citenamefont {Roberts},
  \citenamefont {Luthey-Schulten},\ and\ \citenamefont
  {Goldenfeld}}]{Assaf13a}%
  \BibitemOpen
  \bibfield  {author} {\bibinfo {author} {\bibfnamefont {M.}~\bibnamefont
  {Assaf}}, \bibinfo {author} {\bibfnamefont {E.}~\bibnamefont {Roberts}},
  \bibinfo {author} {\bibfnamefont {Z.}~\bibnamefont {Luthey-Schulten}},\ and\
  \bibinfo {author} {\bibfnamefont {N.}~\bibnamefont {Goldenfeld}},\
  }\href@noop {} {\bibfield  {journal} {\bibinfo  {journal} {Phys. Rev. Lett.}\
  }\textbf {\bibinfo {volume} {111}},\ \bibinfo {pages} {058102} (\bibinfo
  {year} {2013}{\natexlab{a}})}\BibitemShut {NoStop}%
\bibitem [{\citenamefont {Dobramysl}\ and\ \citenamefont
  {T\"auber}(2013)}]{Dobra13}%
  \BibitemOpen
  \bibfield  {author} {\bibinfo {author} {\bibfnamefont {U.}~\bibnamefont
  {Dobramysl}}\ and\ \bibinfo {author} {\bibfnamefont {U.~C.}\ \bibnamefont
  {T\"auber}},\ }\href {https://doi.org/10.1103/PhysRevLett.110.048105}
  {\bibfield  {journal} {\bibinfo  {journal} {Phys. Rev. Lett.}\ }\textbf
  {\bibinfo {volume} {110}},\ \bibinfo {pages} {048105} (\bibinfo {year}
  {2013})}\BibitemShut {NoStop}%
\bibitem [{\citenamefont {Assaf}\ \emph
  {et~al.}(2013{\natexlab{b}})\citenamefont {Assaf}, \citenamefont {Mobilia},\
  and\ \citenamefont {Roberts}}]{Assaf13b}%
  \BibitemOpen
  \bibfield  {author} {\bibinfo {author} {\bibfnamefont {M.}~\bibnamefont
  {Assaf}}, \bibinfo {author} {\bibfnamefont {M.}~\bibnamefont {Mobilia}},\
  and\ \bibinfo {author} {\bibfnamefont {E.}~\bibnamefont {Roberts}},\ }\href
  {https://doi.org/10.1103/PhysRevLett.111.238101} {\bibfield  {journal}
  {\bibinfo  {journal} {Phys. Rev. Lett.}\ }\textbf {\bibinfo {volume} {111}},\
  \bibinfo {pages} {238101} (\bibinfo {year} {2013}{\natexlab{b}})}\BibitemShut
  {NoStop}%
\bibitem [{\citenamefont {Chisholm~{\it et al.}}(2014)}]{Chisholm14}%
  \BibitemOpen
  \bibfield  {author} {\bibinfo {author} {\bibfnamefont {R.~A.}\ \bibnamefont
  {Chisholm~{\it et al.}}},\ }\href {https://doi.org/10.1111/ele.12296}
  {\bibfield  {journal} {\bibinfo  {journal} {Ecol. Lett.}\ }\textbf {\bibinfo
  {volume} {17}},\ \bibinfo {pages} {855} (\bibinfo {year} {2014})}\BibitemShut
  {NoStop}%
\bibitem [{\citenamefont {Ashcroft}\ \emph {et~al.}(2014)\citenamefont
  {Ashcroft}, \citenamefont {Altrock},\ and\ \citenamefont
  {Galla}}]{ashcroft_fixation_2014}%
  \BibitemOpen
  \bibfield  {author} {\bibinfo {author} {\bibfnamefont {P.}~\bibnamefont
  {Ashcroft}}, \bibinfo {author} {\bibfnamefont {P.~M.}\ \bibnamefont
  {Altrock}},\ and\ \bibinfo {author} {\bibfnamefont {T.}~\bibnamefont
  {Galla}},\ }\href {https://doi.org/10.1098/rsif.2014.0663} {\bibfield
  {journal} {\bibinfo  {journal} {J. R. Soc. Interface}\ }\textbf {\bibinfo
  {volume} {11}},\ \bibinfo {pages} {20140663} (\bibinfo {year}
  {2014})}\BibitemShut {NoStop}%
\bibitem [{\citenamefont {Melbinger}\ \emph {et~al.}(2015)\citenamefont
  {Melbinger}, \citenamefont {Cremer},\ and\ \citenamefont
  {Frey}}]{melbinger_emergence_2015}%
  \BibitemOpen
  \bibfield  {author} {\bibinfo {author} {\bibfnamefont {A.}~\bibnamefont
  {Melbinger}}, \bibinfo {author} {\bibfnamefont {J.}~\bibnamefont {Cremer}},\
  and\ \bibinfo {author} {\bibfnamefont {E.}~\bibnamefont {Frey}},\ }\href
  {https://doi.org/10.1098/rsif.2015.0171} {\bibfield  {journal} {\bibinfo
  {journal} {J. R. Soc. Interface}\ }\textbf {\bibinfo {volume} {12}},\
  \bibinfo {pages} {20150171} (\bibinfo {year} {2015})}\BibitemShut {NoStop}%
\bibitem [{\citenamefont {Roberts}\ \emph {et~al.}(2015)\citenamefont
  {Roberts}, \citenamefont {Be'er}, \citenamefont {Bohrer}, \citenamefont
  {Sharma},\ and\ \citenamefont {Assaf}}]{roberts_dynamics_2015}%
  \BibitemOpen
  \bibfield  {author} {\bibinfo {author} {\bibfnamefont {E.}~\bibnamefont
  {Roberts}}, \bibinfo {author} {\bibfnamefont {S.}~\bibnamefont {Be'er}},
  \bibinfo {author} {\bibfnamefont {C.}~\bibnamefont {Bohrer}}, \bibinfo
  {author} {\bibfnamefont {R.}~\bibnamefont {Sharma}},\ and\ \bibinfo {author}
  {\bibfnamefont {M.}~\bibnamefont {Assaf}},\ }\href
  {https://doi.org/10.1103/PhysRevE.92.062717} {\bibfield  {journal} {\bibinfo
  {journal} {Phys. Rev. E}\ }\textbf {\bibinfo {volume} {92}},\ \bibinfo
  {pages} {062717} (\bibinfo {year} {2015})}\BibitemShut {NoStop}%
\bibitem [{\citenamefont {Hufton}\ \emph {et~al.}(2016)\citenamefont {Hufton},
  \citenamefont {Lin}, \citenamefont {Galla},\ and\ \citenamefont
  {McKane}}]{hufton_intrinsic_2016}%
  \BibitemOpen
  \bibfield  {author} {\bibinfo {author} {\bibfnamefont {P.~G.}\ \bibnamefont
  {Hufton}}, \bibinfo {author} {\bibfnamefont {Y.~T.}\ \bibnamefont {Lin}},
  \bibinfo {author} {\bibfnamefont {T.}~\bibnamefont {Galla}},\ and\ \bibinfo
  {author} {\bibfnamefont {A.~J.}\ \bibnamefont {McKane}},\ }\href
  {https://doi.org/10.1103/PhysRevE.93.052119} {\bibfield  {journal} {\bibinfo
  {journal} {Phys. Rev. E}\ }\textbf {\bibinfo {volume} {93}},\ \bibinfo
  {pages} {052119} (\bibinfo {year} {2016})}\BibitemShut {NoStop}%
\bibitem [{\citenamefont {Wienand}\ \emph {et~al.}(2017)\citenamefont
  {Wienand}, \citenamefont {Frey},\ and\ \citenamefont
  {Mobilia}}]{wienand_evolution_2017}%
  \BibitemOpen
  \bibfield  {author} {\bibinfo {author} {\bibfnamefont {K.}~\bibnamefont
  {Wienand}}, \bibinfo {author} {\bibfnamefont {E.}~\bibnamefont {Frey}},\ and\
  \bibinfo {author} {\bibfnamefont {M.}~\bibnamefont {Mobilia}},\ }\href
  {https://doi.org/10.1103/PhysRevLett.119.158301} {\bibfield  {journal}
  {\bibinfo  {journal} {Phys. Rev. Lett.}\ }\textbf {\bibinfo {volume} {119}},\
  \bibinfo {pages} {158301} (\bibinfo {year} {2017})}\BibitemShut {NoStop}%
\bibitem [{\citenamefont {Xue}\ and\ \citenamefont {Leibler}(2017)}]{Xue17}%
  \BibitemOpen
  \bibfield  {author} {\bibinfo {author} {\bibfnamefont {B.~K.}\ \bibnamefont
  {Xue}}\ and\ \bibinfo {author} {\bibfnamefont {S.}~\bibnamefont {Leibler}},\
  }\href {https://doi.org/10.1103/PhysRevLett.119.108103} {\bibfield  {journal}
  {\bibinfo  {journal} {Phys.~Rev.~Lett.}\ }\textbf {\bibinfo {volume} {119}},\
  \bibinfo {pages} {108103} (\bibinfo {year} {2017})}\BibitemShut {NoStop}%
\bibitem [{\citenamefont {Dobramysl}\ \emph {et~al.}(2018)\citenamefont
  {Dobramysl}, \citenamefont {Mobilia}, \citenamefont {Pleimling},\ and\
  \citenamefont {T\"auber}}]{Dobra18}%
  \BibitemOpen
  \bibfield  {author} {\bibinfo {author} {\bibfnamefont {U.}~\bibnamefont
  {Dobramysl}}, \bibinfo {author} {\bibfnamefont {M.}~\bibnamefont {Mobilia}},
  \bibinfo {author} {\bibfnamefont {M.}~\bibnamefont {Pleimling}},\ and\
  \bibinfo {author} {\bibfnamefont {U.~C.}\ \bibnamefont {T\"auber}},\ }\href
  {https://doi.org/10.1088/1751-8121/aa95c7} {\bibfield  {journal} {\bibinfo
  {journal} {J. Phys. A: Math. Theor.}\ }\textbf {\bibinfo {volume} {51}},\
  \bibinfo {pages} {063001} (\bibinfo {year} {2018})}\BibitemShut {NoStop}%
\bibitem [{\citenamefont {Marrec}\ and\ \citenamefont
  {Bitbol}(2018)}]{marrec_quantifying_2018}%
  \BibitemOpen
  \bibfield  {author} {\bibinfo {author} {\bibfnamefont {L.}~\bibnamefont
  {Marrec}}\ and\ \bibinfo {author} {\bibfnamefont {A.-F.}\ \bibnamefont
  {Bitbol}},\ }\href {https://doi.org/10.1016/j.jtbi.2018.08.040} {\bibfield
  {journal} {\bibinfo  {journal} {J. Theor. Biol.}\ }\textbf {\bibinfo {volume}
  {457}},\ \bibinfo {pages} {190} (\bibinfo {year} {2018})}\BibitemShut
  {NoStop}%
\bibitem [{\citenamefont {Wienand}\ \emph {et~al.}(2018)\citenamefont
  {Wienand}, \citenamefont {Frey},\ and\ \citenamefont
  {Mobilia}}]{wienand_eco-evolutionary_2018}%
  \BibitemOpen
  \bibfield  {author} {\bibinfo {author} {\bibfnamefont {K.}~\bibnamefont
  {Wienand}}, \bibinfo {author} {\bibfnamefont {E.}~\bibnamefont {Frey}},\ and\
  \bibinfo {author} {\bibfnamefont {M.}~\bibnamefont {Mobilia}},\ }\href
  {https://doi.org/10.1098/rsif.2018.0343} {\bibfield  {journal} {\bibinfo
  {journal} {J. R. Soc. Interface}\ }\textbf {\bibinfo {volume} {15}},\
  \bibinfo {pages} {20180343} (\bibinfo {year} {2018})}\BibitemShut {NoStop}%
\bibitem [{\citenamefont {West}\ and\ \citenamefont
  {Mobilia}(2020)}]{west2020}%
  \BibitemOpen
  \bibfield  {author} {\bibinfo {author} {\bibfnamefont {R.}~\bibnamefont
  {West}}\ and\ \bibinfo {author} {\bibfnamefont {M.}~\bibnamefont {Mobilia}},\
  }\href {https://doi.org/10.1016/j.jtbi.2019.110135} {\bibfield  {journal}
  {\bibinfo  {journal} {J. Theor. Biol.}\ }\textbf {\bibinfo {volume} {491}},\
  \bibinfo {pages} {110135} (\bibinfo {year} {2020})}\BibitemShut {NoStop}%
\bibitem [{\citenamefont {Taitelbaum}\ \emph {et~al.}(2020)\citenamefont
  {Taitelbaum}, \citenamefont {West}, \citenamefont {Assaf},\ and\
  \citenamefont {Mobilia}}]{taitelbaum_population_2020}%
  \BibitemOpen
  \bibfield  {author} {\bibinfo {author} {\bibfnamefont {A.}~\bibnamefont
  {Taitelbaum}}, \bibinfo {author} {\bibfnamefont {R.}~\bibnamefont {West}},
  \bibinfo {author} {\bibfnamefont {M.}~\bibnamefont {Assaf}},\ and\ \bibinfo
  {author} {\bibfnamefont {M.}~\bibnamefont {Mobilia}},\ }\href
  {https://doi.org/10.1103/PhysRevLett.125.048105} {\bibfield  {journal}
  {\bibinfo  {journal} {Phys. Rev. Lett.}\ }\textbf {\bibinfo {volume} {125}},\
  \bibinfo {pages} {048105} (\bibinfo {year} {2020})}\BibitemShut {NoStop}%
\bibitem [{\citenamefont {Marrec}\ and\ \citenamefont
  {Bitbol}(2020)}]{marrec_resist_2020}%
  \BibitemOpen
  \bibfield  {author} {\bibinfo {author} {\bibfnamefont {L.}~\bibnamefont
  {Marrec}}\ and\ \bibinfo {author} {\bibfnamefont {A.-F.}\ \bibnamefont
  {Bitbol}},\ }\href {https://doi.org/10.1371/journal.pcbi.1007798} {\bibfield
  {journal} {\bibinfo  {journal} {PLOS Computational Biology}\ }\textbf
  {\bibinfo {volume} {16}},\ \bibinfo {pages} {1} (\bibinfo {year}
  {2020})}\BibitemShut {NoStop}%
\bibitem [{\citenamefont {Shibasaki}\ \emph {et~al.}(2021)\citenamefont
  {Shibasaki}, \citenamefont {Mobilia},\ and\ \citenamefont
  {Mitri}}]{shibasaki_exclusion_2021}%
  \BibitemOpen
  \bibfield  {author} {\bibinfo {author} {\bibfnamefont {S.}~\bibnamefont
  {Shibasaki}}, \bibinfo {author} {\bibfnamefont {M.}~\bibnamefont {Mobilia}},\
  and\ \bibinfo {author} {\bibfnamefont {S.}~\bibnamefont {Mitri}},\ }\href
  {https://doi.org/10.1098/rsif.2021.0613} {\bibfield  {journal} {\bibinfo
  {journal} {J. R. Soc. Interface}\ }\textbf {\bibinfo {volume} {18}},\
  \bibinfo {pages} {20210613} (\bibinfo {year} {2021})}\BibitemShut {NoStop}%
\bibitem [{\citenamefont {Taitelbaum}\ \emph {et~al.}(2023)\citenamefont
  {Taitelbaum}, \citenamefont {West}, \citenamefont {Mobilia},\ and\
  \citenamefont {Assaf}}]{taitelbaum2023evolutionary}%
  \BibitemOpen
  \bibfield  {author} {\bibinfo {author} {\bibfnamefont {A.}~\bibnamefont
  {Taitelbaum}}, \bibinfo {author} {\bibfnamefont {R.}~\bibnamefont {West}},
  \bibinfo {author} {\bibfnamefont {M.}~\bibnamefont {Mobilia}},\ and\ \bibinfo
  {author} {\bibfnamefont {M.}~\bibnamefont {Assaf}},\ }\href
  {https://doi.org/10.1103/PhysRevResearch.5.L022004} {\bibfield  {journal}
  {\bibinfo  {journal} {Phys. Rev. Res.}\ }\textbf {\bibinfo {volume} {5}},\
  \bibinfo {pages} {L022004} (\bibinfo {year} {2023})}\BibitemShut {NoStop}%
\bibitem [{\citenamefont {Nowak}(2006)}]{Nowak}%
  \BibitemOpen
  \bibfield  {author} {\bibinfo {author} {\bibfnamefont {M.}~\bibnamefont
  {Nowak}},\ }\href@noop {} {\emph {\bibinfo {title} {Evolutionary Dynamics}}}\
  (\bibinfo  {publisher} {Belknap Press, Cambridge, MA, USA},\ \bibinfo {year}
  {2006})\BibitemShut {NoStop}%
\bibitem [{\citenamefont {Blythe}\ and\ \citenamefont
  {McKane}(2007)}]{blythe_stochastic_2007}%
  \BibitemOpen
  \bibfield  {author} {\bibinfo {author} {\bibfnamefont {R.~A.}\ \bibnamefont
  {Blythe}}\ and\ \bibinfo {author} {\bibfnamefont {A.~J.}\ \bibnamefont
  {McKane}},\ }\href {https://doi.org/10.1088/1742-5468/2007/07/P07018}
  {\bibfield  {journal} {\bibinfo  {journal} {J. Stat. Mech: Theory Exp.}\
  }\textbf {\bibinfo {volume} {2007}},\ \bibinfo {pages} {P07018} (\bibinfo
  {year} {2007})}\BibitemShut {NoStop}%
\bibitem [{\citenamefont {Traulsen}\ and\ \citenamefont
  {Hauert}(2008)}]{traulsen_stochastic_2008}%
  \BibitemOpen
  \bibfield  {author} {\bibinfo {author} {\bibfnamefont {A.}~\bibnamefont
  {Traulsen}}\ and\ \bibinfo {author} {\bibfnamefont {C.}~\bibnamefont
  {Hauert}},\ }\href {https://doi.org/10.48550/arXiv.0811.3538} {\bibfield
  {journal} {\bibinfo  {journal} {arXiv}\ } (\bibinfo {year}
  {2008})}\BibitemShut {NoStop}%
\bibitem [{\citenamefont {Roughgarden}(1979)}]{Roughgarden}%
  \BibitemOpen
  \bibfield  {author} {\bibinfo {author} {\bibfnamefont {J.}~\bibnamefont
  {Roughgarden}},\ }\href@noop {} {\emph {\bibinfo {title} {Theory of
  Population Genetics and Evolutionary Ecology: An Introduction}}}\ (\bibinfo
  {publisher} {Macmillan, New York, USA},\ \bibinfo {year} {1979})\BibitemShut
  {NoStop}%
\bibitem [{\citenamefont {Chuang}\ \emph {et~al.}(2009)\citenamefont {Chuang},
  \citenamefont {Rivoire},\ and\ \citenamefont
  {Leibler}}]{chuang_simpsons_2009}%
  \BibitemOpen
  \bibfield  {author} {\bibinfo {author} {\bibfnamefont {J.~S.}\ \bibnamefont
  {Chuang}}, \bibinfo {author} {\bibfnamefont {O.}~\bibnamefont {Rivoire}},\
  and\ \bibinfo {author} {\bibfnamefont {S.}~\bibnamefont {Leibler}},\ }\href
  {https://doi.org/10.1126/science.1166739} {\bibfield  {journal} {\bibinfo
  {journal} {Science}\ }\textbf {\bibinfo {volume} {323}},\ \bibinfo {pages}
  {272} (\bibinfo {year} {2009})}\BibitemShut {NoStop}%
\bibitem [{\citenamefont {Chuang}\ \emph {et~al.}(2010)\citenamefont {Chuang},
  \citenamefont {Rivoire},\ and\ \citenamefont
  {Leibler}}]{chuang_cooperation_2010}%
  \BibitemOpen
  \bibfield  {author} {\bibinfo {author} {\bibfnamefont {J.~S.}\ \bibnamefont
  {Chuang}}, \bibinfo {author} {\bibfnamefont {O.}~\bibnamefont {Rivoire}},\
  and\ \bibinfo {author} {\bibfnamefont {S.}~\bibnamefont {Leibler}},\ }\href
  {https://doi.org/10.1038/msb.2010.57} {\bibfield  {journal} {\bibinfo
  {journal} {Mol. Syst. Biol.}\ }\textbf {\bibinfo {volume} {6}},\ \bibinfo
  {pages} {398} (\bibinfo {year} {2010})}\BibitemShut {NoStop}%
\bibitem [{\citenamefont {Melbinger}\ \emph {et~al.}(2010)\citenamefont
  {Melbinger}, \citenamefont {Cremer},\ and\ \citenamefont
  {Frey}}]{melbinger_evolutionary_2010}%
  \BibitemOpen
  \bibfield  {author} {\bibinfo {author} {\bibfnamefont {A.}~\bibnamefont
  {Melbinger}}, \bibinfo {author} {\bibfnamefont {J.}~\bibnamefont {Cremer}},\
  and\ \bibinfo {author} {\bibfnamefont {E.}~\bibnamefont {Frey}},\ }\href
  {https://doi.org/10.1103/physrevlett.105.178101} {\bibfield  {journal}
  {\bibinfo  {journal} {Phys. Rev. Lett.}\ }\textbf {\bibinfo {volume} {105}},\
  \bibinfo {pages} {178101} (\bibinfo {year} {2010})}\BibitemShut {NoStop}%
\bibitem [{\citenamefont {Cremer}\ \emph {et~al.}(2012)\citenamefont {Cremer},
  \citenamefont {Melbinger},\ and\ \citenamefont {Frey}}]{cremer_growth_2012}%
  \BibitemOpen
  \bibfield  {author} {\bibinfo {author} {\bibfnamefont {J.}~\bibnamefont
  {Cremer}}, \bibinfo {author} {\bibfnamefont {A.}~\bibnamefont {Melbinger}},\
  and\ \bibinfo {author} {\bibfnamefont {E.}~\bibnamefont {Frey}},\ }\href
  {https://doi.org/10.1038/srep00281} {\bibfield  {journal} {\bibinfo
  {journal} {Sci. Rep.}\ }\textbf {\bibinfo {volume} {2}},\ \bibinfo {pages}
  {281} (\bibinfo {year} {2012})}\BibitemShut {NoStop}%
\bibitem [{\citenamefont {Sanchez}\ and\ \citenamefont
  {Gore}(2013)}]{sanchez_feedback_2013}%
  \BibitemOpen
  \bibfield  {author} {\bibinfo {author} {\bibfnamefont {A.}~\bibnamefont
  {Sanchez}}\ and\ \bibinfo {author} {\bibfnamefont {J.}~\bibnamefont {Gore}},\
  }\bibfield  {journal} {\bibinfo  {journal} {PLoS Biol.}\ }\textbf {\bibinfo
  {volume} {11}},\ \href {https://doi.org/10.1371/journal.pbio.1001547}
  {10.1371/journal.pbio.1001547} (\bibinfo {year} {2013})\BibitemShut {NoStop}%
\bibitem [{\citenamefont {Gokhale}\ and\ \citenamefont
  {Hauert}(2016)}]{Gokhale16}%
  \BibitemOpen
  \bibfield  {author} {\bibinfo {author} {\bibfnamefont {C.~S.}\ \bibnamefont
  {Gokhale}}\ and\ \bibinfo {author} {\bibfnamefont {C.}~\bibnamefont
  {Hauert}},\ }\href {https://doi.org/10.1016/j.tpb.2016.05.005} {\bibfield
  {journal} {\bibinfo  {journal} {Theor. Popul. Biol.}\ }\textbf {\bibinfo
  {volume} {111}},\ \bibinfo {pages} {28} (\bibinfo {year} {2016})}\BibitemShut
  {NoStop}%
\bibitem [{\citenamefont {Tilman}\ \emph {et~al.}(2020)\citenamefont {Tilman},
  \citenamefont {Plotkin},\ and\ \citenamefont {Akcay}}]{Tilman20}%
  \BibitemOpen
  \bibfield  {author} {\bibinfo {author} {\bibfnamefont {A.}~\bibnamefont
  {Tilman}}, \bibinfo {author} {\bibfnamefont {J.}~\bibnamefont {Plotkin}},\
  and\ \bibinfo {author} {\bibfnamefont {E.}~\bibnamefont {Akcay}},\ }\href
  {https://doi.org/10.1038/s41467-020-14531-6} {\bibfield  {journal} {\bibinfo
  {journal} {Nat. Commun.}\ }\textbf {\bibinfo {volume} {11}},\ \bibinfo
  {pages} {915} (\bibinfo {year} {2020})}\BibitemShut {NoStop}%
\bibitem [{\citenamefont {Wang}\ \emph {et~al.}(2023)\citenamefont {Wang},
  \citenamefont {Su}, \citenamefont {Wang},\ and\ \citenamefont
  {Plotkin}}]{Plotkin23}%
  \BibitemOpen
  \bibfield  {author} {\bibinfo {author} {\bibfnamefont {G.}~\bibnamefont
  {Wang}}, \bibinfo {author} {\bibfnamefont {Q.}~\bibnamefont {Su}}, \bibinfo
  {author} {\bibfnamefont {L.}~\bibnamefont {Wang}},\ and\ \bibinfo {author}
  {\bibfnamefont {J.}~\bibnamefont {Plotkin}},\ }\href
  {https://doi.org/10.1073/pnas.2216218120} {\bibfield  {journal} {\bibinfo
  {journal} {Proc. Natl Acad. Sci. USA}\ }\textbf {\bibinfo {volume} {120}},\
  \bibinfo {pages} {e216218120} (\bibinfo {year} {2023})}\BibitemShut {NoStop}%
\bibitem [{\citenamefont {Hernández-Navarro}\ \emph
  {et~al.}(2023)\citenamefont {Hernández-Navarro}, \citenamefont {Asker},
  \citenamefont {Rucklidge},\ and\ \citenamefont {Mobilia}}]{LARM23}%
  \BibitemOpen
  \bibfield  {author} {\bibinfo {author} {\bibfnamefont {L.}~\bibnamefont
  {Hernández-Navarro}}, \bibinfo {author} {\bibfnamefont {M.}~\bibnamefont
  {Asker}}, \bibinfo {author} {\bibfnamefont {A.~M.}\ \bibnamefont
  {Rucklidge}},\ and\ \bibinfo {author} {\bibfnamefont {M.}~\bibnamefont
  {Mobilia}},\ }\href {https://doi.org/10.1098/rsif.2023.0393} {\bibfield
  {journal} {\bibinfo  {journal} {J. R. Soc. Interface}\ }\textbf {\bibinfo
  {volume} {20}},\ \bibinfo {pages} {20230393} (\bibinfo {year}
  {2023})}\BibitemShut {NoStop}%
\bibitem [{\citenamefont {Szolnoki}\ \emph {et~al.}(2014)\citenamefont
  {Szolnoki}, \citenamefont {Mobilia}, \citenamefont {Jiang}, \citenamefont
  {Szczesny}, \citenamefont {Rucklidge},\ and\ \citenamefont
  {M.}}]{Szolnoki14}%
  \BibitemOpen
  \bibfield  {author} {\bibinfo {author} {\bibfnamefont {A.}~\bibnamefont
  {Szolnoki}}, \bibinfo {author} {\bibfnamefont {M.}~\bibnamefont {Mobilia}},
  \bibinfo {author} {\bibfnamefont {L.-L.}\ \bibnamefont {Jiang}}, \bibinfo
  {author} {\bibfnamefont {B.}~\bibnamefont {Szczesny}}, \bibinfo {author}
  {\bibfnamefont {A.~M.}\ \bibnamefont {Rucklidge}},\ and\ \bibinfo {author}
  {\bibfnamefont {P.}~\bibnamefont {M.}},\ }\href
  {https://doi.org/10.1098/rsif.2014.0735} {\bibfield  {journal} {\bibinfo
  {journal} {J. R. Soc. Interface}\ }\textbf {\bibinfo {volume} {11}},\
  \bibinfo {pages} {20140735} (\bibinfo {year} {2014})}\BibitemShut {NoStop}%
\bibitem [{\citenamefont {Perc}\ and\ \citenamefont
  {Szolnoki}(2007)}]{Perc_2007}%
  \BibitemOpen
  \bibfield  {author} {\bibinfo {author} {\bibfnamefont {M.}~\bibnamefont
  {Perc}}\ and\ \bibinfo {author} {\bibfnamefont {A.}~\bibnamefont
  {Szolnoki}},\ }\href {https://doi.org/10.1088/1367-2630/9/8/267} {\bibfield
  {journal} {\bibinfo  {journal} {New Journal of Physics}\ }\textbf {\bibinfo
  {volume} {9}},\ \bibinfo {pages} {267} (\bibinfo {year} {2007})}\BibitemShut
  {NoStop}%
\bibitem [{\citenamefont {He}\ \emph {et~al.}(2011)\citenamefont {He},
  \citenamefont {Mobilia},\ and\ \citenamefont {T\"auber}}]{he2011}%
  \BibitemOpen
  \bibfield  {author} {\bibinfo {author} {\bibfnamefont {Q.}~\bibnamefont
  {He}}, \bibinfo {author} {\bibfnamefont {M.}~\bibnamefont {Mobilia}},\ and\
  \bibinfo {author} {\bibfnamefont {U.~C.}\ \bibnamefont {T\"auber}},\ }\href
  {https://doi.org/10.1140/epjb/e2011-20259-x} {\bibfield  {journal} {\bibinfo
  {journal} {Eur. Phys. J. B}\ }\textbf {\bibinfo {volume} {82}},\ \bibinfo
  {pages} {97} (\bibinfo {year} {2011})}\BibitemShut {NoStop}%
\bibitem [{\citenamefont {Kelsic}\ \emph {et~al.}(2015)\citenamefont {Kelsic},
  \citenamefont {Zhao}, \citenamefont {Vetsigian},\ and\ \citenamefont
  {Kishony}}]{Kelsic_2015}%
  \BibitemOpen
  \bibfield  {author} {\bibinfo {author} {\bibfnamefont {E.}~\bibnamefont
  {Kelsic}}, \bibinfo {author} {\bibfnamefont {J.}~\bibnamefont {Zhao}},
  \bibinfo {author} {\bibfnamefont {K.}~\bibnamefont {Vetsigian}},\ and\
  \bibinfo {author} {\bibfnamefont {R.}~\bibnamefont {Kishony}},\ }\href
  {https://doi.org/doi:10.1038/nature14485} {\bibfield  {journal} {\bibinfo
  {journal} {Nature}\ }\textbf {\bibinfo {volume} {516}},\ \bibinfo {pages}
  {113033} (\bibinfo {year} {2015})}\BibitemShut {NoStop}%
\bibitem [{\citenamefont {Szolnoki}\ and\ \citenamefont
  {Perc}(2015)}]{Szolnoki_2015}%
  \BibitemOpen
  \bibfield  {author} {\bibinfo {author} {\bibfnamefont {A.}~\bibnamefont
  {Szolnoki}}\ and\ \bibinfo {author} {\bibfnamefont {M.}~\bibnamefont
  {Perc}},\ }\href {https://doi.org/10.1088/1367-2630/17/11/113033} {\bibfield
  {journal} {\bibinfo  {journal} {New Journal of Physics}\ }\textbf {\bibinfo
  {volume} {17}},\ \bibinfo {pages} {113033} (\bibinfo {year}
  {2015})}\BibitemShut {NoStop}%
\bibitem [{\citenamefont {Szolnoki}\ and\ \citenamefont
  {Perc}(2016)}]{Szolnoki2016}%
  \BibitemOpen
  \bibfield  {author} {\bibinfo {author} {\bibfnamefont {A.}~\bibnamefont
  {Szolnoki}}\ and\ \bibinfo {author} {\bibfnamefont {M.}~\bibnamefont
  {Perc}},\ }\href {https://doi.org/10.1038/srep38608} {\bibfield  {journal}
  {\bibinfo  {journal} {Scientific Reports}\ }\textbf {\bibinfo {volume} {6}},\
  \bibinfo {pages} {38608} (\bibinfo {year} {2016})}\BibitemShut {NoStop}%
\bibitem [{\citenamefont {Kussell}\ \emph {et~al.}(2005)\citenamefont
  {Kussell}, \citenamefont {Kishony}, \citenamefont {Balaban},\ and\
  \citenamefont {Leibler}}]{Kussell05b}%
  \BibitemOpen
  \bibfield  {author} {\bibinfo {author} {\bibfnamefont {E.}~\bibnamefont
  {Kussell}}, \bibinfo {author} {\bibfnamefont {R.}~\bibnamefont {Kishony}},
  \bibinfo {author} {\bibfnamefont {N.~Q.}\ \bibnamefont {Balaban}},\ and\
  \bibinfo {author} {\bibfnamefont {S.}~\bibnamefont {Leibler}},\ }\href
  {https://doi.org/10.1534/genetics.104.035352} {\bibfield  {journal} {\bibinfo
   {journal} {Genetics}\ }\textbf {\bibinfo {volume} {169}},\ \bibinfo {pages}
  {1807} (\bibinfo {year} {2005})}\BibitemShut {NoStop}%
\bibitem [{\citenamefont {Bernatová}\ \emph {et~al.}(2013)\citenamefont
  {Bernatová}, \citenamefont {Samek}, \citenamefont {Pilát}, \citenamefont
  {Šerý}, \citenamefont {Ježek}, \citenamefont {Jákl}, \citenamefont
  {Šiler}, \citenamefont {Krzyžánek}, \citenamefont {Zemánek},
  \citenamefont {Holá}, \citenamefont {Dvořáčková},\ and\ \citenamefont
  {Růžička}}]{bernatova_following_2013}%
  \BibitemOpen
  \bibfield  {author} {\bibinfo {author} {\bibfnamefont {S.}~\bibnamefont
  {Bernatová}}, \bibinfo {author} {\bibfnamefont {O.}~\bibnamefont {Samek}},
  \bibinfo {author} {\bibfnamefont {Z.}~\bibnamefont {Pilát}}, \bibinfo
  {author} {\bibfnamefont {M.}~\bibnamefont {Šerý}}, \bibinfo {author}
  {\bibfnamefont {J.}~\bibnamefont {Ježek}}, \bibinfo {author} {\bibfnamefont
  {P.}~\bibnamefont {Jákl}}, \bibinfo {author} {\bibfnamefont
  {M.}~\bibnamefont {Šiler}}, \bibinfo {author} {\bibfnamefont
  {V.}~\bibnamefont {Krzyžánek}}, \bibinfo {author} {\bibfnamefont
  {P.}~\bibnamefont {Zemánek}}, \bibinfo {author} {\bibfnamefont
  {V.}~\bibnamefont {Holá}}, \bibinfo {author} {\bibfnamefont
  {M.}~\bibnamefont {Dvořáčková}},\ and\ \bibinfo {author} {\bibfnamefont
  {F.}~\bibnamefont {Růžička}},\ }\href
  {https://doi.org/10.3390/molecules181113188} {\bibfield  {journal} {\bibinfo
  {journal} {Molecules}\ }\textbf {\bibinfo {volume} {18}},\ \bibinfo {pages}
  {13188} (\bibinfo {year} {2013})}\BibitemShut {NoStop}%
\bibitem [{\citenamefont {Pankey}\ and\ \citenamefont
  {Sabath}(2004)}]{pankey_clinical_2004}%
  \BibitemOpen
  \bibfield  {author} {\bibinfo {author} {\bibfnamefont {G.~A.}\ \bibnamefont
  {Pankey}}\ and\ \bibinfo {author} {\bibfnamefont {L.~D.}\ \bibnamefont
  {Sabath}},\ }\href {https://doi.org/10.1086/381972} {\bibfield  {journal}
  {\bibinfo  {journal} {Clin. Infect. Dis.}\ }\textbf {\bibinfo {volume}
  {38}},\ \bibinfo {pages} {864} (\bibinfo {year} {2004})}\BibitemShut
  {NoStop}%
\bibitem [{\citenamefont {Nemeth}\ \emph {et~al.}(2015)\citenamefont {Nemeth},
  \citenamefont {Oesch},\ and\ \citenamefont
  {Kuster}}]{nemeth_bacteriostatic_2015}%
  \BibitemOpen
  \bibfield  {author} {\bibinfo {author} {\bibfnamefont {J.}~\bibnamefont
  {Nemeth}}, \bibinfo {author} {\bibfnamefont {G.}~\bibnamefont {Oesch}},\ and\
  \bibinfo {author} {\bibfnamefont {S.~P.}\ \bibnamefont {Kuster}},\ }\href
  {https://doi.org/10.1093/jac/dku379} {\bibfield  {journal} {\bibinfo
  {journal} {J. Antimicrob. Chemother.}\ }\textbf {\bibinfo {volume} {70}},\
  \bibinfo {pages} {382} (\bibinfo {year} {2015})}\BibitemShut {NoStop}%
\bibitem [{\citenamefont {San~Millan}\ and\ \citenamefont
  {Maclean}(2017)}]{sanmillan2017fitness}%
  \BibitemOpen
  \bibfield  {author} {\bibinfo {author} {\bibfnamefont {A.}~\bibnamefont
  {San~Millan}}\ and\ \bibinfo {author} {\bibfnamefont {R.~C.}\ \bibnamefont
  {Maclean}},\ }\href {https://doi.org/10.1128/microbiolspec.MTBP-0016-2017}
  {\bibfield  {journal} {\bibinfo  {journal} {Microbiol. Spectrum}\ }\textbf
  {\bibinfo {volume} {5}},\ \bibinfo {pages} {5} (\bibinfo {year}
  {2017})}\BibitemShut {NoStop}%
\bibitem [{\citenamefont {Danino}\ and\ \citenamefont
  {Shnerb}(2018)}]{danino_fixation_2018}%
  \BibitemOpen
  \bibfield  {author} {\bibinfo {author} {\bibfnamefont {M.}~\bibnamefont
  {Danino}}\ and\ \bibinfo {author} {\bibfnamefont {N.~M.}\ \bibnamefont
  {Shnerb}},\ }\href {https://doi.org/10.1016/j.jtbi.2018.01.004} {\bibfield
  {journal} {\bibinfo  {journal} {J. Theor. Biol.}\ }\textbf {\bibinfo {volume}
  {441}},\ \bibinfo {pages} {84} (\bibinfo {year} {2018})}\BibitemShut
  {NoStop}%
\bibitem [{\citenamefont {Danino}\ \emph {et~al.}(2018)\citenamefont {Danino},
  \citenamefont {Kessler},\ and\ \citenamefont
  {Shnerb}}]{danino_stability_2018}%
  \BibitemOpen
  \bibfield  {author} {\bibinfo {author} {\bibfnamefont {M.}~\bibnamefont
  {Danino}}, \bibinfo {author} {\bibfnamefont {D.~A.}\ \bibnamefont
  {Kessler}},\ and\ \bibinfo {author} {\bibfnamefont {N.~M.}\ \bibnamefont
  {Shnerb}},\ }\href {https://doi.org/10.1016/j.tpb.2017.11.003} {\bibfield
  {journal} {\bibinfo  {journal} {Theor. Popul. Biol.}\ }\textbf {\bibinfo
  {volume} {119}},\ \bibinfo {pages} {57} (\bibinfo {year} {2018})}\BibitemShut
  {NoStop}%
\bibitem [{\citenamefont {Danino}\ \emph {et~al.}(2016)\citenamefont {Danino},
  \citenamefont {Shnerb}, \citenamefont {Azaele}, \citenamefont {Kunin},\ and\
  \citenamefont {Kessler}}]{danino_effect_2016}%
  \BibitemOpen
  \bibfield  {author} {\bibinfo {author} {\bibfnamefont {M.}~\bibnamefont
  {Danino}}, \bibinfo {author} {\bibfnamefont {N.~M.}\ \bibnamefont {Shnerb}},
  \bibinfo {author} {\bibfnamefont {S.}~\bibnamefont {Azaele}}, \bibinfo
  {author} {\bibfnamefont {W.~E.}\ \bibnamefont {Kunin}},\ and\ \bibinfo
  {author} {\bibfnamefont {D.~A.}\ \bibnamefont {Kessler}},\ }\href
  {https://doi.org/10.1016/j.jtbi.2016.08.029} {\bibfield  {journal} {\bibinfo
  {journal} {J. Theor. Biol.}\ }\textbf {\bibinfo {volume} {409}},\ \bibinfo
  {pages} {155} (\bibinfo {year} {2016})}\BibitemShut {NoStop}%
\bibitem [{\citenamefont {Hidalgo}\ \emph {et~al.}(2017)\citenamefont
  {Hidalgo}, \citenamefont {Suweis},\ and\ \citenamefont
  {Maritan}}]{hidalgo_species_2017}%
  \BibitemOpen
  \bibfield  {author} {\bibinfo {author} {\bibfnamefont {J.}~\bibnamefont
  {Hidalgo}}, \bibinfo {author} {\bibfnamefont {S.}~\bibnamefont {Suweis}},\
  and\ \bibinfo {author} {\bibfnamefont {A.}~\bibnamefont {Maritan}},\ }\href
  {https://doi.org/10.1016/j.jtbi.2016.11.002} {\bibfield  {journal} {\bibinfo
  {journal} {J. Theor. Biol.}\ }\textbf {\bibinfo {volume} {413}},\ \bibinfo
  {pages} {1} (\bibinfo {year} {2017})}\BibitemShut {NoStop}%
\bibitem [{\citenamefont {Horsthemke}\ and\ \citenamefont
  {René}(1984)}]{horsthemke_lefever}%
  \BibitemOpen
  \bibfield  {author} {\bibinfo {author} {\bibfnamefont {W.}~\bibnamefont
  {Horsthemke}}\ and\ \bibinfo {author} {\bibfnamefont {L.}~\bibnamefont
  {René}},\ }in\ \href@noop {} {\emph {\bibinfo {booktitle} {Noise-induced
  transitions}}}\ (\bibinfo  {publisher} {Springer},\ \bibinfo {year} {1984})\
  p.\ \bibinfo {pages} {258–292}\BibitemShut {NoStop}%
\bibitem [{\citenamefont {Bena}(2006)}]{bena2006}%
  \BibitemOpen
  \bibfield  {author} {\bibinfo {author} {\bibfnamefont {I.}~\bibnamefont
  {Bena}},\ }\href {https://doi.org/10.1142/S0217979206034881} {\bibfield
  {journal} {\bibinfo  {journal} {Int. J. Mod. Phys. B}\ }\textbf {\bibinfo
  {volume} {20}},\ \bibinfo {pages} {2825} (\bibinfo {year}
  {2006})}\BibitemShut {NoStop}%
\bibitem [{\citenamefont {Ridolfi}\ \emph {et~al.}(2011)\citenamefont
  {Ridolfi}, \citenamefont {D'Odorico},\ and\ \citenamefont
  {Laio}}]{Ridolfi2011}%
  \BibitemOpen
  \bibfield  {author} {\bibinfo {author} {\bibfnamefont {L.}~\bibnamefont
  {Ridolfi}}, \bibinfo {author} {\bibfnamefont {P.}~\bibnamefont {D'Odorico}},\
  and\ \bibinfo {author} {\bibfnamefont {F.}~\bibnamefont {Laio}},\ }\href@noop
  {} {\emph {\bibinfo {title} {Noise-induced Phenomena in the Environmental
  Sciences}}}\ (\bibinfo  {publisher} {Cambridge University Press, Cambridge,
  U.K.},\ \bibinfo {year} {2011})\BibitemShut {NoStop}%
\bibitem [{\citenamefont {Gardiner}(2002)}]{Gardiner}%
  \BibitemOpen
  \bibfield  {author} {\bibinfo {author} {\bibfnamefont {C.~W.}\ \bibnamefont
  {Gardiner}},\ }\href@noop {} {\emph {\bibinfo {title} {Handbook of Stochastic
  Methods}}}\ (\bibinfo  {publisher} {Springer, USA},\ \bibinfo {year}
  {2002})\BibitemShut {NoStop}%
\bibitem [{\citenamefont {van Kampen}(1992)}]{VanKampen}%
  \BibitemOpen
  \bibfield  {author} {\bibinfo {author} {\bibfnamefont {N.~G.}\ \bibnamefont
  {van Kampen}},\ }\href@noop {} {\emph {\bibinfo {title} {Stochastic Processes
  in Physics and Chemistry}}}\ (\bibinfo  {publisher} {North-Holland,
  Amsterdam},\ \bibinfo {year} {1992})\BibitemShut {NoStop}%
\bibitem [{\citenamefont {Pinsky}\ and\ \citenamefont
  {Karlin}(2011)}]{Pinsky2011-px}%
  \BibitemOpen
  \bibfield  {author} {\bibinfo {author} {\bibfnamefont {M.~A.}\ \bibnamefont
  {Pinsky}}\ and\ \bibinfo {author} {\bibfnamefont {S.}~\bibnamefont
  {Karlin}},\ }in\ \href@noop {} {\emph {\bibinfo {booktitle} {An Introduction
  to Stochastic Modeling}}}\ (\bibinfo  {publisher} {Elsevier},\ \bibinfo
  {year} {2011})\ pp.\ \bibinfo {pages} {90--92}\BibitemShut {NoStop}%
\bibitem [{\citenamefont {Antal}\ and\ \citenamefont
  {Scheuring}(2006)}]{antal_fixation_2006}%
  \BibitemOpen
  \bibfield  {author} {\bibinfo {author} {\bibfnamefont {T.}~\bibnamefont
  {Antal}}\ and\ \bibinfo {author} {\bibfnamefont {I.}~\bibnamefont
  {Scheuring}},\ }\href {https://doi.org/10.1007/s11538-006-9061-4} {\bibfield
  {journal} {\bibinfo  {journal} {Bull. Math. Biol.}\ }\textbf {\bibinfo
  {volume} {68}},\ \bibinfo {pages} {1923} (\bibinfo {year}
  {2006})}\BibitemShut {NoStop}%
\bibitem [{\citenamefont {Gibson}\ and\ \citenamefont
  {Bruck}(2000)}]{gibson_efficient_2000}%
  \BibitemOpen
  \bibfield  {author} {\bibinfo {author} {\bibfnamefont {M.~A.}\ \bibnamefont
  {Gibson}}\ and\ \bibinfo {author} {\bibfnamefont {J.}~\bibnamefont {Bruck}},\
  }\href {https://doi.org/10.1021/jp993732q} {\bibfield  {journal} {\bibinfo
  {journal} {J. Phys. Chem. A}\ }\textbf {\bibinfo {volume} {104}},\ \bibinfo
  {pages} {1876} (\bibinfo {year} {2000})}\BibitemShut {NoStop}%
\bibitem [{\citenamefont {Anderson}(2007)}]{anderson_modified_2007}%
  \BibitemOpen
  \bibfield  {author} {\bibinfo {author} {\bibfnamefont {D.~F.}\ \bibnamefont
  {Anderson}},\ }\href {https://doi.org/10.1063/1.2799998} {\bibfield
  {journal} {\bibinfo  {journal} {J. Chem. Phys.}\ }\textbf {\bibinfo {volume}
  {127}},\ \bibinfo {pages} {214107} (\bibinfo {year} {2007})}\BibitemShut
  {NoStop}%
\bibitem [{\citenamefont {Spalding}\ \emph {et~al.}(2017)\citenamefont
  {Spalding}, \citenamefont {Doering},\ and\ \citenamefont
  {Flierl}}]{Spalding17}%
  \BibitemOpen
  \bibfield  {author} {\bibinfo {author} {\bibfnamefont {C.}~\bibnamefont
  {Spalding}}, \bibinfo {author} {\bibfnamefont {C.}~\bibnamefont {Doering}},\
  and\ \bibinfo {author} {\bibfnamefont {G.}~\bibnamefont {Flierl}},\ }\href
  {https://doi.org/10.1103/PhysRevE.96.042411} {\bibfield  {journal} {\bibinfo
  {journal} {Phys. Rev. E}\ }\textbf {\bibinfo {volume} {96}},\ \bibinfo
  {pages} {042411} (\bibinfo {year} {2017})}\BibitemShut {NoStop}%
\bibitem [{\citenamefont {Davis}(1984)}]{davis_piecewise-deterministic_1984}%
  \BibitemOpen
  \bibfield  {author} {\bibinfo {author} {\bibfnamefont {M.~H.~A.}\
  \bibnamefont {Davis}},\ }\href {https://www.jstor.org/stable/2345677}
  {\bibfield  {journal} {\bibinfo  {journal} {J. R. Stat. Soc. B}\ }\textbf
  {\bibinfo {volume} {46}},\ \bibinfo {pages} {353} (\bibinfo {year}
  {1984})}\BibitemShut {NoStop}%
\bibitem [{\citenamefont {Assaf}\ and\ \citenamefont
  {Meerson}(2017)}]{assaf2017}%
  \BibitemOpen
  \bibfield  {author} {\bibinfo {author} {\bibfnamefont {M.}~\bibnamefont
  {Assaf}}\ and\ \bibinfo {author} {\bibfnamefont {B.}~\bibnamefont
  {Meerson}},\ }\href {https://doi.org/10.1088/1751-8121/aa669a} {\bibfield
  {journal} {\bibinfo  {journal} {J. Phys. A: Math. Theor.}\ }\textbf {\bibinfo
  {volume} {50}},\ \bibinfo {pages} {263001} (\bibinfo {year}
  {2017})}\BibitemShut {NoStop}%
\bibitem [{\citenamefont {Mobilia}\ and\ \citenamefont {Assaf}(2010)}]{MA10}%
  \BibitemOpen
  \bibfield  {author} {\bibinfo {author} {\bibfnamefont {M.}~\bibnamefont
  {Mobilia}}\ and\ \bibinfo {author} {\bibfnamefont {M.}~\bibnamefont
  {Assaf}},\ }\href {https://doi.org/10.1209/0295-5075/91/10002} {\bibfield
  {journal} {\bibinfo  {journal} {EPL}\ }\textbf {\bibinfo {volume} {91}},\
  \bibinfo {pages} {10002} (\bibinfo {year} {2010})}\BibitemShut {NoStop}%
\bibitem [{\citenamefont {Assaf}\ and\ \citenamefont {Mobilia}(2010)}]{AM10}%
  \BibitemOpen
  \bibfield  {author} {\bibinfo {author} {\bibfnamefont {M.}~\bibnamefont
  {Assaf}}\ and\ \bibinfo {author} {\bibfnamefont {M.}~\bibnamefont
  {Mobilia}},\ }\href {https://doi.org/10.1088/1742-5468/2010/09/P09009}
  {\bibfield  {journal} {\bibinfo  {journal} {J. Stat. Mech.}\ }\textbf
  {\bibinfo {volume} {2010}},\ \bibinfo {pages} {P09009} (\bibinfo {year}
  {2010})}\BibitemShut {NoStop}%
\bibitem [{\citenamefont {Cremer}\ \emph {et~al.}(2009)\citenamefont {Cremer},
  \citenamefont {Reichenbach},\ and\ \citenamefont {Frey}}]{cremer_edge_2009}%
  \BibitemOpen
  \bibfield  {author} {\bibinfo {author} {\bibfnamefont {J.}~\bibnamefont
  {Cremer}}, \bibinfo {author} {\bibfnamefont {T.}~\bibnamefont
  {Reichenbach}},\ and\ \bibinfo {author} {\bibfnamefont {E.}~\bibnamefont
  {Frey}},\ }\bibfield  {journal} {\bibinfo  {journal} {New J. Phys.}\ }\textbf
  {\bibinfo {volume} {11}},\ \href
  {https://doi.org/10.1088/1367-2630/11/9/093029}
  {10.1088/1367-2630/11/9/093029} (\bibinfo {year} {2009})\BibitemShut
  {NoStop}%
\bibitem [{\citenamefont {Reichenbach}\ \emph {et~al.}(2007)\citenamefont
  {Reichenbach}, \citenamefont {Mobilia},\ and\ \citenamefont
  {Frey}}]{reich2007}%
  \BibitemOpen
  \bibfield  {author} {\bibinfo {author} {\bibfnamefont {T.}~\bibnamefont
  {Reichenbach}}, \bibinfo {author} {\bibfnamefont {M.}~\bibnamefont
  {Mobilia}},\ and\ \bibinfo {author} {\bibfnamefont {E.}~\bibnamefont
  {Frey}},\ }\href {https://doi.org/10.1038/nature06095} {\bibfield  {journal}
  {\bibinfo  {journal} {Nature}\ }\textbf {\bibinfo {volume} {448}},\ \bibinfo
  {pages} {1046} (\bibinfo {year} {2007})}\BibitemShut {NoStop}%
\bibitem [{\citenamefont {Asker}\ \emph
  {et~al.}(2023{\natexlab{a}})\citenamefont {Asker}, \citenamefont
  {{Hern\'andez-Navarro}},\ and\ \citenamefont {Mobilia}}]{videos}%
  \BibitemOpen
  \bibfield  {author} {\bibinfo {author} {\bibfnamefont {M.}~\bibnamefont
  {Asker}}, \bibinfo {author} {\bibfnamefont {L.}~\bibnamefont
  {{Hern\'andez-Navarro}}},\ and\ \bibinfo {author} {\bibfnamefont
  {M.}~\bibnamefont {Mobilia}},\ }\bibfield  {journal} {\bibinfo  {journal}
  {Figshare}\ }\href {https://doi.org/10.6084/m9.figshare.23553066}
  {10.6084/m9.figshare.23553066} (\bibinfo {year}
  {2023}{\natexlab{a}})\BibitemShut {NoStop}%
\bibitem [{\citenamefont {Assaf}\ and\ \citenamefont {Mobilia}(2011)}]{AM11}%
  \BibitemOpen
  \bibfield  {author} {\bibinfo {author} {\bibfnamefont {M.}~\bibnamefont
  {Assaf}}\ and\ \bibinfo {author} {\bibfnamefont {M.}~\bibnamefont
  {Mobilia}},\ }\href {https://doi.org/10.1016/j.jtbi.2011.01.025} {\bibfield
  {journal} {\bibinfo  {journal} {J. Theor. Biol.}\ }\textbf {\bibinfo {volume}
  {275}},\ \bibinfo {pages} {93} (\bibinfo {year} {2011})}\BibitemShut
  {NoStop}%
\bibitem [{\citenamefont {West}\ \emph {et~al.}(2023)\citenamefont {West},
  \citenamefont {Adler}, \citenamefont {Gallaher}, \citenamefont {Strobl},
  \citenamefont {Brady-Nicholls}, \citenamefont {Brown}, \citenamefont
  {Roberson-Tessi}, \citenamefont {Kim}, \citenamefont {Noble}, \citenamefont
  {Viossat}, \citenamefont {Basanta},\ and\ \citenamefont
  {Anderson}}]{adaptive_therapy}%
  \BibitemOpen
  \bibfield  {author} {\bibinfo {author} {\bibfnamefont {J.}~\bibnamefont
  {West}}, \bibinfo {author} {\bibfnamefont {F.}~\bibnamefont {Adler}},
  \bibinfo {author} {\bibfnamefont {J.}~\bibnamefont {Gallaher}}, \bibinfo
  {author} {\bibfnamefont {M.}~\bibnamefont {Strobl}}, \bibinfo {author}
  {\bibfnamefont {R.}~\bibnamefont {Brady-Nicholls}}, \bibinfo {author}
  {\bibfnamefont {J.}~\bibnamefont {Brown}}, \bibinfo {author} {\bibfnamefont
  {M.}~\bibnamefont {Roberson-Tessi}}, \bibinfo {author} {\bibfnamefont
  {E.}~\bibnamefont {Kim}}, \bibinfo {author} {\bibfnamefont {R.}~\bibnamefont
  {Noble}}, \bibinfo {author} {\bibfnamefont {Y.}~\bibnamefont {Viossat}},
  \bibinfo {author} {\bibfnamefont {D.}~\bibnamefont {Basanta}},\ and\ \bibinfo
  {author} {\bibfnamefont {A.~R.}\ \bibnamefont {Anderson}},\ }\href
  {https://doi.org/10.7554/eLife.84263} {\bibfield  {journal} {\bibinfo
  {journal} {eLife}\ }\textbf {\bibinfo {volume} {12}},\ \bibinfo {pages}
  {e84263} (\bibinfo {year} {2023})}\BibitemShut {NoStop}%
\bibitem [{\citenamefont {Asker}\ \emph
  {et~al.}(2023{\natexlab{b}})\citenamefont {Asker}, \citenamefont
  {{Hern\'andez-Navarro}},\ and\ \citenamefont {Mobilia}}]{data}%
  \BibitemOpen
  \bibfield  {author} {\bibinfo {author} {\bibfnamefont {M.}~\bibnamefont
  {Asker}}, \bibinfo {author} {\bibfnamefont {L.}~\bibnamefont
  {{Hern\'andez-Navarro}}},\ and\ \bibinfo {author} {\bibfnamefont
  {M.}~\bibnamefont {Mobilia}},\ }\bibfield  {journal} {\bibinfo  {journal}
  {Research Data Leeds Repository}\ }\href {https://doi.org/10.5518/1371}
  {10.5518/1371} (\bibinfo {year} {2023}{\natexlab{b}})\BibitemShut {NoStop}%
\end{thebibliography}%
\newpage\hbox{}\thispagestyle{empty}\newpage
\onecolumngrid
\section*{Appendix}
\label{sec:SM}

\setcounter{figure}{0}
\setcounter{equation}{0}
\renewcommand{\thefigure}{S\arabic{figure}}
\renewcommand{\theequation}{S\arabic{equation}}

In this 
appendix, we provide some further technical details and supplementary information in support
of the results discussed in the main text. We also provide additional information concerning the mean-field, Moran and PDMP approximations used in the main text, the simulation methods, we 
illustrate our main findings by discussing typical sample paths, and briefly discuss 
the generalisation
of the model to correlated/anticorrelated EV.

\section*{SM1. Table of parameters and physical quantities of interest}
\label{SM1}

{
The various parameters of the model and physical quantities of interest are summarised in the following table.}

\begin{center}

\begin{table}[h!]
\centering
\caption{Table of parameters and physical quantities of interest.}
\begin{tabular}{|l|l|}
\hline
\multicolumn{1}{|c|}{\textbf{Parameter / Quantity}}                                                                           & \multicolumn{1}{c|}{\textbf{Notation / Definition}} \\ \hline
Population size                                                                                                    & $N$                                    \\ \hline
Population size of type $i \in \{R, S\}$                                                             & $N_i$                             \\ \hline
Population size of type $i \in \{R, S\}$                                                             & $f_i$                             \\ \hline
Selection strength / strength of toxin environmental variability                                                   & $s$                                    \\ \hline
Dichotomous Markov Noise variable for noise of type $\alpha \in \{T, K\}$                                          & $\xi_\alpha$                           \\ \hline
Carrying capacity of the microbial community                                                                                       & $K$                                    \\ \hline
Largest value that $K$ can take                                                                    & $K_+$                                  \\ \hline
Smallest value that $K$ can take                                                                    & $K_-$                                  \\ \hline
Arithmetic mean of $K_{\pm}$                                                            & $K_0\equiv \frac{K_++K_-}{2}$                                  \\ \hline
Scaled difference in carrying capacity values / strength of nutrient environmental variability                     & $\gamma\equiv \frac{K_+-K_-}{2K_0}$                               \\ \hline
Average switching rate and half of the inverse  correlation time of  $\xi_\alpha$, with $\alpha \in \{T, K\}$                                        & $\nu_\alpha$                           \\ \hline
Switching rate from $+$ into $-$ ($+$) or from $-$ into $+$ ($-$) of the environmental noise $\xi_\alpha$, with $\alpha \in \{T, K\}$ & $\nu_\alpha^\pm$                       \\ \hline
Switching bias of the environmental noise $\xi_\alpha$, with $\alpha \in \{T, K\}$                                            & $\delta_\alpha\equiv \frac{\nu_{\alpha}^- -\nu_{\alpha}^+}{2\nu_{\alpha}}$                        \\ \hline
Fraction of resistant bacteria out of total population                                                             & $x$                                    \\ \hline
Average fitness of current population                                                                              & $\overline{f}$                         \\ \hline
Birth ($+$) and death ($-$) rates for bacteria of type $i \in \{R,S\}$                                                & $T_i^\pm$                         \\ \hline
Birth ($+$) and death ($-$) rates for the  Moran approximation                                                                & $\widetilde{T}_R^\pm\equiv \frac{T_R^{\pm}T_S^{\mp}}{N}$ (constant  $N$)                     \\ \hline

Starting fraction of resistant bacteria                                                                            & $x_0$                                  \\ \hline
Fixation probability in state $\xi_T$ with Moran approximation (initially $n$ bacteria of type $R$)                                                                           & $\phi_n^{\xi_T}$                   \\ \hline
Mean fixation time in state $\xi_T$ with Moran approximation (initially $n$ bacteria of type $R$)                                                                          & $\tau_n^{\xi_T}$                   \\ \hline
Fixation probability with the Moran approximation ($x_0=0.5$)                                                                    & $\phi_{\text{MA}}$                     \\ \hline
Mean fixation time with the Moran approximation ($x_0=0.5$)                                                                     & $\tau_{\text{MA}}$                     \\ \hline
Fixation probability of strain $R$                                                                                              & $\phi$                                 \\ \hline
Coexistence probability                                                                                            & $\eta$                                 \\ \hline
Mean population size                                                                                               & $\langle N \rangle$                    \\ \hline
Modal population size                                                                                              & $\hat{N}$                              \\ 
\hline
Equilibrium fraction of $R$ in coexistence state                                                                           & $x^*$                    \\ \hline
Fraction of resistant strain $R$ (regardless of coexistence/fixation)                                                                          & $\langle x \rangle$ ($\langle x \rangle\approx x^*$ when $\eta\approx 1$)                   \\ \hline
Population density function of the $N$-PDMP                                                                          & $p(N)$                    \\ \hline
Mean population size of type $R$                                                                          & $\langle N_R \rangle$                   \\ \hline

\end{tabular}
\end{table}
\end{center}

\section*{SM2. $T$-switching coexistence: fluctuation amplitude and time scale}
\label{SM2}
The selection strength $s$ does not only shape the dynamics of the composition, see Eq.~(7), but it also determines the amplitude of the $T$-EV fluctuations, i.e. its variance, which is linked to stronger coexistence in the fast-switching regime (see Sec.~III). To show this, we first consider the normalised fitness of the resistant subpopulation
\begin{equation}
\label{eq:resistantratio}
    \frac{f_R}{\overline{f}}=\frac{1}{x+(1-x)\exp(\xi_T s)}.
    \nonumber
\end{equation}
Since we here focus on how $s$ may shape coexistence through the toxin level fluctuation size, and coexistence dominates in the fast toxin-switching regime, we assume $\nu_T\gg1$. As discussed in the main text, in the fast-switching regime, the per capita growth rate 
(normalised fitness) of strain $R$ averaged over the stationary distribution of $\xi_T$ reads:
\begin{equation}
    \avg{\frac{f_R}{\overline{f}}}=\frac{1-\delta_T}{2} \frac{1}{x+(1-x)\exp(-s)}+\frac{1+\delta_T}{2} \frac{1}{x+(1-x)\exp(s)}.
    \nonumber
\end{equation}
To derive its variance, we also have to compute the average of its square as
\begin{equation}
    \avg{\left(\frac{f_R}{\overline{f}}\right)^2}=\frac{1-\delta_T}{2} \frac{1}{(x+(1-x)\exp(-s))^2}+\frac{1+\delta_T}{2} \frac{1}{(x+(1-x)\exp(s))^2}.
    \nonumber
\end{equation}
Combining both, we obtain the variance of the normalised resistant fitness due to the environmental fluctuations in the toxin level
\begin{equation}
\begin{aligned}
    \text{var}\left(\frac{f_R}{\overline{f}}\right)&=\avg{\left(\frac{f_R}{\overline{f}}\right)^2}-\avg{\frac{f_R}{\overline{f}}}^2\\
    &=\frac{(1-\delta_T^2)}{4}\left(\frac{(\exp(2s)-1)(1-x)}{(x+(1-x)\exp(s))(1+x(\exp(s)-1))}\right)^2.
\end{aligned}
\nonumber
\end{equation}
A similar analysis for the sensitive strain, which has a normalised fitness $\exp(\xi_T s)$ times that of the resistant strain, provides
\begin{equation}
    \text{var}\left(\frac{f_S}{\overline{f}}\right)=\frac{(1-\delta_T^2)}{4}\left(\frac{(\exp(2s)-1)x}{(x+(1-x)\exp(s))(1+x(\exp(s)-1))}\right)^2.
    \nonumber
\end{equation}
In both cases, we conclude that the variance (arising from the $T$-EV) indeed increases with the selection strength $s$. Therefore, $s$ does shape the strength of coexistence.

Note that the amplitude of the fluctuations of the carrying capacity $K$-EV (at no bias $\delta_K=0$, for simplicity) increases with $\gamma$. However, in this case, the larger the $K$-EV the weaker the coexistence, as the harsh state $K_{-}\rightarrow0$ further promotes extinction. In conclusion, and as discussed in Sec.~IV~B and Fig.~6, the environmental variability can either promote coexistence ($T$-EV) or jeopardise it ($K$-EV), and both parameters $s$ and $\gamma$ determine long-lived coexistence.

{To discuss the coexistence timescale,} we consider the small $s$ regime by expanding  
the numerator and denominator of the right-hand-side of  Eq.~(9) to order ${\cal O}(s^2)$, which yields
\begin{equation}
\begin{aligned}
\label{eq:MFsmalls}
    \dot{x}&\approx\frac{x(1-x)}{2}\left[\frac{(1+\delta_T) (1-(1+s+s^2/2))}{x+(1-x)(1+s+s^2/2)} + \frac{(1-\delta_T)(1-(1-s+s^2/2))}{x+(1-x)(1-s+s^2/2)}\right]\\
    &\approx \frac{x(1-x)}{2}\left[\frac{(1+\delta_T)(-s-s^2/2)(1-s(1-x))+(1-\delta_T)(s-s^2/2)(1+s(1-x))}{1-s^2(1-x)}\right]\\
    &\approx -s^2x(1-x)\left[x-\left(\frac{1}{2}-\frac{\delta_T}{s}\right)\right]=-s^2x(1-x)\left[x-x^*\right],
\end{aligned}
\end{equation}
where $x^*=\frac{1}{2}-\frac{\delta_T}{s}$ is the coexistence
equilibrium under $s\ll1$. The equilibrium $x^*$ is physical (and stable) only when $-\frac{s}{2}<\delta_T<\frac{s}{2}$; i.e. only in the special case of almost symmetric switching ($\delta_T={\cal O}(s)$ or smaller). 
From Eq.~\eqref{eq:MFsmalls}, the coexistence equilibrium is approached slowly, with a relaxation time on the order of $\sim 1/s^2$; thus, taking into account demographic noise, the expected fixation time is $\tau\sim e^{\avg{N}s^2}$ when $\avg{N}s^2\gg 1$.

\section*{SM3. Moran approximation: probability and mean fixation time}
\label{SM3}
\subsection{Moran fixation probability}

The exact Moran fixation probability $\phi\left(N_R^0,N,s,\xi_{T}\right)$ for the resistant strain to take over the entire population of constant size $N$ in the {\it static} toxin environment $\xi_T$, starting with \(N_R^0\) resistant individuals, is~\cite{Ewens, Gardiner,VanKampen,antal_fixation_2006}
\begin{equation}
    \phi\left(N_R^0,N,s,\xi_{T}\right)=\frac{1+\sum_{k=1}^{N_R^0-1}\prod_{i=1}^{k}\gamma\left(i,N,s,\xi_{T}\right)}{1+\sum_{k=1}^{N-1}\prod_{i=1}^{k}\gamma\left(i,N,s,\xi_{T}\right)},~\text{for}~\gamma\left(N_R^0,N\right)\equiv\frac{\widetilde{T}^-_R\left(N_R^0,N,s,\xi_{T}\right)}{\widetilde{T}^+_R\left(N_R^0,N,s,\xi_{T}\right)}~\text{and}~N_R^0=1, 2,..., N.
    \label{SuppEq:GenFixProb}
\end{equation}

The effective Moran transition rates are \(\widetilde{T}^+_R=T^+_RT^-_S/N\) and \(\widetilde{T}^-_R=T^-_RT^+_S/N\) (see Sec.~III)~\cite{wienand_evolution_2017,wienand_eco-evolutionary_2018}, which read
\begin{equation}
\nonumber
\begin{aligned}
 \widetilde{T}^+_R\left(x,N,s,\xi_{T}\right)&=\frac{x(1-x)}{x+(1-x)e^{s\xi_{T}}}N,\text{~and}\\
 \widetilde{T}^-_R\left(x,N,s,\xi_{T}\right)&=\frac{x(1-x)e^{s\xi_{T}}}{x+(1-x)e^{s\xi_{T}}}N,
\label{SuppEq:transrates}
\end{aligned}
\end{equation}
where \(x\equiv N_R/N\) as in the main text. Hence, our particular factor \(\gamma\) yields
\begin{align}
 \nonumber
 \gamma\left(s,\xi_{T}\right)&=e^{s\xi_{T}}.
\label{SuppEq:gamma}
\end{align}

The general Moran fixation probability Eq.~\eqref{SuppEq:GenFixProb} is valid for fixed \(N\) and time-independent \(\widetilde{T}^\pm_R\), i.e. fixed \(\xi_{T}=\pm1\) (static environments). Therefore, we compute our particular resistant fixation probability in static environments as
\begin{equation}
    \phi\left(N_R^0,N,s,\xi_{T}\right)=\frac{\sum_{k=0}^{N_R^0-1}\left(e^{s\xi_{T}}\right)^{k}}{\sum_{k=0}^{N-1}\left(e^{s\xi_{T}}\right)^{k}}=\frac{1-e^{sN_R^0\xi_{T}}}{1-e^{sN\xi_{T}}},
    \label{SuppEq:ParticularFixProb}
\end{equation}
where, in the last step, we use that this is a finite geometric series. As noted in the main text, we always use a starting resistant fraction \(x_0=0.5\), a starting resistant population of \(N_R^0=N/2=K_0/2\). For brevity, in the main text, the fixation probability $\phi\left(N/2,N,s,\xi_{T}\right)$ is thus denoted by 
$\phi_{\rm MA}{\big|_{\xi_T}}$.

In the regime \(\nu_{T}\gg1\), the effective Moran transition rates become time-independent; see Eq.~(8). However, in this case, as the \(\gamma\) factor shows a complex dependency on the resistant fraction \(x\), the fixation probability
$\phi_{\rm MA}{\big|_{\xi_T}}$ in the main text is computed numerically through Eq.~\eqref{SuppEq:GenFixProb}, as shown in Eq.~(12).

\subsection{Moran unconditional Mean Fixation Time (MFT)}\label{A2}

When the Moran `birth-death' transition rates are time-independent,  the unconditional MFT \(t\left(N_R^0,N,s,\xi_{T}\right)\) in a population of constant size $N$ in the {\it static} toxin environment $\xi_T$, and consisting initially of $N_R^0$ resistant individuals,   
reads~\cite{Ewens,Gardiner,VanKampen,antal_fixation_2006}:
\begin{equation}
    \begin{aligned}
    \tau\left(N_R^0,N,s,\xi_{T}\right)=-\tau_1\left(N,s,\xi_{T}\right)\sum_{k=N_R^0}^{N-1}\prod_{i=1}^{k}\gamma\left(i,N,s,\xi_{T}\right)+
    \sum_{k=N_R^0}^{N-1}\sum_{n=1}^{k}\frac{\prod_{m=n+1}^{k}\gamma\left(m,N,s,\xi_{T}\right)}{\widetilde{T}^+_R\left(n,N,s,\xi_{T}\right)},\\
    \text{for}~N_R^0=1, 2,..., N.
    \end{aligned}
    \label{SuppEq:GenMeanAbsTime}
    \nonumber
\end{equation}
where \(\gamma\left(N_R^0,N,s,\xi_{T}\right)\) is defined as in Eq.~\eqref{SuppEq:GenFixProb}; \(\widetilde{T}^\pm_R\left(N_R^0,N,s,\xi_{T}\right)\) are defined either in Eq.~\eqref{SuppEq:transrates} or Eq.~(8) for static or fast toxin switching environments, respectively; and
\begin{equation}
    \begin{aligned}
    \tau_1\left(N,s,\xi_{T}\right)=\frac{\sum_{k=1}^{N-1}\sum_{n=1}^{k}\frac{\prod_{m=n+1}^{k}\gamma\left(m,N,s,\xi_{T}\right)}{\widetilde{T}^+_R\left(n,N,s,\xi_{T}\right)}}{1+\sum_{k=1}^{N-1}\prod_{i=1}^{k}\gamma\left(i,N,s,\xi_{T}\right)}
    \end{aligned}
    \label{SuppEq:1ResMeanAbsTime}
\end{equation}
is the unconditional MFT starting with a single resistant individual. In our examples,  we always use  \(x_0=0.5, N_R^0=N/2=K_0/2\). For brevity, in the main text, the MFT $\tau\left(N/2,N,s,\xi_{T}\right)$ is thus denoted by 
$\tau_{\rm MA}{\big|_{\xi_T}}$.

As for the fixation probability in the main text, we can obtain analytical expressions of $\tau_{\rm MA}{\big|_{\xi_T}}$ in the very slow and fast toxin switching regimes ($\nu_T\ll 1$ and $\nu_T\gg 1$). For the former, we have to substitute $\phi_{\rm MA}{\big|_{\xi_T}}$ in Eq.~(11) by the MFT expression of Eq.~\eqref{SuppEq:GenMeanAbsTime}. And, for the latter, we have to substitute each $\widetilde{T}^\pm_R$ by its corresponding average across the stationary $\xi_T$ distribution, i.e. $\langle\widetilde{T}^\pm_R\rangle$.

\section*{SM4. Simulation methods \& environmental noise}
\label{SM4} 
To study the stochastic behaviour of the full model presented here, we perform {\it exact} stochastic numerical simulations~\cite{gibson_efficient_2000, anderson_modified_2007}. Simulations start at an initial time \(t=t_0=0\) with an initial toxin level \(\xi_T(t_0)\) and resourcce level \(\xi_K(t_0)\) always at stationarity (with \(\left<\xi_{T/K}(t_0)\right>=\delta_{T/K}\)), initial populations \(N_R(t_0)=N_S(t_0)=K_0/2=(K_+ + K_-)/4\), and we take into account all the possible reactions that can take place.
In the case of the full model this means: (1) the four possible birth or death reactions with rates \{\(T^+_{R}(t)\), \(T^-_{R}(t)\), \(T^+_{S}(t)\), \(T^-_{S}(t)\)\} that depend on the variables \{\(N_R(t)\), \(N_S(t)\), \(K(t)\), \(\xi_T(t)\)\} and the constant parameter \(s\); (2) the nutrient level switches stochastically with constant rate \(\nu_K^{\pm}\) for the state \(K(t)=K_{\pm}\); and (3) the toxin level switches stochastically with constant rate \(\nu_T^{\pm}\) for the state \(\xi_T(t)=\pm1\). We perform efficient stochastic simulations by implementing the Next Reaction Method~\cite{gibson_efficient_2000, anderson_modified_2007}. 

{We have determined the standard errors for the data sets shown in Figs. 2, 4, and 8, in each case on 
 the sample mean over $10^3$ realisations. In the main text, we show the standard error on the mean for the case $\delta_T=0.0$ in Fig. 2(b) where we expect the highest variance in simulation result.
This illustrates that typical error bars in  simulation results reported in Figs. 2, 4, and 8
do not ever exceed the size of the markers. Hence to maintain readability of the figures, and without loss of information, we have omitted error bars in the figures others than Fig. 2(b) for $\delta_T=0.0$.}

Regarding Fig.~5, the direct numerical implementation of Eqs.~(19) and~(20) to predict fixation and coexistence probabilities is not feasible, because calculating \(\phi(N)\) and the coexistence probability for each integer \(N\in[K_-=400,K_+=2000]\) is computationally too expensive. Therefore, we capitalise on the exact Moran results for static environments (see Eqs.~\eqref{SuppEq:ParticularFixProb} and~\eqref{SuppEq:GenMeanAbsTime}), and the $N$-PDMP PDF \(p(N)\) (see Eq.~(17) and Fig.~4), to provide the theoretical prediction for the  fixation and coexistence probabilities. For the first regime at very low $\nu_K$  (Fig.~5(a,d)), the starting environment is unlikely to switch, and the distribution of the total population is bimodal; see left inset in Fig.~4. The fixation and coexistence probabilities of Fig.~5(d) are then computed as the weighted average of the Moran probabilities for the static environment cases \(N=K_\pm\), with weights \(\left(1\pm\delta_K\right)/2\). For the regime of Fig.~5(b,e), irrespective of the starting environment, the population size will, on average, spend significant time on both \(N=K_-\) and \(K_+\). Since relative demographic fluctuations are larger at \(K_-\), the simulated fixation and coexistence probabilities of Fig.~5(b) are captured in Fig.~5(e) by a Moran process under total population \(N=K_-\). Finally, for high $\nu_K$, the environmental noise averages out and the behaviour of the system corresponds to that of a Moran process at \(N=\mathcal{K}\); see right inset in Fig.~4.

{
An additional comment on environmental noise is in order. As explained in the main text, we have chosen to model environmental variability using dichotomous Markov noise (DMN).
This choice allows us to make mathematical progress while keeping the theoretical modelling close to laboratory experimental conditions. In fact, the properties of DMN and their relations with other forms of noises have been extensively studied \cite{taitelbaum2023evolutionary,horsthemke_lefever,bena2006,Ridolfi2011}.
The advantage of modelling environmental variability with DMN is well illustrated by considering the time-varying carrying capacity
\[K(t)=K_0[1+\gamma \xi_K(t)],\]
where $\gamma\equiv (K_+-K_-)/(2K_0)$ and $0<K_0\equiv (K_++K_-)/2$ (with $0<\gamma<1$ and $0<K_0<\infty$). 
When $\xi_K(t)\in\{-1,1\}$ is a dichotomous noise, $K(t)\in [K_-,K_+]$ is bounded and always physical, and 
the evolution of the community is fully characterised by the master equation of Eq.~(6) that can be simulated {\it exactly} by the  methods described above. However, if $\xi_K(t)$ was an unbounded noise of continuous range $(-\infty,\infty)$, e.g. a 
Gaussian white noise (zero correlation time) or an Ornstein-Uhlenbeck noise (finite correlation time), we would face a number of  challenging problems:\\
(i) The carrying capacity would no longer be bounded, and could take unphysical (negative) values (when $\xi_K<-1/\gamma$).\\
(ii) The evolutionary process resulting from the coupling of the birth-death process defined by Eq.~(4) with the dynamics of 
$K(t)$ driven by an (unbounded or bounded) environmental noise with continuous range  would generally result in a {\it non-Markovian} process~\cite{taitelbaum2023evolutionary}. There would therefore be no corresponding master equation governing the population dynamics. The  mathematical analysis would thus be difficult, and even simulating the  process would be a  challenging task since the above methods could not be used directly. Circumventing these issues generally
 requires some approximations (e.g. truncating a Gaussian noise, finding a way to restore the Markov property) whose accuracy, validity and implications may be difficult to control, see e.g. \cite{roberts_dynamics_2015, taitelbaum2023evolutionary, Assaf13b} and in particular
 the discussion in the Supplemental Material of \cite{taitelbaum2023evolutionary}.

\section*{SM5. Stationary probability density function of the $x$-PDMP}
\label{SM5} 
The  joint stationary PDF
of the $x$-PDMP 
is labeled by $\rho_{\pm}(x)\equiv \rho(x,\xi_T=\pm)$, and satisfies~\cite{horsthemke_lefever,bena2006,Ridolfi2011}
\begin{equation}
\label{eq:PDMPde}
    \partial_t \rho_\pm = -\partial_x(\dot{x}_\pm\rho_\pm)-\nu_\pm \rho_\pm + \nu_\mp \rho_\mp,
\end{equation}
where $\dot{x}_{\pm}\equiv x(1-x)(1-e^{\pm s})/[
    x+(1-x)e^{\pm s}]
    $,
and the marginal stationary PDF of the $x$-PDMP defined by Eq.~(7)
is $\rho(x)\equiv \rho_+(x)+\rho_-(x)$.

Following \cite{horsthemke_lefever, Ridolfi2011}, from  Eq.~\eqref{eq:PDMPde}, we can define $J_\pm=\dot{x}_\pm \rho_\pm + \int^x_0{(\nu_\pm \rho_\pm - \nu_\mp \rho_\mp)~\text{d}x'}$ as the probability flux of the system, and then rewrite Eq.~\eqref{eq:PDMPde} as $\partial_t \rho_\pm = -\partial_x J_\pm$. Note that by definition the $x$-PDMP ignores
 intrinsic noise, hence $x=0,1$ states cannot be reached. We thus set   probability flux at the boundaries to zero as natural boundary conditions (BCs)~\cite{Gardiner}. Since we want to derive the stationary joint PDF \(\rho\left(t\rightarrow\infty\right)\)
 under zero-current BCs, we use 
$\rho =\rho_+ + \rho_-$ and take $\partial_t \rho = -\partial_x (J_+ + J_-) = 0$. In this case, we obtain $J_+ + J_- = 0$, where the flux boundary conditions set the integration constant to $0$. From this relation we can find that $\rho_\pm = \frac{-\dot{x}_\mp \rho_\mp}{\dot{x}_\pm}$. It is  useful to  
introduce the auxiliary variable $q \equiv \rho_+ - \rho_-$, and write $J_+ + J_- = 0$ as
$\dot{x}_+ \frac{\rho+q}{2} + \dot{x}_- \frac{\rho - q}{2} = 0$. After rearranging for $q$ we can substitute this into our expression for $\rho_\pm$ to find $\rho_\pm = \frac{\rho \pm q}{2}=\frac{\pm \dot{x}_\mp}{\dot{x}_- - \dot{x}_+}\rho$. The equation for the PDMP density of the resistant fraction \(x\) at quasi-stationary coexistence then reads
\begin{equation}
\nonumber
    \partial_x \left(\frac{\dot{x}_- \dot{x}_+}{\dot{x}_- - \dot{x}_+}\rho\right) + \frac{\dot{x}_- \dot{x}_+}{\dot{x}_- - \dot{x}_+}\rho \left(\frac{\nu_+}{\dot{x}_+} + \frac{\nu_-}{\dot{x}_-} \right) = 0.
\end{equation}
Multiplying by \(-1\), rearranging, and integrating, leads to
\begin{equation}
    \ln{\left(\frac{\rho}{\frac{1}{-\dot{x}_+} + \frac{1}{\dot{x}_-}}\right)} - \ln{\left(C\right)} = \int^x{\left(\frac{\nu_+}{-\dot{x}_+} - \frac{\nu_-}{\dot{x}_-} \right)~\text{d}x'},
    \label{SuppEq:xPDMP0}
\end{equation}
where $C$ is an integration constant, with \[\frac{1}{-\dot{x}_+}=\frac{\frac{e^s}{x}+\frac{1}{1-x}}{e^s-1}\text{~~and~~}\frac{1}{\dot{x}_-}=\frac{\frac{1}{x}+\frac{e^s}{1-x}}{e^s-1},\] see Eq.~(7) with \(\xi_{T}=\pm1\). Eq.~\eqref{SuppEq:xPDMP0} yields
\begin{equation}
    \ln{\left(\frac{\rho}{C\frac{e^s+1}{e^s-1}\frac{1}{x(1-x)}}\right)} = \frac{\nu_{+}\left(e^s\ln{\left(x\right)}-\ln{\left(1-x\right)}\right)-\nu_{-}\left(\ln{\left(x\right)}-e^s\ln{\left(1-x\right)}\right)}{e^s-1},
    \nonumber
    \label{SuppEq:xPDMP1}
\end{equation}
and the normalised $x$-PDMP stationary probability density function then becomes
\begin{equation}
    \rho = \frac{\Gamma\left(\lambda+\mu\right)}{\Gamma\left(\lambda\right)\Gamma\left(\mu\right)}x^{\lambda-1}(1-x)^{\mu-1},
    \label{SuppEq:xPDMP}
\end{equation}
which corresponds to the well-known {\it beta distribution} with \[\lambda\equiv\frac{\nu_+e^s-\nu_-}{e^s-1}\text{~~and~~}\mu\equiv\frac{\nu_-e^s-\nu_+}{e^s-1}.\] Under the change of variables \(\nu_\pm=\nu_{T}\left(1\mp\delta_{T}\right)\), and after some algebra, the exponents read \[\lambda=\nu_{T}\left(1-\delta_{T}~\text{coth}\left(\frac{s}{2}\right)\right)\text{~~and~~}\mu=\nu_{T}\left(1+\delta_{T}~\text{coth}\left(\frac{s}{2}\right)\right),\] giving the probability density Eq.~\eqref{SuppEq:xPDMP} in terms of the environmental parameters and the selection bias. Fig.~\ref{fig:xPDMPdistributions} shows the predictions of Eq.~\eqref{SuppEq:xPDMP} and its excellent match with simulation data.

As mentioned in Sec.~V~A, a complementary characterisation of the coexistence phase is provided by the modal value, denoted by $\hat{x}$, of the PDF of the $x$-PDMP, derived in Eq.~\eqref{SuppEq:xPDMP}. As the $x$-PDMP density is a beta distribution, its modal value is
\begin{equation}
    \hat{x} = \frac{\lambda-1}{\lambda+\mu-2}=\frac{1}{2}\left(1-\frac{\nu_T}{\nu_T-1}\delta_T\coth\left(\frac{s}{2}\right)\right).
    \label{SuppEq:modalxPDMP}
\end{equation}
Note that the $x$-PDMP distribution is unimodal only for $\lambda,\mu>1$, i.e. $\nu_T(1-\left|\delta_T\right|)>1$. However, we use $\hat{x}$ only as an analytical proxy for the expected value of $x$ in the coexistence and fixation-coexistence crossover regions (coexistence probability $0<\eta\leq1$), which coincide with the unimodal regime. For pure coexistence $\eta\rightarrow1$, observed at $\nu_T\rightarrow\infty$ and $\delta_T\neq\pm1$, the modal $\hat{x}$ value reduces to the mean field expression $x^*$ of Eq.~(10).

As shown in Fig.~\ref{fig:avgx-with-xmode}, the modal value $\hat{x}$ approximates the unconditional expected $R$ fraction $\langle x\rangle$ (red crosses) in the regime $\nu_T>1/(1-\left|\delta_T\right|)$, where there is non-zero coexistence probability ($0<\eta<1$). 

In  the fixation regime where $\nu_T<1/(1-\left|\delta_T\right|)$, we find a higher (lower) fixation probability for the resistant strain than for the sensitive one when $\delta_T$ is negative (positive), with the fraction of $R$ (hence $\phi$) approaching one (Fig.~\ref{fig:avgx-with-xmode}, left) or zero (Fig.~\ref{fig:avgx-with-xmode}, right) when $\nu_T\approx 1$. 
The rationale is that, for very small $\nu_T$, the strain that fixates is set by the initial toxin state, as we expect no toxin switches before fixation has occurred; the probability to start at $\xi_T=\pm1$ is $(1\pm\delta_T)/2$. When $\nu_T$ is increased further the system experiences both environmental states, and the 
the toxin bias $\delta_T$ sets which strain is more likely to fixate. For larger $\nu_T>1/(1-\left|\delta_T\right|)$, coexistence takes over as the result of the self-averaging of the transition rates over the stationary $\xi_T$ distribution.

\begin{figure}
    \centering
    \includegraphics[width=\linewidth]{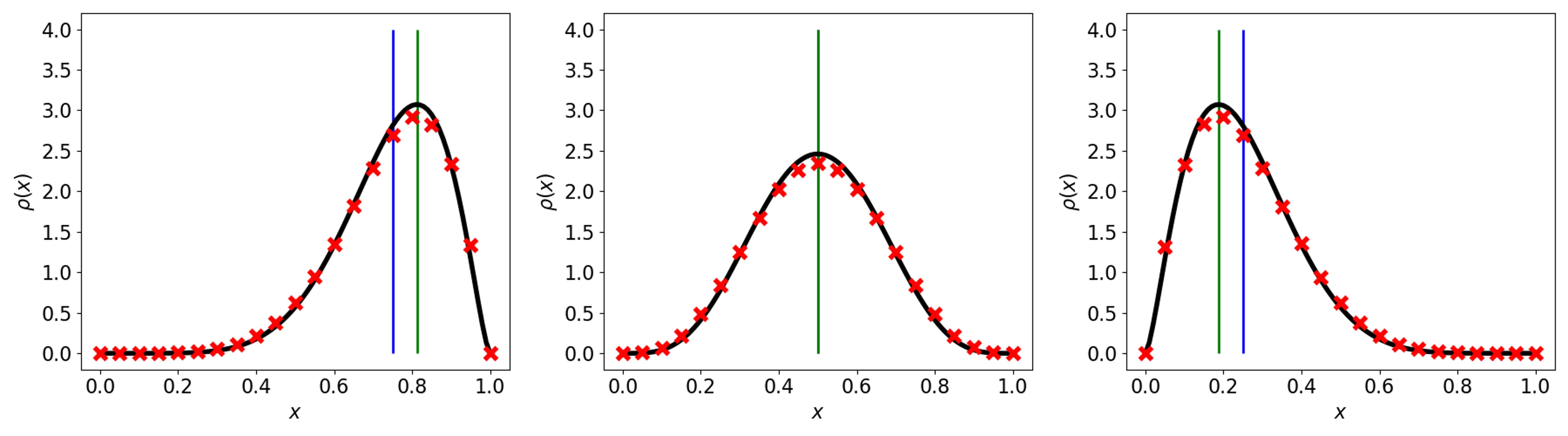}
    \caption{Predictions of $x$-PDMP stationary density $\rho(x)$ against $x$ for $\delta_T\in\{-0.5, 0.0, 0.5\}$ from left to right. Theoretical predictions (solid lines) from Eq.~\eqref{SuppEq:xPDMP} are in excellent agreement with simulation results ($\times$). Blue vertical line shows predicted mean $x^*$ from Eq.~(10) and green vertical line shows predicted mode $\hat{x}$ from Eq.~\eqref{SuppEq:modalxPDMP}. Parameters used are $s=10$, $\nu_T=5$, and fixed $K_0=100$. Simulation results have been averaged over $10^3$ realisations.}
    \label{fig:xPDMPdistributions}
\end{figure}
\begin{figure}
    \centering
    \includegraphics[width=0.49\linewidth]{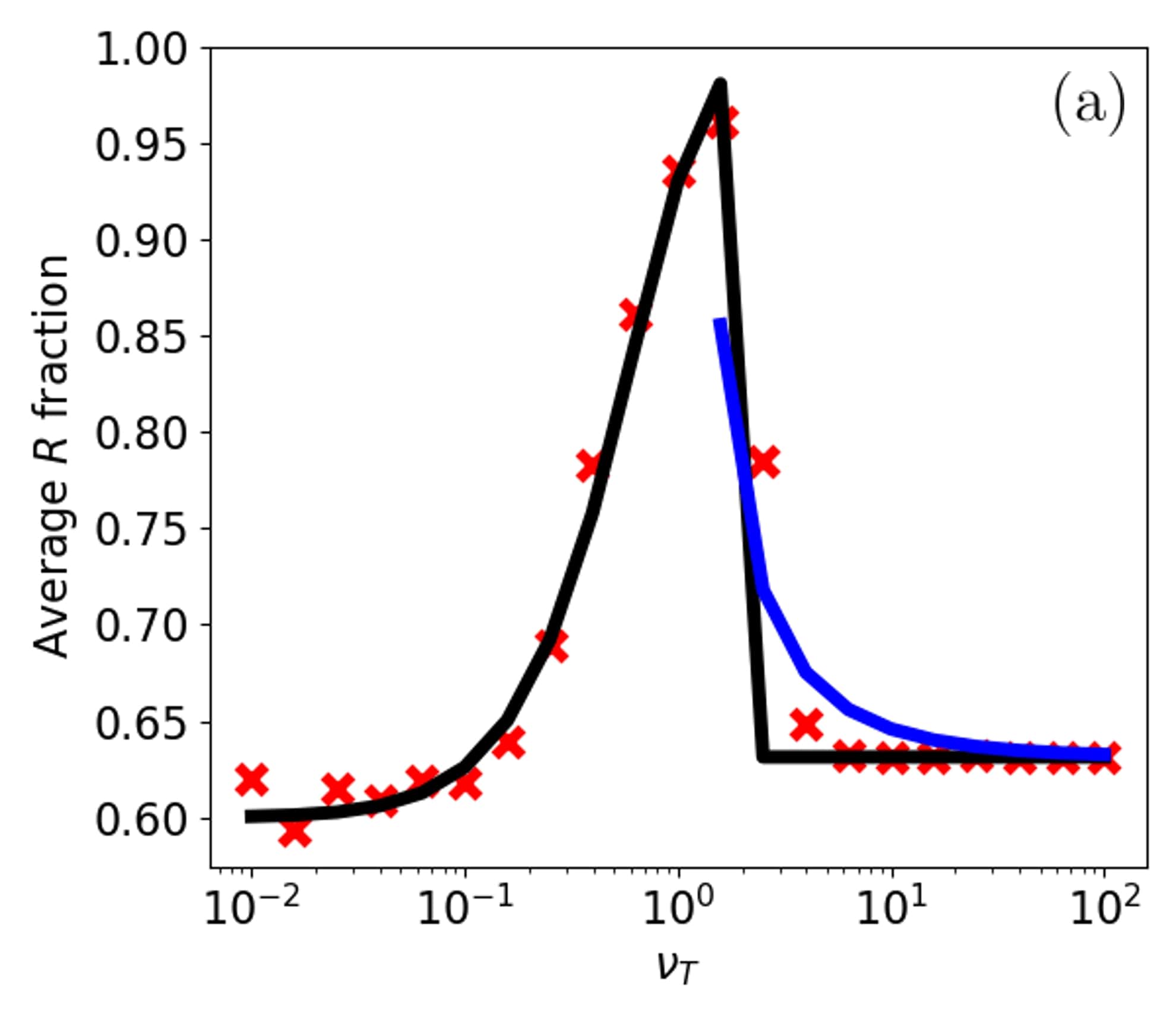}
    \includegraphics[width=0.49\linewidth]{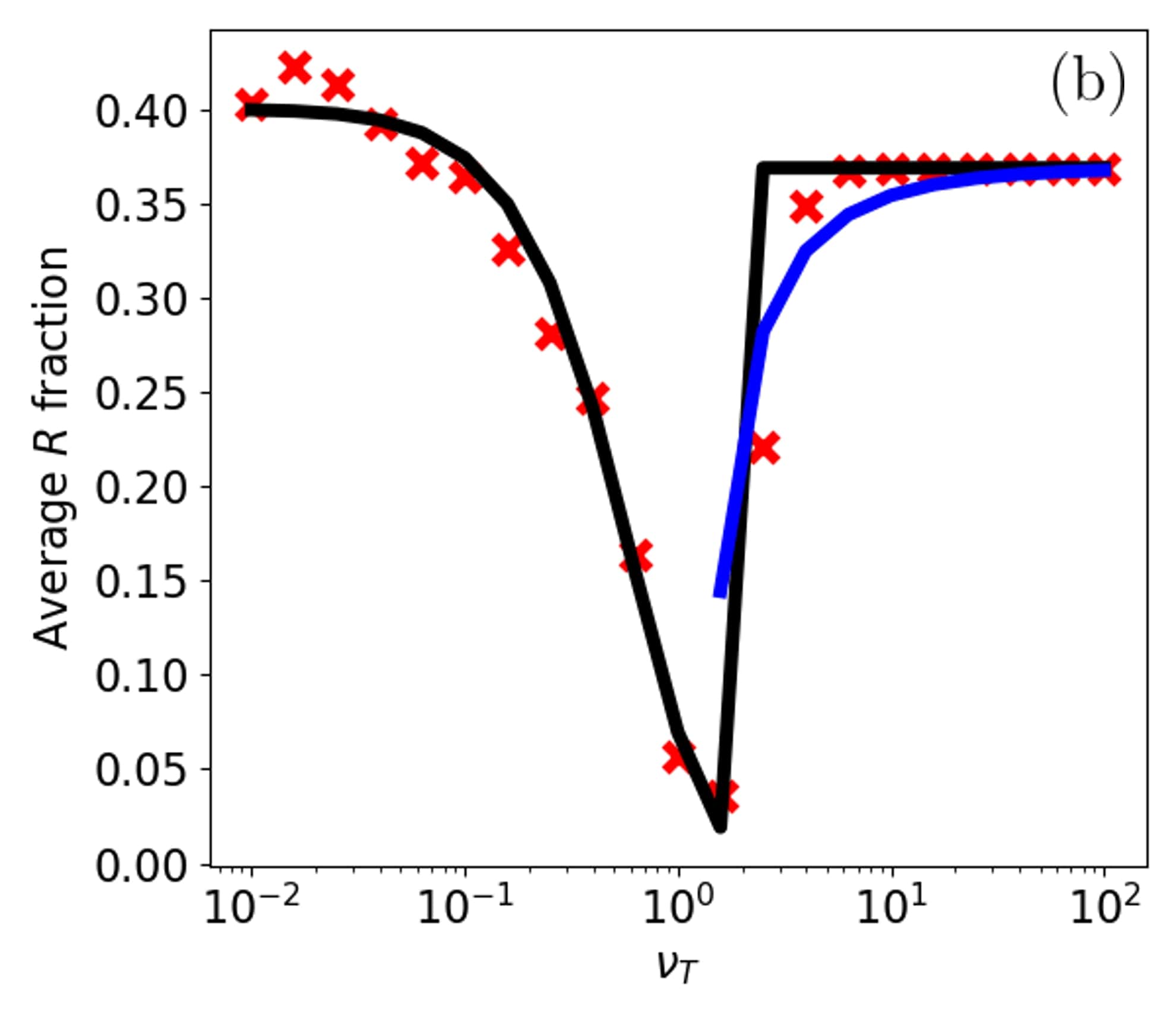}
    \caption{Average $R$ fraction unconditional on coexistence against $\nu_T$ for $\delta_T\in\{-0.2, 0.2\}$ from left to right. Black line $\avg{x}$ from Eq.~(22), blue line from $\hat{x}$ Eq.~\eqref{SuppEq:modalxPDMP}, and simulation results ($\times$). Simulation results obtained from a ``horizontal cut'' at fixed $\delta_T$ of Fig.~7, where we now count the overall fraction of $R$ (no longer conditioned to coexistence). There is a nontrivial dependence on $\nu_T$, matching that which is seen in Fig.~8(b) where we find peaks of fixation / extinction. Other parameters are: $K_0=500$, $\gamma=0.5$, $\nu_K=10$, $\delta_K=0.0$, and $s=2$.}
    \label{fig:avgx-with-xmode}
\end{figure}
\section*{SM6. Coexistence under $K$-EV and final fixation}
\label{SM6}
In Fig.~5(a-c) we find  that the region of coexistence  grows when $\nu_K$ is increased.
This behaviour is expected for an effective population size that would increase with $\nu_K$, as suggested by the MFT that increase with the population size (Fig.~2(c)). However, this seems at odds
the average population size $\avg{N}$ decreasing with $\nu_K$ as shown by 
Fig.~4.
A more suitable characterisation of the influence of $\nu_K$ on coexistence phase is thus provided by the modal value $\hat{N}$
of the $N$-PDMP PDF, given by Eq.~(21). Indeed, Fig.~\ref{fig:modalNvsnuK} illustrates that $\hat{N}$, unlike $\langle N\rangle$, 
increases $\nu_K$ for $\delta_K<\gamma$ in line with the results  reported in Fig.~5(a-c).

For a complete characterisation of  the coexistence regime we now briefly discuss the final state that is attained after after a long transient ensuring that a large fluctuation drive the system from the long-lived metastable coexistence to one of the two absorbing states where a single strain takes over the entire population~\cite{MA10,AM10}. Fig.~\ref{fig:fixationincoexistence} illustrates the full fixation outcome diagram (under no $K$-EV for simplicity), even in the phase of  long-lived coexistence. Fixation in this regime is determined by the sign of $\delta_T$, with a sharp transition at $\delta_T=0$. This is because fixation occurs most likely in the absorbing state  closest to the coexistence equilibrium $x^*$ given by Eq.~(10): here, 
the toxin bias 
 eventually imposes the fixation of $S$ ($\delta_T>0$) or $R$ ($\delta_T<0$).

\begin{figure}
    \centering
    \includegraphics[width=0.5\linewidth]{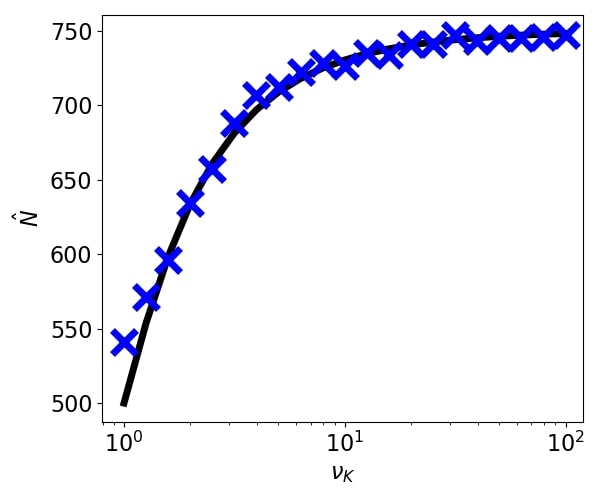}
    \caption{Modal value $\hat{N}$ of the PDF Eq.~(17) against $\nu_K$ using Eq.~(21) (solid line) and points from simulations averaged over $10^2$ realisations ($\times$) obtained by tracking the modal value of the QPSD. Parameters used are $K_0=1000$, $\gamma=0.5$, $\delta_K=0.0$, $\nu_T=10.0$, $\delta_T=0.0$. As discussed in Sec.~IV~B, we find that for $\delta_K < \gamma$ there is an increase of $\hat{N}$ with $\nu_K$ up to $\mathcal{K}=750$ (fast-switching limit), thus allowing for a greater range of $\delta_T$ values that give coexistence.}
    \label{fig:modalNvsnuK}
\end{figure}
\begin{figure}
    \centering
    \includegraphics[width=0.5\linewidth]{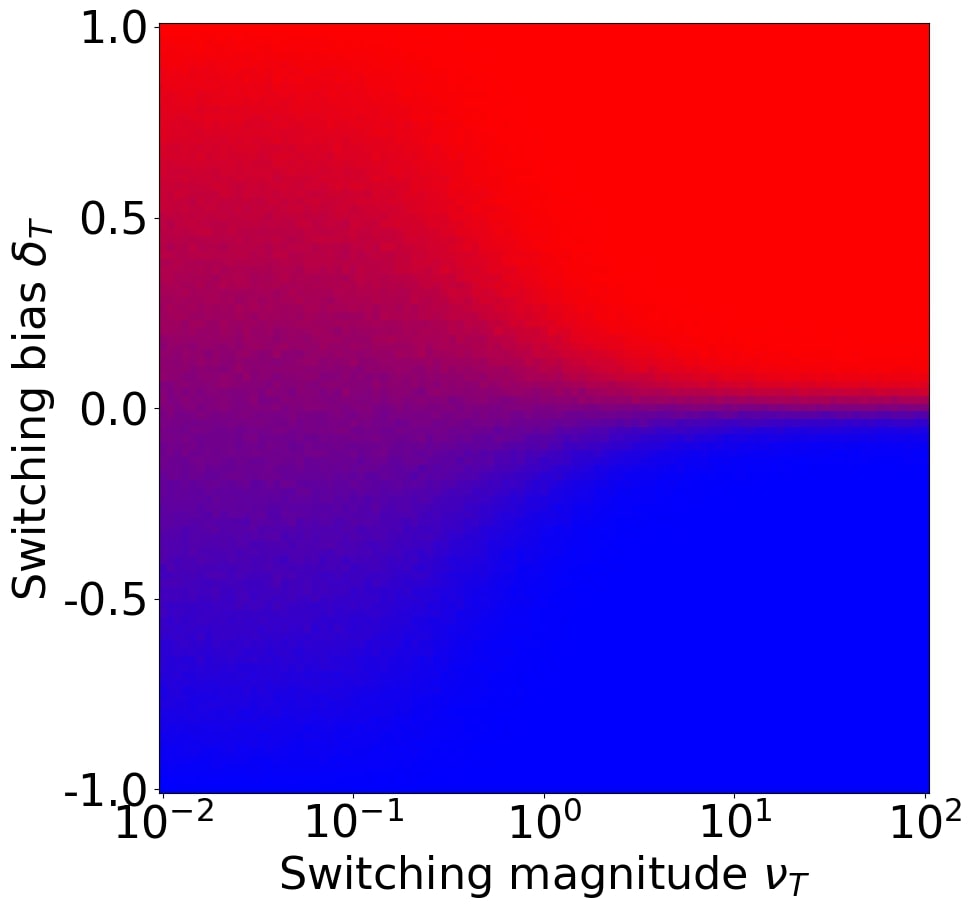}
    \caption{Fixation in coexistence region. The system fixates within the coexistence region at time $t\approx400=8\avg{N}$ in the most extreme case. The corresponding fixation-coexistence diagram is given by Fig.~3(b). Parameters used: $K_0=50$, $\gamma=0.0$ (no $K$-EV), and $s=1.0$. 
    {Color (greyscale) coding is the same as in Fig.~3.}
    }
    \label{fig:fixationincoexistence}
\end{figure}
\section*{SM7. Videos of sample paths}
\label{SM7} 
\begin{figure}
    \centering
    \includegraphics[width=0.49\linewidth]{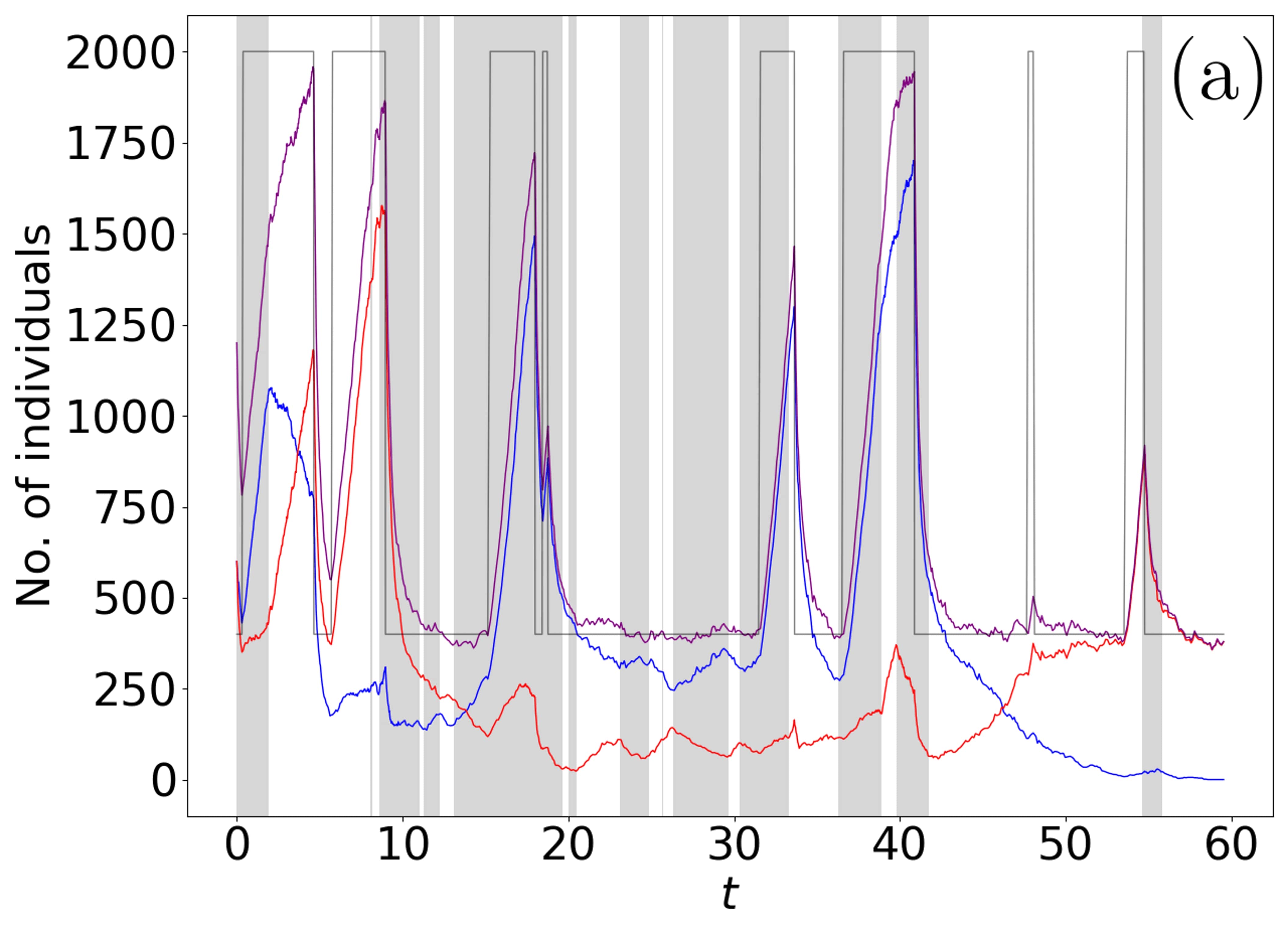}
    \includegraphics[width=0.49\linewidth]{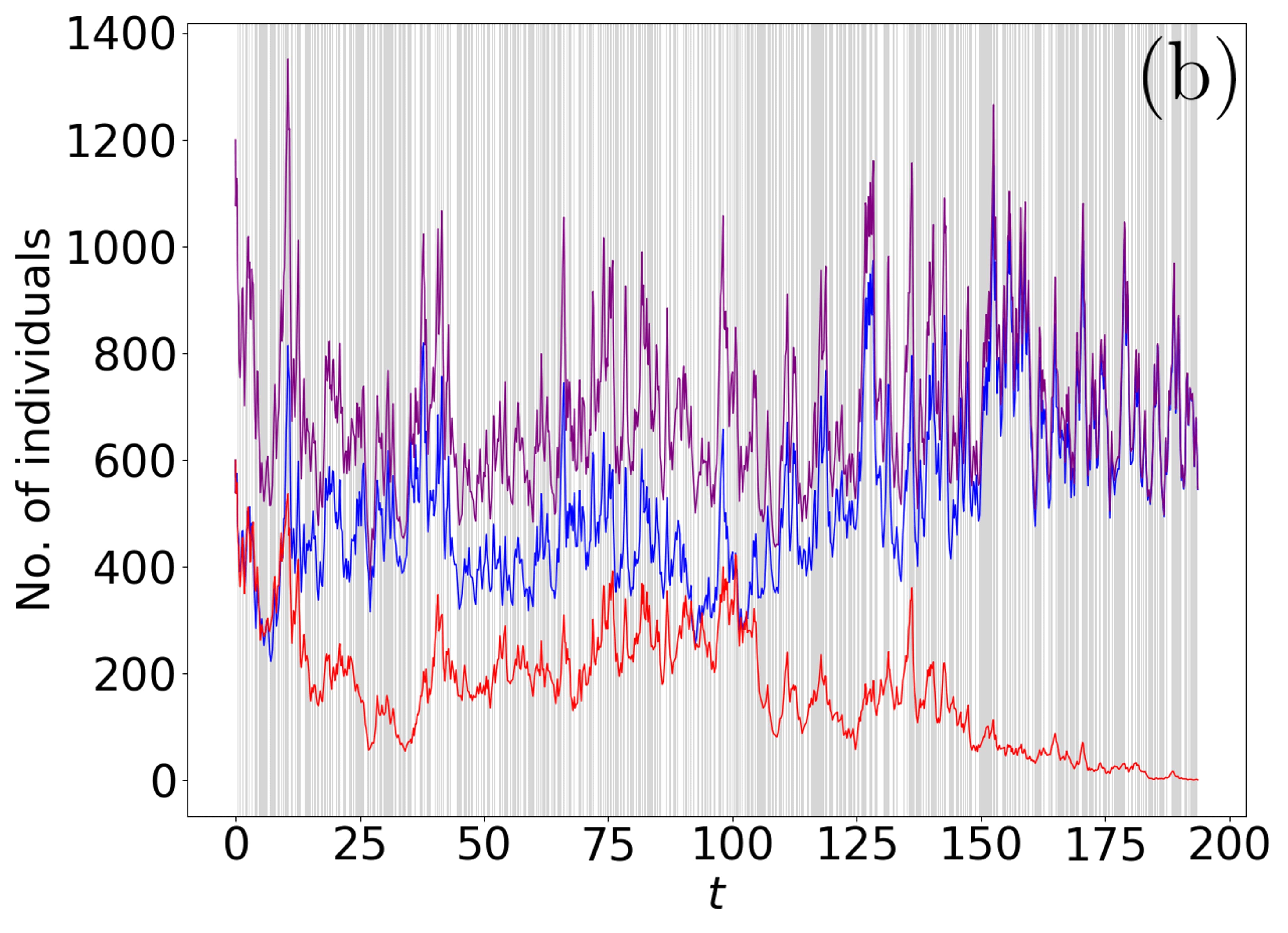}
    \includegraphics[width=0.49\linewidth]{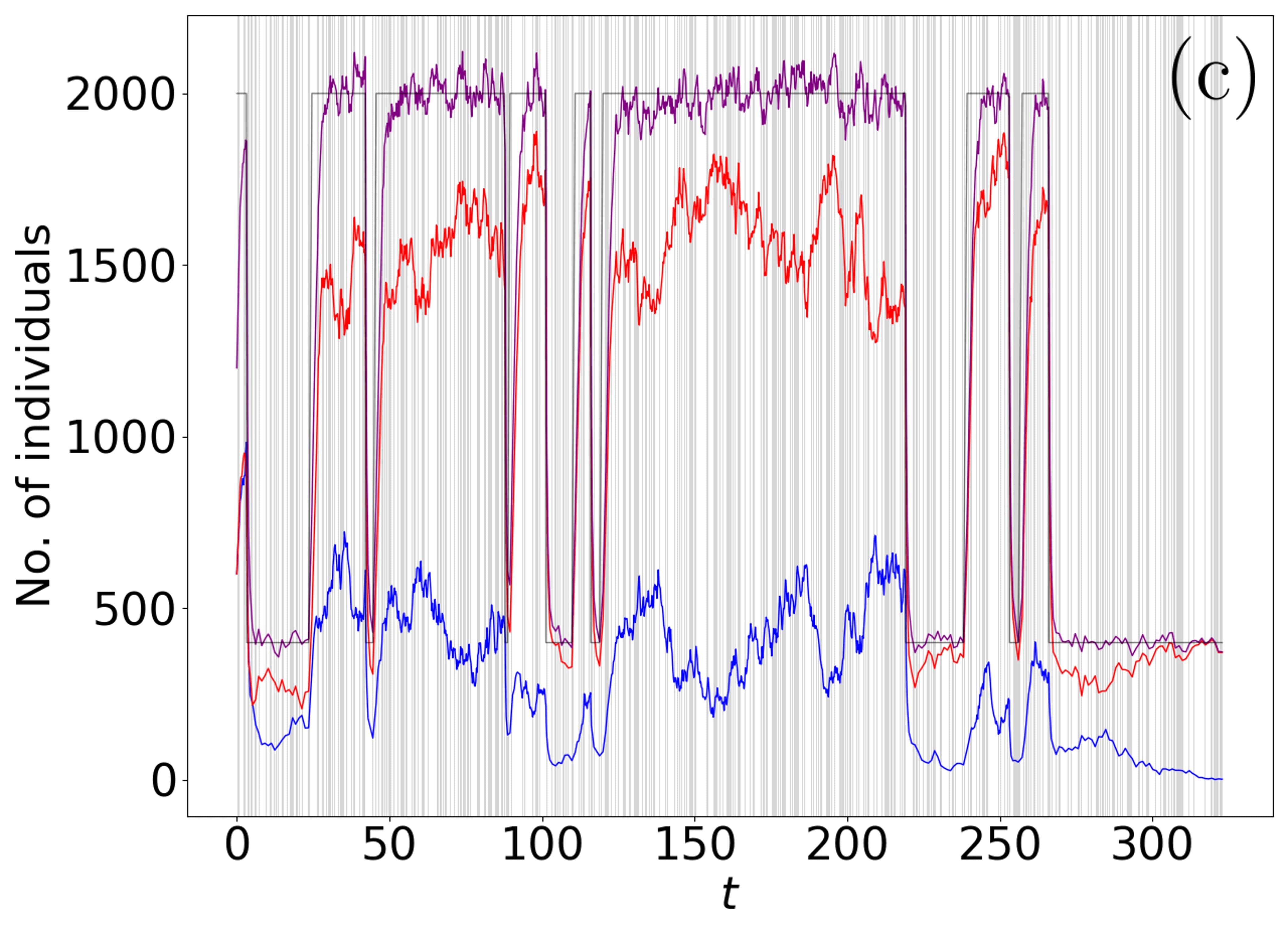}
    \includegraphics[width=0.49\linewidth]{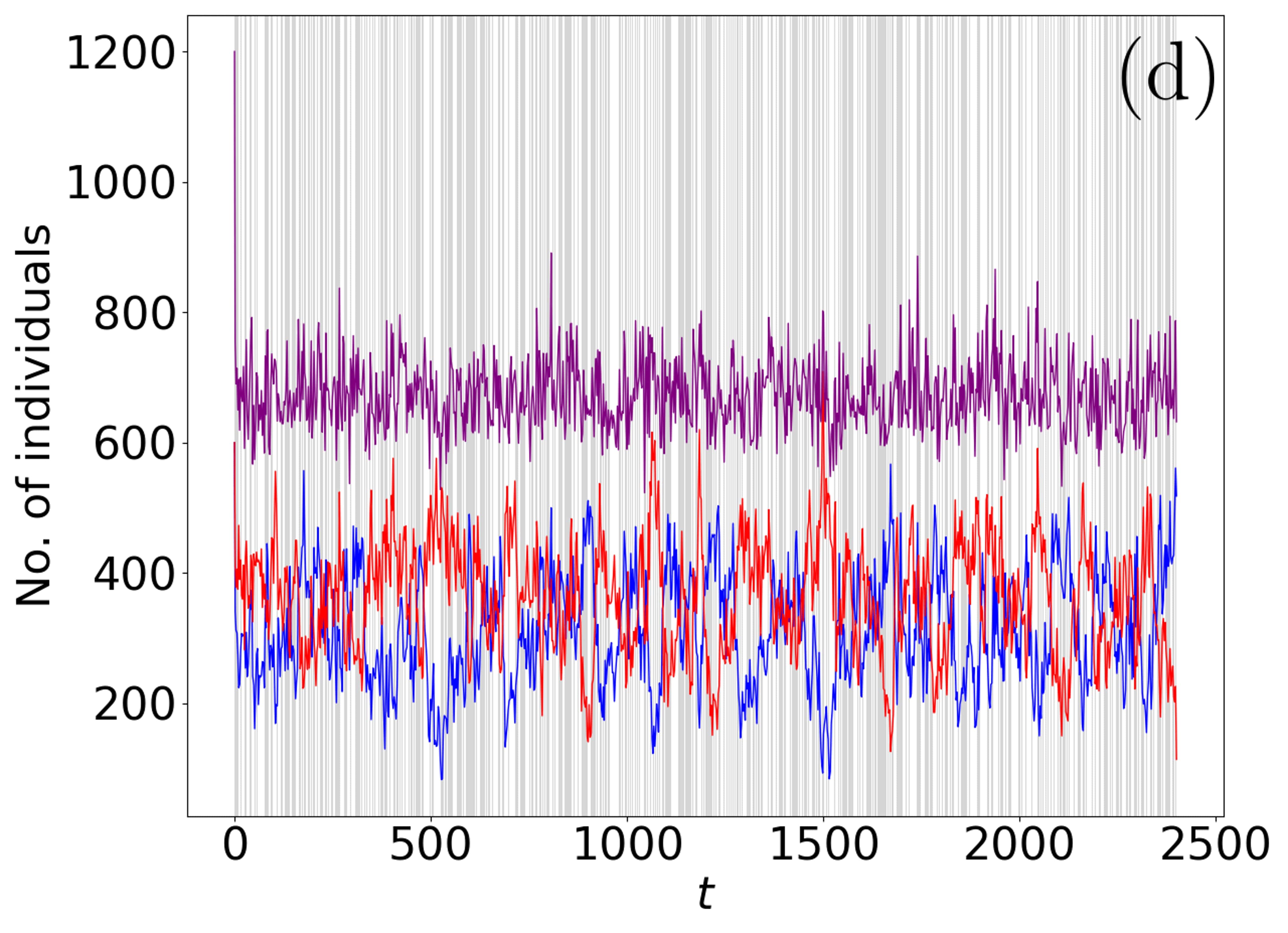}
    \includegraphics[width=0.49\linewidth]{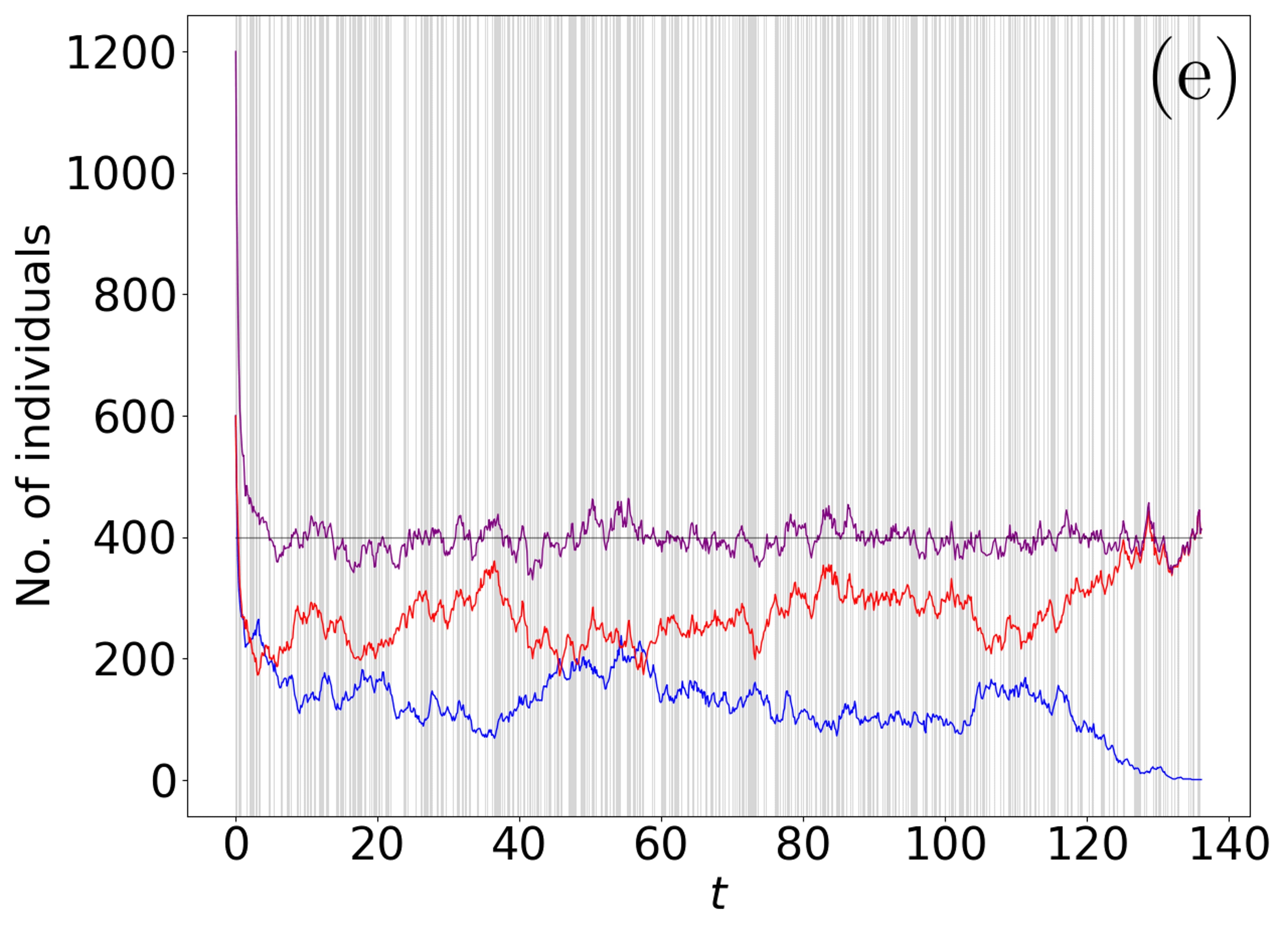}
    \includegraphics[width=0.49\linewidth]{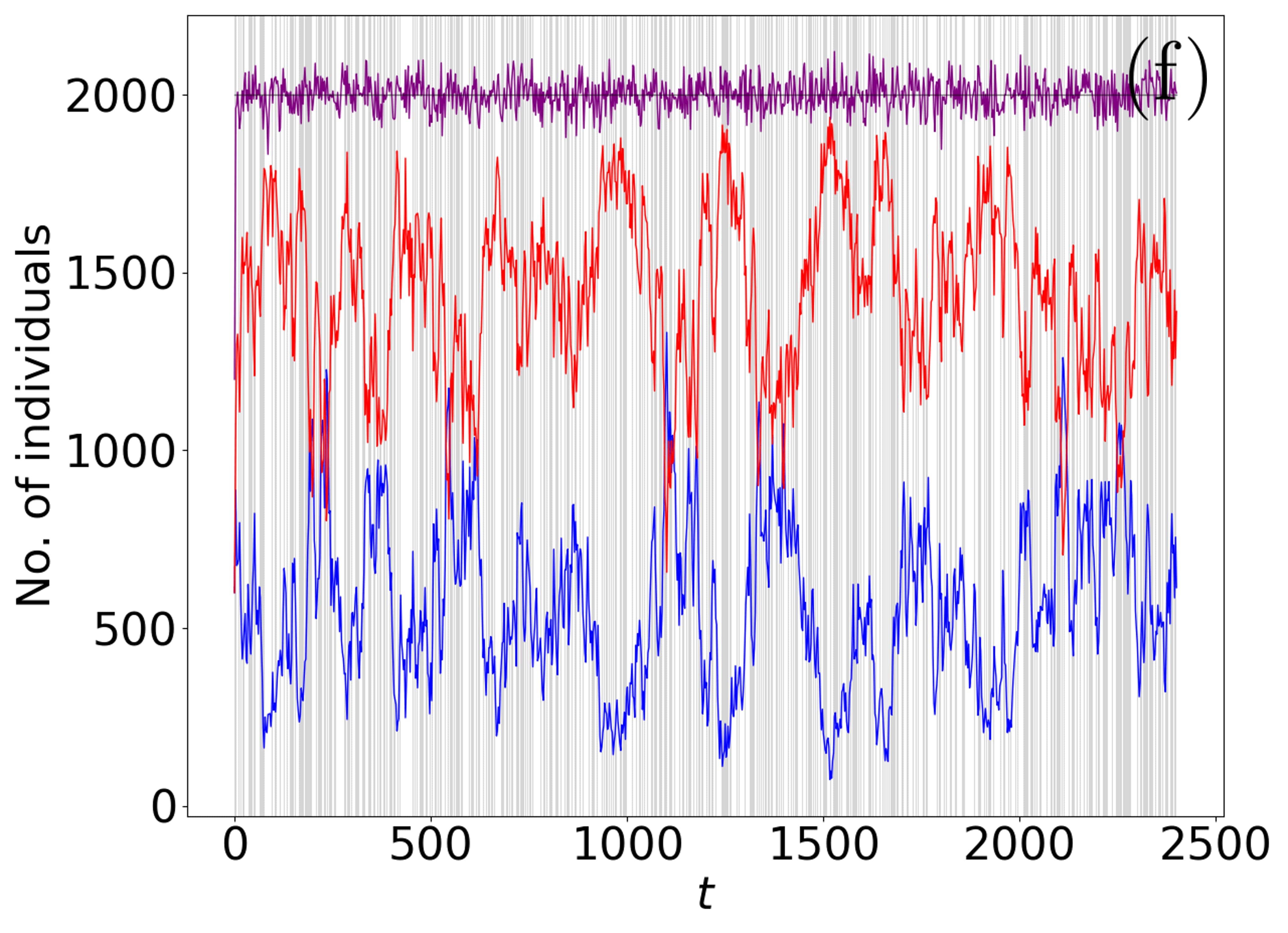}
    \caption{
    Typical sample paths of $N, N_R$ and $N_S$ with colours magenta, blue, and red, respectively. Carrying capacity $K(t)$ is plotted in grey except in (d) where $K$-switching is very fast. Grey background indicates high toxin ($\xi_T=-1$) while white background indicates low toxin ($\xi_T=+1)$. Parameters are given by the quadruple $(s, \nu_T, \delta_T, \nu_K)$, with $\delta_K=0.0$, $K_{-}=400$ and $K_{+}=2000$ in all panels, runs until either fixation occurs or $t=2K_0(1+\gamma\delta_K)=2400$. (a) $(0.5, 0.5, 0.0, 0.2)$, (b) $(0.5, 10.0, -0.15, 5.0)$, (c) $(0.5, 10.0, 0.15, 0.05)$, (d) $(0.5, 10.0, 0.0, 50.0)$, (e)-(f) $(0.5, 10.0, 0.1, 5\times 10^{-6})$. In (b) $R$ fixates; in (a,c,e) there is fixation of $S$; in (d,f) there is long-lived coexistence; see text for details.}
    \label{fig:samplepaths}
\end{figure}

In this section we provide example trajectories for representative realisations of the full model under both fluctuations in the toxin level and in the carrying capacity. The corresponding movies are provided online in \cite{videos}.

Fig.~S5 captures the dynamics of the system in six example realisations under different EV parameters $(\nu_T,\delta_T,\nu_K)$  for a selection strength $s=0.5$, which sets the magnitude of $T$-EV; and mean carrying capacity $K_0=1200$ with $\gamma=2/3$ ($K_{-}=400$ and $K_{+}=2000$), setting the magnitude of $K$-EV. For simplicity, we keep $\delta_K=0$.

- The example of Fig.~S5(a) 
illustrates an unbiased fluctuating environment with intermediate $T$-switching rate
 (white background: low toxin level, grey: high toxin level; $\nu_T=0.5$, $\delta_T=0$) and $K$-switching rate (solid black line; $\nu_K=0.2$). We notice  that total population $N$ rapidly attains  quasi-stationarity well before a fixation event  occurs ($S$ fixation in this example).
 
- In Fig.~S5(b), we show faster environmental $T$ and $K$ switching, with $\nu_T=10$ and $\nu_K=5$. There is also a bias towards the harsh $T$ state favouring the strain $R$ ($\delta_T=-0.15$, high toxin level)
that is responsible for a robust offset in strain abundance eventually leading to 
extinction of $S$ and fixation of $R$.

In this example, the  $K$-EV switching rate is sufficiently high to ensure its self-averaging: in this fast $K$ switching regime, the population size is distributed about the effective carrying capacity  $\mathcal{K}= K_0(1-\gamma^2)/(1-\gamma\delta_K)$ (here $N\approx \mathcal{K}=667$), see  Fig.~4 and Sec.~IV~A.

- Fig.~S5(c) illustrates the case of fast $T$-EV ($\nu_T=10$)
with a bias towards the mild/low $T$ state favouring the strain $S$ ($\delta_T=0.15$). In this example, the $K$-EV switching rate is low ($\nu_K=0.05$)
and the population size $N$ follows $K(t)$ and fluctuates about $K_-$ and $K_+$.
The $T$-EV bias ($\delta_T>0$) is here responsible for a systematic
offset in the strain abundance, with $N_S>N_R$, resulting in the fixation of $S$ and extinction of $R$, see Fig.~2(a). In this example, $K$-EV is responsible for larger demographic fluctuations in the environmental state $\xi_K=-1$, where $N\approx K_-$ and $S$ fixation is more likely than when $N\approx K_+$ ($\xi_K=+1$).

- Fig.~S5(d) shows typical trajectories in the long-lived coexistence phase  in the case of unbiased fast switching $T$-EV and $K$-EV, with $\nu_T=10, \delta_T=0$ and $\nu_K=50$. This set of parameters essentially corresponds (here $\nu_K=50$ instead of $\nu_K=5$) to a point in the bright green region in the diagram of Fig.~5(c), where $\eta\approx 1$
and long-lived coexistence is almost certain.
As in panel (b), the fast $K$ switching (solid black line not shown in Fig.~S5(d))
leads to fluctuations of the total population size about ${\cal K}$, i.e.
$N\approx \mathcal{K}=667$. In this example $x^*=1/2$, see  Fig.~7, and the number of $R$ and $S$ cells fluctuates about their averages: $\avg{N_R}\approx \avg{N_S}\approx {\cal K}/2\approx 333.5$, see Sec.~V~B.

- Fig.~S5(e,f) illustrate the 
dynamics under fast $T$-EV ($\nu_T=10$) and extremely slow $K$-switching rate  ($\nu_K=5\times10^{-6}$), with an initial carrying capacity $K(0)=K_-$ in (e) and $K(0)=K_+$ (f). There is also a small bias towards the mild $T$ state favouring $S$ ($\delta_T=0.1$). This choice of parameters corresponds to a point in the faint green region in the diagram of Fig.~5(a), where $0<\eta<1$
and long-lived coexistence is possible but not certain. In panel (e), the population size fluctuates about $K(0)=K_-$ ($N\approx K_-$) and is not able to sustain long-lived coexistence: demographic fluctuations yield $S$ fixation in a time $t\lesssim 2K_-=800$. In panel (f), we have $N\approx K_+$ and, as demographic fluctuations are unable to cause the fixation/extinction of either strain in a time less than $2\avg{N}=2K_+=4000$ (for clarity, the time series in Fig.~S5 have been truncated at $t=2400$), we have long-lived coexistence.

In all examples in the panels of Fig.~S5,
the population size $N$ reaches quasi-stationarity (settles in the QPSD) a significant time before the
fixation of one strain and extinction of the other, in line with the considerations underpinning Eqs.~(19) and~(20).

\section*{SM8. Fully correlated and anti-correlated environmental variability}
\label{SM8} 
\begin{figure}
    \centering
    \includegraphics[width=0.95\linewidth]{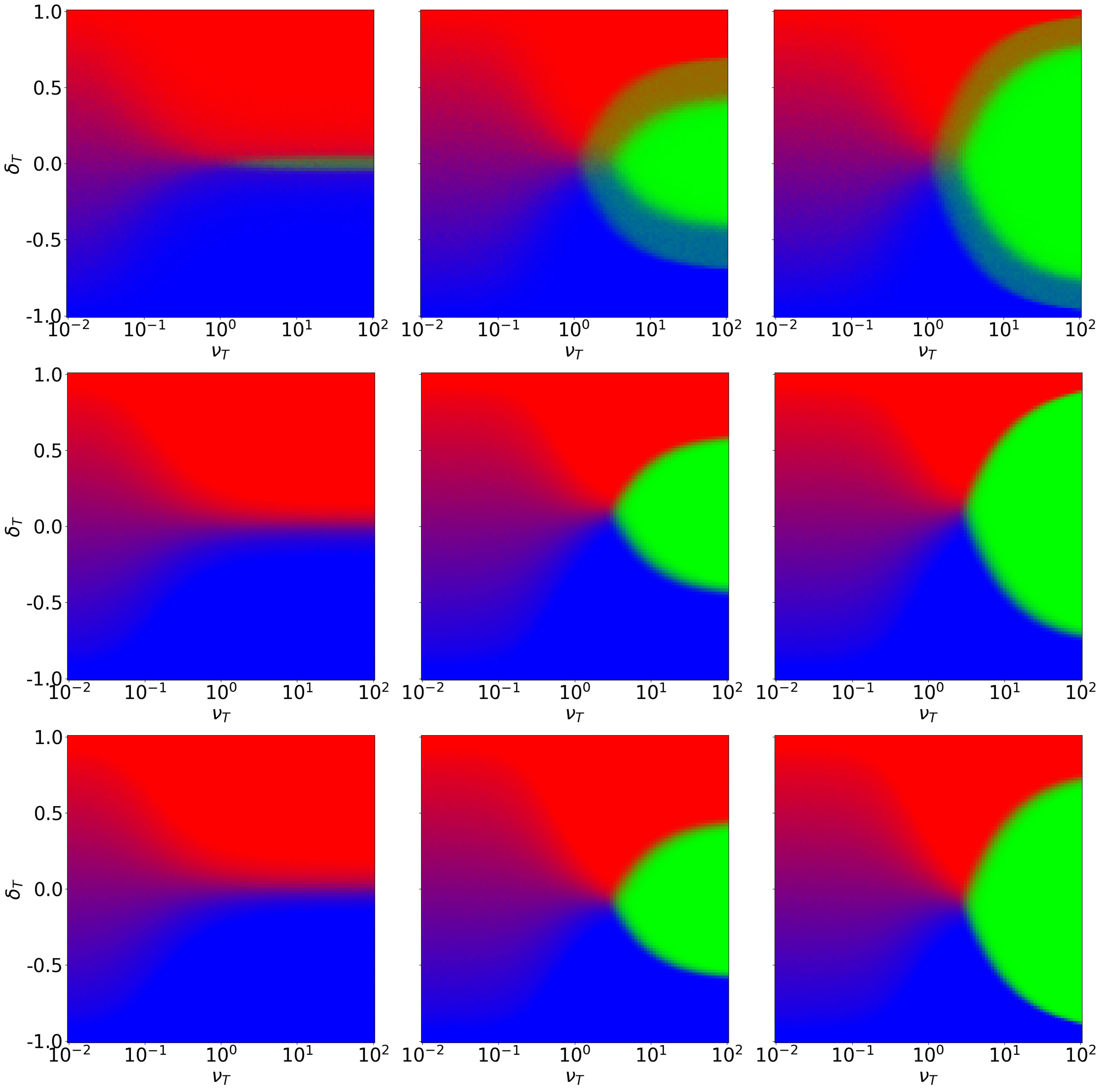}
    \caption{Top to bottom: uncorrelated, fully correlated and fully anticorrelated case heatmap plots from $10^3$ realisations with $\gamma=0.9$ (correlated: $\widetilde{\gamma}=0.9$, anticorrelated: $\widetilde{\gamma}=-0.9$). Left to right we have $s\in\{0.2, 2, 20\}$ and use $K_0=1000$. In the uncorrelated case we have $\nu_K=0.0001$ and $\delta_K=0.0$. Considering the correlated case, in the coexistence region, for $\delta\to+1$ and fast-switching $\nu\gtrsim 1$ the effective carrying capacity increases. Therefore, we see the coexistence region extended further towards $\delta=1$. For $\delta\to-1$ and $\nu\gtrsim 1$ we find the converse where the effective carrying capacity decreases and thus the coexistence region shrinks away from $\delta=-1$. A similar logic applies to the anticorrelated case; see text. {Color (greyscale) coding is the same as in Fig.~3.}}
    \label{fig:correlatedcomparison}
\end{figure}
In the case of fully correlated/anti-correlated $T$-EV and $K$-EV, 
environmental variability is no longer twofold since 
we have $\xi\equiv\xi_T$ and $\xi_K=\xi$ (fully correlated  EV) or $\xi_K=-\xi$ (fully anti-correlated  EV). The switching carrying capacity can thus be written as \[K(t)=K_0[1+\widetilde{\gamma} \xi(t)],\] where $\widetilde{\gamma}=\gamma$ in the fully correlated case, and $\widetilde{\gamma}=-\gamma$ when $T$-EV and $K$-EV are fully anti-correlated. For instance, this implies that, in the correlated case, the environmental state $\xi=+1$ corresponds to $f_s=e^{s}>1$ and $K=K_+$ (low toxin level, abundant resources), while $\xi=-1$ is associated with $f_S=e^{-s}<1$ and $K=K_-$ (high toxin level, scarce resources). As said, under fully correlated/anti-correlated $T/K$-EV, environmental variability is no longer twofold: $\xi$ simultaneously drives the level of toxin and the abundance of resources. Hence, we can characterise the effect of fully correlated/anti-correlated EV in terms of $\nu\equiv \nu_T$ and $\delta\equiv \delta_T$, and the fully anti-correlated case is related to completely correlated EV via $\widetilde{\gamma}\to -\widetilde{\gamma}$. An example of fully-anticorrelated EV modelled in terms of a dichotomous process driving the level of toxins and resources in the context of competitive exclusion is considered in~\cite{shibasaki_exclusion_2021}.

Fig.~\ref{fig:correlatedcomparison} shows the comparison between the uncorrelated $T$-EV and $K$-EV studied in the main text (top row), and the fully correlated (middle row) and fully anti-correlated (bottom row) cases reported here, all under $K_0=1000$ and $\gamma=0.9$, and for different selection strengths $s\in\{0.2,2,20\}$ (left to right columns). For the uncorrelated case,  the parameters of $K$-EV are independent from those of $T$-EV parameters,
and are here chosen to be $(\nu_K,\delta_K)=(10^{-4},0)$, i.e. unbiased slow-switching $K$-EV (similar to the example of Fig.~5(a)). 

Since the fully correlated and anti-correlated cases (middle and bottom rows)
are mirror images through the {horizontal axis (symmetric with respect to $\delta_T=0$)}
and  under a red-blue colour change, we focus on the correlated case only (middle row). For this case, $\widetilde{\gamma}=\gamma=0.9$, with $\xi_K=\xi_T\equiv\xi$, $\nu_K=\nu_T$, and $\delta_K=\delta_T$. In the fully correlated case, we observe that both the blue (resistant fixation) and the bright green (coexistence) regions shift upwards (to higher toxin level biases $\delta_T$) as the selection strength increases (from left to right column). We can understand this phenomenon in light of the $T$-correlated $K$-EV fluctuations. This is, since $\delta_K=\delta_T$, a lower value of $\delta_T$ in the diagrams implies longer cumulative periods in the harsh toxin level, but also in the low carrying capacity environment. Therefore, lower $\delta_T$ provides the selective advantage to resistant strain at the same time that it shrinks the total population. Demographic fluctuations being stronger in smaller populations, correlated $T$-EV and $K$-EV thus provide higher $R$ fixation probability (blue region shifted up), as well as lower coexistence probability (green region shifted up). Moreover, since total population increases with the bias towards positive carrying (here $\delta_K=\delta_T$), see Sec.~IV~A,
the MFT thus increases with $\delta_T$ (see Fig.~2(c)), and the coexistence probability thus shifts upwards. The magnitude of the upwards shift in the correlated case, is  small but increases with selection strength $s$ that  increases the amplitude of the $T$-EV fluctuations. 

In summary, we obtain the same qualitative results for fully correlated/anti-correlated $T/K$-EV as when $\xi_{T/K}$ are independent (uncorrelated environmental noise, twofold EV), with some minor quantitative differences, as shown in Fig.~\ref{fig:correlatedcomparison} and discussed above. We conclude that the similar behaviour observed for uncorrelated and (anti-)correlated $T/K$-EV indicates that our findings are robust against the detailed model specifications: the results are expected to be  valid for the general case of twofold environmental variability where $T/K$-EV are neither completely independent nor fully correlated/anti-correlated.
\end{document}